\documentclass[11pt,a4paper,onecolumn,twoside,openany]{book}

\usepackage[latin1]{inputenc}
\usepackage{latexsym}
\usepackage{amsmath}
\usepackage{exscale}
\usepackage{flafter}
\usepackage[nonstop]{feynmf}
\usepackage{hhline}
\usepackage{amssymb}
\usepackage{dcolumn}
\usepackage{hhline}
\usepackage{epsfig}
\usepackage{rotating}
\usepackage{array}
\usepackage{tabularx}
\usepackage{units}
\usepackage{nicefrac}
\usepackage{axodraw}

\newcommand{\tr}{\mathrm{Tr}}
\begin{document}\setlength{\unitlength}{1mm}
\thispagestyle{empty}
\begin{center}
  {\bf\Large B-Meson Properties from Modified Sum Rule Analyses}\\[1.5cm]
{\bf\large Tobias Kleinschmidt}\\[0.5cm]
{Institut f\"ur Theoretische Physik}\\
{Philosophenweg 16}\\
{69120 Heidelberg}\\
{Germany}\\[1.5cm]
{\bf\large Abstract}\\[0.5cm]
\end{center}
We introduce two approaches to analyses of Borel sum rules. In the
first method, we analyze the sum rules in the limit
$M^2\rightarrow \infty$. Here, the sum rules become very sensitive
to the chosen threshold $s_0$. We fix the threshold by setting the
daughter sum rule equal to the meson mass in this limit. The
second method introduces a Borel mass dependent threshold
$s_0(M^2)$. We choose functions $s_0(M^2)$ such, that the
corresponding sum rule is not dependent on $M^2$ anymore. The
relevant hadronic parameter is extracted at the most stable
function $s_0(M^2)$. The two methods differ in the notion of
quark-hadron duality. Whereas the first method emphasizes errors
coming from the simple duality ansatz, the second method is
constructed such as to extract hadronic properties from the region
where the duality approximation is fulfilled best. We use these
modifications of the sum rule approach to extract values for the
B-meson decay constant $f_B$, the semi-leptonic form factor
$f^+(0)$ at zero momentum transfer and the strong coupling
$g^{}_{B^*\!B\pi}$. The results coincide with other sum rule
analyses.

\pagenumbering{roman}
\tableofcontents
\newpage
\pagenumbering{arabic}
\setcounter{chapter}{0}

\chapter{Introduction}

In 1979 M. Shifman, A. Vainshtein and V. Zakharov proposed the so
called QCD sum rules \cite{Shifman:1979bx,Shifman:1979by} - a
method to relate hadronic properties with QCD calculations, where
non-perturbative effects are systematically included. With these
sum rules, from now on referred to as SVZ sum rules, one is able
to express static hadronic observables like masses and widths in
terms of a few universal phenomenological parameters, the vacuum
condensates of quark and gluon fields. Today, sum rules and their
various derivatives are by no means the only approach to predict
hadronic parameters. Other methods are calculations in heavy quark
effective theories and chiral perturbation theory. Compared to sum
rules they can only be applied to limited regions of hadronic
physics - systems involving heavy quarks and only light quarks,
respectively. Further methods are lattice calculations and various
quark models. The advantages and disadvantages of lattice results
are discussed below.
\newline

The basic idea of the SVZ sum rules is to represent hadrons by
currents of quark fields, that have the same quantum numbers as
the corresponding hadrons, and to construct the two-point
correlator of a current and its hermitian conjugate. This
structure is then evaluated in the euclidean region for large
spacelike momenta, where the high virtualities of the quarks
ensure that the main contribution to the two-point function can be
extracted from perturbative calculations. Corrections are added by
use of Wilson's operator product expansion (OPE), separating the
hard and soft regime of the correlation function to take large
distance dynamics into account. This gives rise to a truncated
power series in $\frac{\mu^2}{Q^2}$, where $\mu$ is the scale
introduced in the OPE to separate short and long distance
dynamics. By the use of \textit{dispersion relations} the
two-point function is then related to the hadronic spectral
density. Finally, a so called \textit{Borel transformation} is
applied to improve the convergence of the power series on one side
and suppress higher states and continuum contribution of the
hadronic spectral density on the other side. The SVZ sum rules
will be derived and discussed more intensively in section
\ref{SVZ} of this thesis.
\newline

Over the last 25 years, the initial SVZ sum rules have further
been developed and improved. They have been applied to three-point
functions, baryonic currents and external fields and were used to
determine the masses of the light quarks, decay constants, form
factors, couplings and magnetic moments of mesons and baryons.
They were extended to investigate properties of hadronic matter at
high temperatures and densities and many more systems (see
\cite{Radyushkin:1998du,Colangelo:2000dp} for some examples and
applications of various sum rules). Besides the 2-point SVZ sum
rules, we will make use of the so called light-cone sum rules
(LCSR) in this thesis. Here, the relevant non-local matrix
elements will be parameterized in form of wave functions instead
of the expansion in local vacuum condensates. This method turned
out to be quite useful to extract hadronic properties from
reactions including three particles. We will derive the LCSR in
section \ref{sec-lcsr}.
\newline

One of the major advantage of the sum rule approach is its
flexibility. Sum rules can be applied to a variety of problems and
give reliable predictions of hadronic parameters. Compared to
lattice results, which became more and more important in the last
years, sum rules do not give very accurate results - usual
uncertainties lie in the range of 10\% to 20\%. Nowadays, lattice
results have smaller errors and will certainly achieve higher
accuracy in the near future, when unquenched calculations will be
feasible within an acceptable period of time. But until now,
lattice calculations still include relatively high masses for the
light quarks, which are related to the length of the lattice.
Thus, the results have to be extrapolated to the small masses of
the light quarks to give physical predictions, which reduces the
accuracy of lattice calculations. Furthermore, they only give
\textit{brute force} numerical results, where the fundamental
physical dynamics are hidden beyond the extracted numbers and
barely observable. On the other side, the simple analytical sum
rule method allows physical insight into the systems they are
applied to and an interpretation of the results - at the expense
of accurate numerical results. Furthermore, the uncertainties of
the results can be traced back to uncertainties of several
ingredients of the sum rules, like the error coming from the
truncation of the condensate expansion of the correlation function
and the assumption of \textit{quark-hadron duality}, which is
necessary to compensate for the scarce knowledge of the hadronic
spectral density. Another advantage of the sum rule approach is
the fact, that it can be continued to Minkowski space in an
analytical way, whereas this continuation of the numerical lattice
results obtained in euclidean space sometimes is problematic.
Thus, although they suffer from some inaccuracies, sum rules are
still an important method of calculating hadronic properties.
\newline

In this thesis we will propose two modifications to the usual sum
rule calculations of the decay constant $f_{B}$, the form factor
$f^+_B(0)$ at zero momentum transfer $q^2=0$, and the coupling
$g^{}_{B^*\!B\pi}$ of the heavy-light B-meson. To have an precise
measurement, or at least an accurate prediction of these
parameters is of high importance. The weak decay constant $f_{B}$
contributes to calculations of $B-\overline{B}$-mixing amplitudes
and is needed in determinations of the CKM-matrix element
$V_{ub}$, when it is extracted from leptonic decays of the
B-meson. It contributes as well to the sum rule calculations of
the form factor $f^+_B(q^2)$ and the coupling $g^{}_{B^*\!B\pi}$.
The form factor can also be used in predictions of the CKM
element. Knowledge of the fundamental parameter $V_{ub}$ is of
greatest importance in today's physics, since it contributes to
the question of unitarity of the CKM-matrix and thus tests the
standard model.
\newline

This thesis is structured as follows. In the next chapter, we will
introduce the basic methods and theoretical tools. Different sum
rules and their ingredients will be discussed. Furthermore, some
of the crucial points of the sum rules will be stressed, like the
dual relation of the hadronic spectral function to the
perturbative QCD calculation.
\newline

In chapter \ref{decay-constant}, we will start with briefly
reviewing the sum rule analysis for the weak decay constant $f_B$.
We will basically follow an approach from Jamin and
Lange\cite{Jamin:2001fw}, using the running b-quark mass in the
$\overline{MS}$-scheme. Compared to analyses using the pole mass
of the quark, this shows an improved convergence of the
perturbative expansion in the strong coupling $\alpha_s$. In most
sum rule approaches, the hadronic parameter is extracted at rather
ambiguous values of the squared Borel mass $M^2$ and the threshold
$s_0$. In the sections \ref{sec-u-tu-infinity} and \ref{s(u)}, we
will introduce two new methods to extract a distinct value of the
hadronic parameter.
\newline

In section \ref{sec-u-tu-infinity}, we will evaluate the sum rule
in the limit $M^2\rightarrow\infty$. This is related to an
approach proposed several times by
Radyushkin\cite{Nesterenko:1982gc,Szczepaniak:1998sa}. Hereby,
uncertainties coming from the truncation of the expansion in the
condensates essentially vanish and the threshold parameter $s_0$
can be fixed to one distinct value, thus leading to a unique
numerical result for $f_B$. The drawback of this method lies in
the emphasis of uncertainties coming from the simple duality
ansatz.
\newline

Section \ref{s(u)} introduces another possible modification. Here,
we will introduce a Borel mass dependent threshold $s_0(M^2)$
instead of a constant $s_0$. With this at hand, one can find
functions $s_0(M^2)$, for which the sum rule is constant in a
given window of the Borel mass. In this case, also the ratio of
the sum rule and its first derivative, which is often used to get
a reading point $(\hat{M}^2,\hat{s}_0)$, is constant and equal to
the squared mass of the ground state of the hadronic side. For
each value of the sum rule, we find a corresponding function
$s_0(M^2)$. We will give arguments to extract the value of the
hadronic parameter, for which the function $s_0(M^2)$ is most
stable.
\newline

In the chapters \ref{sec-Form Factor} and \ref{coupling} we will
apply the two methods to the analyses of the B-meson form factor
$f^+_B(0)$ at zero momentum transfer $q^2=0$ and the strong
coupling $g^{}_{B^*\!B\pi}$. The LCSR of these parameters are
proportional to the weak decay constant $f_B$ and, in the case of
the coupling, also to the decay constant of the $B^*$-meson
$f_{B^*}$. We expect some cancellations of the intrinsic errors of
the two methods, when the ratio with the decay constants, analyzed
within the same methods, is taken. Our extracted results are close
to the values obtained in other sum rule approaches. Chapter
\ref{Conclusion} will conclude.

\chapter{Theoretical Foundations}

Quantum Chromodynamics is described by the SU(3) Yang-Mills
Lagrangian:
\begin{equation}
\mathcal{L}_{QCD}=-\frac{1}{4}(G_{\mu\nu})^2+\sum\limits_f
\overline{\psi}_f(iD\!\!\!\!\slash-m_f)\psi_f,
\end{equation}
where $G_{\mu\nu}$ is in the adjoint representation of the gauge
group and $\psi_f$ in the fundamental - $f$ indicating the
different quark-flavors. Although $\mathcal{L}_{QCD}$ is made out
of quarks, rather than hadrons, it is believed to incorporate all
hadronic physics. However, due to confinement, individual quarks
are never observed experimentally, and thus perturbative
calculations, which start from this Lagrangian are limited.
Calculations involving the free quark propagator are clearly only
valid far off-shell, which is one of the features of the SVZ sum
rules, where the perturbative coefficients of the operator product
expansion are evaluated at large virtualities of the quarks.
\newline

To lower scales the running coupling $\alpha_s(\mu)$ increases and
the perturbative expansion breaks down. It even reaches a pole at
a scale of order $\Lambda_{QCD}\approx 200\;MeV$ and perturbative
expressions are only valid down to energies of about $1\;GeV$.
With increasing separation of two color charges, the interaction
becomes so strong, that the potential between the charges cannot
be described by the exchange of single gluons anymore. A whole
cloud of gluons will develop between the charges and one has to
describe physics at this scales with methods of (chromo-) electric
and magnetic fields. A perturbative approach is not feasible in
this region.
\newline

However, for the treatment of hadronic parameters perturbative QCD
is still applicable, since the average separations of the
color-charged partons in the hadrons are not too large. One can
treat them perturbatively and add corrections to the expressions
by considering interactions of the partons with the QCD-vacuum.
The QCD-vacuum is described by slowly fluctuating fields and can
be included in the calculations by introduction of the so called
\textit{condensates}. With the sum rules, a systematic treatment
was introduced by Shifman, Vainshtein and Zakharov in 1979. The
SVZ sum rules will be derived in the next section.
\newline

Throughout this thesis we will use the
$\overline{MS}$-renormalization scheme with dimensional
regularization in perturbative calculations. Adopted expressions
from other authors are also obtained using this scheme. The
coefficients of the corresponding Gell-Mann-Low function for the
running coupling are listed in appendix \ref{app-rg}. Being a
mass-independent scheme - the $\beta$-function is not a function
of the quark masses - the application of this scheme leads to
difficulties in treating the heavy quarks, since they do not
decouple in non-physical quantities. If one tries to express the
running coupling constant at a certain scale as a function of the
coupling at another scale, one has to apply \textit{matching
conditions} if the mass-threshold of a heavy quark lies between
the two scales. This is also described in appendix \ref{app-rg}.

\section{The SVZ sum rules}\label{SVZ}

The basic object of the SVZ sum rules is the two-point correlation
function of a gauge invariant quark current and its hermitian
conjugate:
\begin{equation}
\Pi_{AB}(q)=i\!\int\!
d^4\!x\,e^{iqx}\langle0|T\{j_A(x),j^{\dag}_B(0)\}|0\rangle.
\end{equation}

\begin{figure}[t]
\begin{center}
\begin{minipage}[t]{4cm}
  \epsfig{file=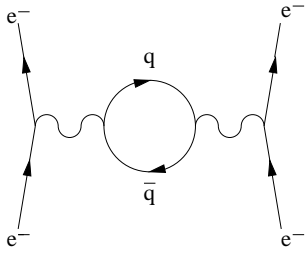,width=4cm}
\end{minipage}
\end{center}
\caption{\label{fig-eescatt}\small Appearance of the quark-loop in
$e^-e^-\rightarrow e^-e^-$-scattering. At large negative values of
the momentum $q^2$
 of the virtual photon, the quarks can be represented by their free
propagator.} \vspace{3mm}
\end{figure}

Here, $j_A(x)=\overline{\psi}_f(x)\Gamma_A\psi_{f'}(x)$ is a quark
field bilinear with Lorentz structure and flavor content chosen
such, that it reflects the quantum numbers of the interpolated
hadron. The currents are injected into the vacuum to avoid long
distance dynamics, which would occur in case of an initial or
final hadronic state. A typical process, which is relevant for
investigation of properties of the $\rho$-meson, is shown in
figure \ref{fig-eescatt}. Here, the T-product of two vector
currents is considered. In this case, due to current conservation
$\partial_{\mu}j^{\mu}=0$, one can explicitly factor out the
kinematical transverse Lorentz structure:
\begin{equation}
\Pi_{\mu\nu}(q)=(q_{\mu}q_{\nu}-q^2g_{\mu\nu})\Pi(q^2).
\end{equation}

In general one can always decompose the initial n-point function
into various terms, where the kinematical structure is factored
out, and then proceed with the objects containing the dynamics.
\newline

The idea of the sum rules is to evaluate the correlation function
in two different ways. On one side, $\Pi(q^2)$ is calculated at
large negative momentum transfer $Q^2\equiv-q^2\gg\Lambda_{QCD}$,
where it can be shown to be dominated by short distance dynamics
\cite{Colangelo:2000dp}. This will be done within the framework of
Wilson's operator product expansion (OPE), which gives additional
corrections to the pure perturbative part in form of an expansion
in vacuum condensates. On the other side a complete set of
hadronic states will be inserted in the correlator to obtain
hadronic matrix elements. Usually, the sum rules are setup to
extract values for these matrix elements or hadronic properties
which they contain, like decay or coupling constants. The two
sides are related to each other by the use of dispersion
relations. A Borel transformation is applied to both sides to
improve the stability. It suppresses higher states on the hadronic
side and removes arbitrary subtraction terms that might appear in
the dispersion relations. In the following we will derive the sum
rules for the B-meson decay constant $f_B$.

\subsection{The Hadronic Side of the Sum Rules}
Starting point is the two-point function of two pseudoscalar
currents of a bottom and a massless quark:
\begin{equation}\label{tpc}
\Pi(q^2)= i\!\int\!
d^4\!x\,e^{iqx}\langle0|T\{\bar{q}(x)i\gamma_5b(x),\bar{b}(0)i\gamma_5
q(0)\}|0\rangle.
\end{equation}
It was shown by K\"allen and Lehmann that any two-point correlator
$\Pi(q^2)$ is an analytic function in the complex $q^2$-plane with
possible singularities and a branch cut on the real positive
$q^2$-axis. The singularities and the cut appear at momentum
transfers $q^2$, where intermediate states go on-shell. For
negative real $q^2<0$, $\Pi(q^2)$ is always real and fulfills
Schwartz' reflection principle:
\begin{equation}\label{pitopiconj}
\Pi(z)=\Pi^*(z^*).
\end{equation}
This identity can be analytically continued to the whole complex
$q^2$-plane. With this at hand we can represent $\Pi(Q^2)$, where
$Q^2=-q^2$, by a dispersion integral, relating the two-point
function at negative momentum transfer to its imaginary part at
positive $q^2$. We consider the integral over the contour depicted
in figure \ref{fig-DispRel}:
\begin{figure}[t]
\begin{center}
\begin{minipage}[b]{80mm}
\setlength{\unitlength}{.7mm}
\begin{picture}(100,100)
\put(13,50){\line(1,0){47}} \put(50,13){\vector(0,1){74}}
\qbezier[35](78.284,78.284)(66.569,90)(50,90)
\qbezier[35](50,90)(33.431,90)(21.716,78.284)
\qbezier[35](21.716,78.284)(10,66.569)(10,50)
\qbezier[35](10,50)(10,33.431)(21.761,21.716)
\qbezier[35](21.761,21.761)(33.431,10)(50,10)
\qbezier[35](50,10)(66.569,10)(78.284,21.761)
\put(85,50){\oval(70,7)[l]}
\qbezier[35](88,55)(88,66.269)(78.284,78.284)
\qbezier[35](88,45)(88,33.431)(78.284,21.761)
\qbezier[5](85,53.5)(88,53.5)(88,55)
\qbezier[5](85,46.5)(88,46.5)(88,45)
\put(60,75){\framebox(7,7){$q^2$}}
\multiput(60,50)(1.5,0){18}{\line(1,0){.7}}
\put(87,50){\vector(1,0){1}} \put(55,50){\circle*{1.2}}
\put(25,50){\circle*{1.2}}\put(60,50){\circle*{1.2}}
\put(25,51){\makebox(0,0)[b]{$\Pi(Q^2)$}}
\end{picture}
\end{minipage}
\end{center}
\vspace{-1.5cm} \caption{\label{fig-DispRel}\small Contour of the
integral relating $\Pi(Q^2)$ at negative momentum transfers to its
imaginary part at positive $q^2$. The circle is taken to infinity.
The dots and the dashed line at the positive $q^2$-axis indicate
bound states and multi-particle states.} \vspace{3mm}
\end{figure}
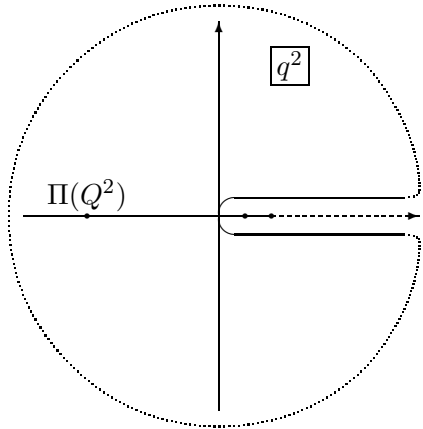

\begin{equation}
\Pi(Q^2)=\frac{1}{2\pi
i}\int\limits_\mathcal{C}dz\frac{\Pi(z)}{z+Q^2}.
\end{equation}
If $\Pi(z)$ vanishes sufficiently fast at $|z|\rightarrow\infty$,
the integration over the circle does not contribute and we are
left with
\begin{equation}
\Pi(Q^2)=\frac{1}{2\pi
i}\int\limits_0^\infty\frac{\Pi(s+i\varepsilon)-\Pi(s-i\varepsilon)}{s+Q^2}ds.
\end{equation}
With use of (\ref{pitopiconj}) the numerator can be replaced by
the imaginary part of the correlation function and we get the
dispersion relation:
\begin{equation}\label{dispersionrelation}
\Pi(Q^2)=\frac{1}{\pi}\int\limits_0^\infty\frac{Im\Pi(s+i\varepsilon)}{s+Q^2}ds
=\int\limits_0^\infty\frac{\frac{1}{\pi}Im\Pi(s)}{s+Q^2-i\varepsilon}ds
\end{equation}
In most cases the imaginary part does not vanish in the limit
$s\rightarrow\infty$ and one has to introduce subtraction terms to
make the integral finite. In our case the two-point function has a
mass dimension of two and an asymptotic behavior of $\Pi(s)\propto
s^2$ in the limit $s\rightarrow\infty$. We therefore setup the
dispersion relation for $\Pi(q^2)/q^4$ instead of $\Pi(q^2)$:
\begin{equation}
\Pi(Q^2)=Q^4\int\limits_0^\infty\frac{\frac{1}{\pi}Im\Pi(s)}
{s^2(s+Q^2)}ds=\int\limits_0^\infty\frac{\frac{1}
{\pi}Im\Pi(s)}{s+Q^2}ds-\int\limits_0^\infty\frac{\frac{1}{\pi}Im\Pi(s)}{s}ds+
Q^2\int\limits_0^\infty\frac{\frac{1}{\pi}Im\Pi(s)}{s^2}ds
\end{equation}
Instead of (\ref{dispersionrelation}), we now have an additional
polynomial in $Q^2$ rendering the initial integral finite. The
application of a Borel transformation (see section
\ref{boreltransformation}) will remove the subtraction terms and
exponentially suppress higher contributions to the spectral
function.
\newline

We can get an expression for the imaginary part of the correlator
$\Pi(q^2)$ by use of the optical theorem.  It relates the
imaginary part of a forward scattering amplitude with the sum over
all intermediate states - and thus with the total cross section
for particle production of these states. Inserting in (\ref{tpc})
a complete set of states yields
\begin{equation}
2 Im \Pi(q^2)=\sum\limits_n\int \! d\tau_n
(2\pi)^4\delta^{(4)}(q-p_n)\langle0|\bar{q}i\gamma_5b|n\rangle\langle
n|\bar{b}i\gamma_5q|0\rangle.
\end{equation}
Here, $|n\rangle$ is a single-/multiparticle state with quantum
numbers of the quark current and $d\tau_n$ indicates the
integration over the corresponding phase space. In our case, the
B-meson is the lowest lying state. The corresponding
meson-to-vacuum matrix element defines the decay-constant $f_B$,
\begin{equation}\label{fb-definition}
\langle0|\bar{q}i\gamma_5b|B\rangle=\frac{m_B^2f_B}{m_b},
\end{equation}
where $m_b$ is the b-quark mass and $m_B$ the B-meson mass. Using
this definition and integrating out the phase space, we get
\begin{equation}
\frac{1}{\pi}Im\Pi(q^2)=\frac{m_B^4f_B^2}{m_b^2}\delta(q^2-m_B^2)
+higher\,states.
\end{equation}
Here, \textit{higher states} stands for the contribution of higher
lying resonances (radial excitations) of the B-meson and the
continuum contribution of multi-particle states. Very little is
known of the spectral density, so for the time being, we write
\begin{equation}\label{hadronicspectralfunction}
\frac{1}{\pi}Im\Pi(q^2)=\frac{m_B^4f_B^2}{m_b^2}\delta(q^2-m_B^2)
+\rho^h(q^2)\Theta(s_0^h-q^2),
\end{equation}
where the step function is explicitly written out to emphasize
that the onset of the contribution of the higher states and the
continuum $\rho^h(q^2)$ is at $s_0^h>m_B^2$. This ansatz,
\textit{first resonance plus continuum}, for the hadronic spectral
density is typically used in most applications of sum rules. Only
the lowest lying state enters explicitly the calculations, whereas
the hadronic continuum will be approximated by the perturbative
results (see section \ref{duality}). Plugging
(\ref{hadronicspectralfunction}) in the dispersion relation
(\ref{dispersionrelation}), we get
\begin{equation}\label{hadronicside}
\Pi(Q^2)=\frac{m_B^4f_B^2}{m_b^2(m_B^2+Q^2)}+
\int\limits_{s_0^h}^\infty\!\frac{\rho^h(s)}{s+Q^2}\,ds+\cdots,
\end{equation}
where the ellipses stand for subtraction terms necessary to make
the integral finite.
\newline

Equation (\ref{hadronicside}) is the hadronic side of the sum
rules. On the other side, the correlation function $\Pi(Q^2)$ will
be derived by means of perturbative QCD. In the next section we
will calculate the simple quark loop and introduce corrections to
it, coming from an expansion in vacuum condensates.

\subsection{The QCD-Side of the Sum Rules}
\begin{figure}[t]
\begin{center}
\epsfig{file=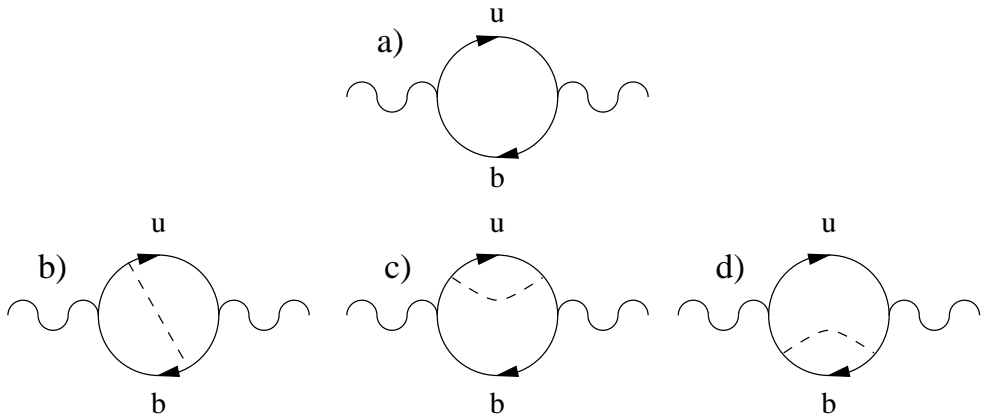,width=13cm}
\end{center}
 \caption{\label{fig-quarkloop}\small The free quark-loop
(a) and O($\alpha_s$)-corrections to it (b-d).}
\vspace{3mm}
\end{figure}

We start with calculating the contribution of the simple
quark-loop (figure \ref{fig-quarkloop}$\,$a). This means, we
contract the quark fields in the correlator and replace them by
the free quark propagators. In the chiral limit of massless
$u,d$-quarks, we get
\begin{equation}\label{eq-quark-loop}
i\Pi(q^2)=-3\!\int \! \!
\frac{d^4\!k}{(2\pi)^4}\tr\left\{(i\gamma_5)
\frac{i(q\!\!\!\slash-k\!\!\!\slash)}{(q-k)^2+i\varepsilon}
(i\gamma_5)\frac{i(k\!\!\!\slash+m_b)}{k^2-m_b^2+i\varepsilon}\right\}
\end{equation}
times an overall delta function. The '$3$' arises from summation
over color indices. Instead of calculating this amplitude
directly, we again setup a dispersion relation. By inspecting the
poles in (\ref{eq-quark-loop}), it can be seen that the correlator
$\Pi(q^2)$ develops an imaginary part from $q^2\geq m_b^2$ on, and
we can write:
\begin{equation}\label{dr-ql}
\Pi(Q^2)=\int\limits_{m_b^2}^\infty\frac{\frac{1}{\pi}Im\Pi(s)}{s+Q^2}ds
\end{equation}
We calculate the imaginary part of (\ref{eq-quark-loop}) by
applying the Cutcosky cutting rules to it. We get the
discontinuity across the branch cut on the positive $q^2$-axis by
replacement of the propagators with delta-functions\footnote{This
is a purely mathematical treatment to calculate the imaginary part
of $\Pi(q^2)$ - putting the quarks on-shell is certainly not a
physical procedure.}.
\begin{equation}
Disc\,\Pi(q^2)=\frac{-12}{i}\!\int\!\!\frac{d^4\!k}{(2\pi)^4}(k^2-qk)
(2\pi i)\delta((q-k)^2)(2\pi i)\delta(k^2-m_b^2)
\end{equation}
After integration and using $Disc\,\Pi(q^2)=2iIm\,\Pi(q^2)$ we get
\begin{equation}
Im\,\Pi(q^2)=\frac{3}{8\pi}\frac{(m_b^2-q^2)^2}{q^2}\label{Ci}
\end{equation}
Inserting this expression in (\ref{dr-ql}) the integral becomes
divergent. Again, this will be cured later by application of the
Borel transformation.
\newline

First order corrections in the strong coupling to the quark-loop
(figure \ref{fig-quarkloop} b-d) are well known and we adopt the
expressions from \cite{Khodjamirian:1998ji}:
\begin{equation}
\Delta
Im\Pi(s)=\frac{\alpha_s}{2\pi^2}\frac{(m_b^2-s)^2}{s}c^1(s),
\end{equation}
with
\begin{eqnarray}\label{c-fo}
c^1(s)&=
&\frac{9}{4}+2Li_2\left(\frac{m_b^2}{s}\right)+\log\frac{s}{m_b^2}
\log\frac{s}{s-m_b^2}
+\frac{3}{2}\log\frac{m_b^2}{s-m_b^2}\nonumber \\
&&+\log\frac{s}{s-m_b^2}+\frac{m_b^2}{s}\log\frac{s-m_b^2}{m_b^2}
+\frac{m_b^2}{s-m_b^2}\log\frac{s}{m_b^2}
\end{eqnarray}
Here, the dilogarithmic function $Li_2(x)$ satisfies
\begin{equation}
Li_2(x)=-\int\limits_0^x\frac{\log(1-t)}{t}dt.
\end{equation}
O($\alpha_s^2$)-corrections to the heavy-light system were
calculated recently by Chetyrkin and
Steinhauser\cite{Chetyrkin:2001je} in a semi-analytic way. These
three-loop corrections are available as a mathematica package.
\newline

Up to now, the two-point correlator was calculated only within
perturbation theory. One cannot expect to reproduce hadronic
properties by only considering free quark propagators in the
calculations. At some point perturbative QCD breaks down and
non-perturbative objects have to be taken into account. This will
be illustrated in the next section.

\subsubsection{Appearance of non-perturbative Contributions}

\begin{figure}[t]
\begin{center}
\epsfig{file=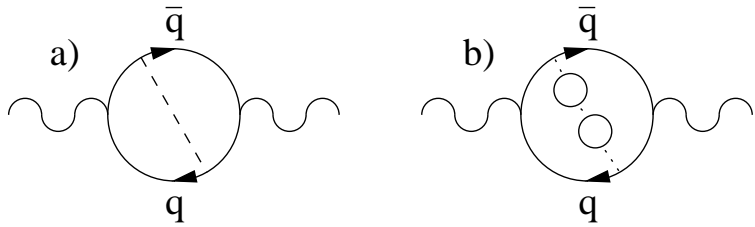,width=10cm}
\end{center}
\caption{\label{fig-adler}\small Replacement of the
free gluon propagator by the effective \textit{bubble chain}.}
\vspace{3mm}
\end{figure}

If one considers higher order corrections to the quark loop, the
appearance of non-perturbative corrections can be seen. We
illustrate this by considering corrections to the simple loop of
massless quarks appearing in figure \ref{fig-eescatt}, following
Shifman\cite{Shifman:1998rb}. The Adler function is defined as the
first derivative of the two-point function:
\begin{equation}
D(Q^2)=-4\pi^2Q^2\frac{d\Pi(Q^2)}{dQ^2}.
\end{equation}
With this definition the loop without $\alpha_s$-corrections is
unity and the inclusion of a gluon-exchange gives:
\begin{equation}
\Delta D(Q^2)=\alpha_s(Q^2)\!\int\limits_{0}^{\infty}\! dk^2\,
F(k^2,Q^2)=\frac{\alpha_s}{\pi}.
\end{equation}
In higher orders one has to replace the coupling $\alpha_s$ by the
running coupling $\alpha_s(k^2)$ in the integral. This replacement
takes the increase of the strength of the coupling constant into
account, if only small momenta $k^2$ are exchanged by the gluon.
It is equivalent to replace the free gluon propagator by the full
one, carrying the effective charge (to one-loop accuracy):
\begin{equation}
\alpha_s(k^2)=\frac{\alpha_s(Q^2)}{1+\frac{\alpha_s(Q^2)}{4\pi}\,b_0\log
\frac{k^2}{Q^2}}.
\end{equation}
The resulting feynman diagrams are often referred to as
bubble-chain graphs. The insertion of the effective charge leads
to a resummation of graphs of this type (see figure
\ref{fig-adler}).
Inserting the running coupling, the integrand becomes infinite at
the Landau pole $k^2=\Lambda^2$:
\begin{equation}
\Lambda^2=Q^2e^{-\frac{4\pi}{b_0\alpha_s}}.
\end{equation}
Dividing the integral in two parts at a scale $\mu^2\gg\Lambda^2$,
the IR-part can be separated:
\begin{equation}
\Delta D(Q^2)_{IR}=\alpha_s(Q^2)\int\limits_0^{\mu^2}
\!dk^2\frac{1}{1+\frac{\alpha_s(Q^2)}{4\pi}\,b_0\log
\frac{k^2}{Q^2}}\,F(k^2,Q^2).
\end{equation}
There is no prescription for how to skip the pole, thus the
integral cannot be solved in an unambiguous way. However, using
the small $k^2$-behavior of
$F(k^2,Q^2)\rightarrow\frac{2}{\pi}\frac{k^2}{Q^4}$, one can
deduce the scaling-behavior
\begin{equation}
\Delta D(Q^2)_{IR}\propto\frac{\Lambda^4}{Q^4}.
\end{equation}
Thus, the low momentum contribution to the Adler function scales
like $\frac{\Lambda^4}{Q^4}$ with an ambiguous numerical
coefficient. However, since the Adler function is an observable,
this term has to be cancelled by some non-perturbative object of
mass-dimension four. In the operator product expansion, which will
be introduced in the following, such non-perturbative effects are
systematically taken into account by an expansion in vacuum
condensates. The vacuum expectation value of the squared field
strength,
$\langle0|\frac{\alpha_s}{\pi}G^{\mu\nu}G_{\mu\nu}|0\rangle$, may
be seen as the corresponding object incorporating the
non-perturbative effects of the bubble chain.

\subsubsection{Wilson Operator Product Expansion}

The idea of Wilson's operator product
expansion\cite{Wilson:1969zs} is to factor out the long-distance
part below the scale $\mu$ into the non-perturbative vacuum
expectation values of gauge and Lorentz invariant local operators,
the vacuum condensates. These are accompanied by $\mu$-dependent
coefficients, which can be calculated perturbatively. This means
that any correlation function will be split into a perturbative
expansion, taking into account the coulomb interactions and an
expansion in condensates, which bear the long distance
interactions of the quark and gluon fields.
\newline

The effect of the product of two operators separated by a small
distance $x$ can be described as the effect of a local operator of
the same quantum numbers. This local operator creates the same
disturbance as the product and can be written in form of a linear
combination in some basis:
\begin{equation}
O(x)O(0)=\sum\limits_nC_n(x)O_n(0).
\end{equation}
The resulting operators $O_n$ are purely local ones, whereas the
coefficients $C_n(x)$, which are c-numbers, now depend on the
separation $x$.
\newline

In the case of the correlator of two quark currents, we have to
expand the product of the currents into a series of local
operators:
\begin{equation}
j^\mu(x) {j^\nu}^\dag(0)=\sum\limits_nC^{\mu\nu}_n(x)O_n(0).
\end{equation}
Since we are interested in the vacuum expectation value of the
currents, only the local operators $O_n$ which are Lorentz and
gauge invariant, contribute to the expansion. Taking the Fourier
transform we can write:
\begin{equation}\label{condensateexpansion}
i\!\int\!\!
d^4\!x\,e^{iqx}\langle0|T\{j^\mu(x),{j^\nu}^{\dag}(0)\}|0\rangle=
C_1^{\mu\nu}(q^2)\cdot\mathrm{1}+C_{\bar{q}q}^{\mu\nu}(q^2)
\langle\bar{q}q\rangle+C_{GG}^{\mu\nu}(q^2)
\langle\frac{\alpha_s}{\pi}G_{\mu\nu}G^{\mu\nu}\rangle+\cdots.
\end{equation}
Here, $\mathrm{1}$ stands for the unity operator, so the
coefficient $C^{\mu\nu}_1(q^2)$ is just the perturbative result of
the quark loop. The condensates do not depend on the momentum
transfer and also do not depend on quantum numbers or any other
structure of the correlator. They are universal non-perturbative
objects. The two condensates which are written explicitly in the
expansion are the quark condensate of mass dimension three and the
four dimensional gluon condensate. The quark condensate is the
order parameter of chiral symmetry breaking. It is always
accompanied by a quark mass and therefore effectively a dimension
four object. The ellipses stand for higher dimensional
condensates. The dimensions of the coefficients can be extracted
directly from dimensional analysis. If the kinematical structure
is factored out - in case of two conserved vector currents this is
just the transverse Lorentz structure $(q_\mu
q_\nu-q^2g_{\mu\nu})$ - it becomes clear that the condensates are
suppressed by $(Q^2)^{-d/2}$, where $d$ is the dimension of the
relevant condensate. Thus the expansion in the dimensions of the
condensates is suitable, if the correlator is evaluated at large
negative momentum transfer.
\newline

The correlation function $\Pi(q^2)$ does not depend on any
renormalization scale $\mu$, so the expansion on the r.h.s has to
obey the Callan-Symanzik equation. We introduce a normalization
point $\mu$ such, that it separates the hard and the soft regime
of the correlator. The soft modes of the vacuum fluctuations are
then described by the condensates, which become scale-dependent,
whereas the hard modes are calculated perturbatively. One chooses
a rather safe scale, where the coupling $\alpha_s(\mu)$ is
somewhat smaller than unity, so that the perturbative expansion
and the calculation of the Wilson coefficients is valid.
\newline

\begin{figure}[t]
 \begin{center}
 \epsfig{file=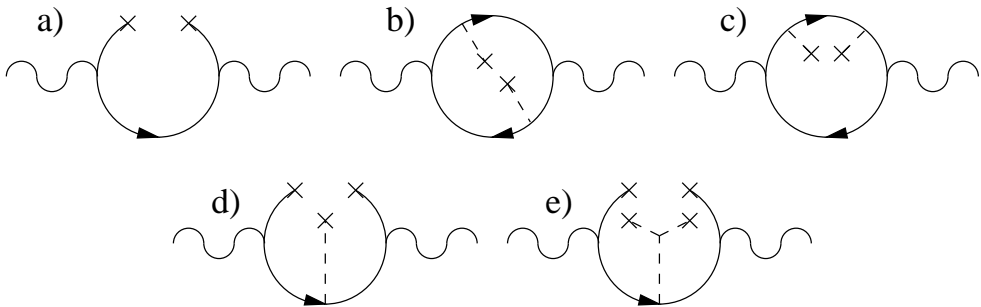,width=13cm}
 \end{center}
\caption{\label{fig-condensates}\small Relevant Feynman diagrams
for calculation of the coefficients of the condensates up to
dimension 6. Appearance of the quark condensate (a), the gluon
condensate (b,c), the mixed quark-gluon condensate (d) and the
four-quark condensate (e).} \vspace{3mm}
\end{figure}

The coefficients of the condensates are calculated by cutting the
lines of the low momentum quark or gluon. Thereby, the momentum of
this particle is set to zero and the fields are treated as
external. In the case we are interested in, the heavy-light
B-meson, the relevant diagrams up to dimension 6 are depicted in
figure \ref{fig-condensates}. The crosses indicate the insertion
of the vacuum condensates.
\newline

In our case of a pseudoscalar current of one bottom and a massless
quark we have to start with (\ref{tpc}). To calculate the Wilson
coefficient of the quark condensate $\langle\bar{q}q\rangle$,
$C_{\bar{q}q}(q^2)$, we contract the b-quarks in the T-product and
replace it with the free quark propagator, whereas the light
quarks are treated as external fields (see figure
\ref{fig-condensates}$\,$a). We expand the quark-field $q(x)$
around zero, to get an expression in the local condensates:
\begin{eqnarray}
q(x)=q(0)+x^\mu \overrightarrow{D}_\mu q(0)+\cdots \\
\bar{q}(x)=\bar{q}(0)+\bar{q}(0)\overleftarrow{D}_\mu
x^\mu+\cdots\label{equ-quark-field-expansion} .
\end{eqnarray}
The ellipses stand for higher derivatives which contribute to
condensates of higher dimensions. The second term vanishes by the
equations of motion for the dirac fields and we are left with:
\begin{equation}
\Pi(q^2)= \int\!\! \frac{d^4\!k}{(2\pi)^4}
d^4\!x\,e^{i(q-k)x}\frac{m_b}{k^2-m_b^2}\langle0|\bar{q}q|0\rangle.
\end{equation}
The term proportional to $\langle0|\bar{q}(x)k\!\!\!\slash
q(0)|0\rangle$ gives contributions of order $O(m_q)$, when the
expansion (\ref{equ-quark-field-expansion}) is inserted and we
neglect its contribution. Evaluating the integral we get
\begin{equation}
C_{\bar{q}q}(q^2)=\frac{m_b}{q^2-m_b^2}.
\end{equation}
Besides the four-dimensional condensates in
(\ref{condensateexpansion}), there are further contributions to
the sum rules from the condensates depicted in figure
\ref{fig-condensates}$\,$d and \ref{fig-condensates}$\,$e. The
dimension five mixed quark-gluon condensate is usually
parameterized by:
\begin{equation}
\langle g_s\bar{q}\sigma^{\mu\nu}\frac{\lambda^a}{2}
G^a_{\mu\nu}q\rangle=m_0^2\langle\bar{q}q\rangle.
\end{equation}
The quark-gluon condensate appears, if one expands the non-local
quark condensate. Using Fock-Schwinger gauge, $x_\mu A^\mu(x)=0$,
to make the path ordered integral between the two quark fields
unity, one gets an expansion of quark condensates with covariant
derivatives between them. The condensate
$\langle\bar{q}D^2q\rangle$ is directly related to the mixed
quark-gluon condensate. Thus $m_0^2$ stands for the ratio
$\langle\bar{q}D^2q\rangle/\langle\bar{q}q\rangle$, which can be
interpreted as average virtuality of the quark fields.
\newline

The four quark condensates can be factorized into two two-quark
condensates in the limit of large number of colors (corrections
come with $\frac{1}{N_c^2}$):
\begin{equation}
\langle O_{4q}\rangle=\alpha_s\langle\bar{q}q\rangle^2.
\end{equation}
The calculation of the coefficients of the higher dimensional
condensates for the heavy-light pseudoscalar currents can be found
in \cite{Reinders:1985sr}. In \cite{Jamin:1993se}, Jamin and
M\"unz calculated the coefficients to the condensates up to
dimension 5 to all orders in the quark masses. It is found that
long distance contributions in form of mass logarithms of the
light quark mass $m^n\log m^2/\mu^2$, which appear in the
calculations starting at order $n=3$, can be absorbed in the
condensates, if non-normal-ordered condensates are used in the
operator product expansion. This fact was later used to obtain a
first order correction in the strong coupling to the quark
condensate \cite{Jamin:2001fw}.
\newline

Throughout this thesis, we neglect light quark masses, and we
adopt the results from Aliev \cite{Aliev:1983ra} for the
coefficients contributing to the heavy-light correlator:
\begin{eqnarray}
C_{GG}(q^2)&=&-\frac{1}{12}\frac{1}{q^2-m_b^2}\\
C_{\bar{q}Gq}(q^2)&=&-\frac{m_b}{2(q^2-m_b^2)^2}
\left(1+\frac{m_b^2}{2(q^2-m_b^2)}\right)\\
C_{4q}(q^2)&=&-\frac{16\pi}{27}\frac{1}{(q^2-m_b^2)^2}
\left(1+\frac{m_b^2}{2(q^2-m_b^2)}-\frac{m_b^4}{2(q^2-m_b^2)^2}\right).
\end{eqnarray}
Gathering all the coefficients of the expansion we get for the
QCD-part of the correlation function $\Pi(Q^2)$ up to dimension 6
in the condensates:
\begin{eqnarray}\label{QCD-part}
\Pi(Q^2)=\int\limits_{m_b^2}^\infty\!\frac{\rho^{QCD}(s)}
{(s+Q^2)}\,ds\!\!&-&\!\!\frac{1}{(m_b^2+Q^2)}
\left[m_b\langle\bar{q}q\rangle+\frac{m_b}{2(m_b^2+Q^2)}
\left(1-\frac{m_b}{2(m_b^2+Q^2)}\right)m_0^2
\langle\bar{q}q\rangle\right.+ \nonumber \\
&&\left.\frac{16\pi}{27(m_b^2+Q^2)}\left(1-\frac{m_b^2}
{2(m_b^2+Q^2)}-\frac{m_b^4}{2(m_b^2+Q^2)^2}
\right)\alpha_s\langle\bar{q}q\rangle^2\right]
\end{eqnarray}
where
\begin{equation}\label{rhoQCD}
\rho^{QCD}(s)=\frac{3}{8\pi^2}\frac{(m_b^2-s)^2}{s}
\left(1+\frac{4\alpha_s}{3\pi}c^1(s)\right),
\end{equation}
and $c^1(s)$ is given in (\ref{c-fo}).
\newline

The values of the condensates do not depend on the channel one is
investigating. We take the following typical numerical values for
the condensates.
\newline

The value of the quark condensate can be extracted from the
Gell-Mann-Oakes-Renner relation obtained by current algebra:
\begin{equation}\label{equ-gell-mann}
(m_u+m_d)\langle\bar{u}u+\bar{d}d\rangle=-m_\pi^2f_\pi^2,
\end{equation}
 and we take from \cite{Jamin:2002ev}:
\begin{equation}
\langle\bar{q}q\rangle(2 \;GeV)=-(267\pm16)^3 \;MeV^3.
\end{equation}
The gluon condensate is extracted from sum rules themselves, thus
one channel is sacrificed. Instead of giving a prediction to a
hadronic matrix element, this value is taken by experiment and the
condensate adjusted such, that the sum rules reproduce it. From
charmonium sum rules we get:
\begin{equation}
\langle0|\frac{\alpha_s}{\pi}G^{\mu\nu}G_{\mu\nu}|0\rangle=0.012
\;GeV^4
\end{equation}
This value and the next two parameters are taken from
Belyaev\cite{Belyaev:1982sa}:
\begin{eqnarray}
m_0^2(1 \;GeV)&=&(0.8\pm0.2) \;GeV^2\\
\alpha_s\langle\bar{q}q\rangle^2&=&8\cdot10^{-5}\;GeV^6.
\end{eqnarray}
We did not quote an error on the gluon and the four quark
condensate. We will later take an uncertainty of $100\%$ on these
values, which will still be numerically negligible.
\newline

The operator product expansion is an expansion in the soft
non-perturbative condensates. This power series is only valid up
to a certain dimension, where contribution from \textit{direct
instantons} appear. These short distance vacuum fluctuations give
rise to an additional exponential term and the expansion in the
condensates breaks down. However, the instanton contribution is
numerically negligible.

\subsection{Equating the two Sides}

\subsubsection{The Borel Transformation}\label{boreltransformation}
After evaluating the two-point correlator in two different ways,
the phenomenological, hadronic part and the QCD-part are equaled.
To improve the stability, Shifman, Vainshtein and Zakharov
proposed a so called Borel transformation. Applying it to the two
parts will remove the subtraction terms of the dispersion
relations and improve the convergence of the power-series
expansion at the same time. The algebraic operator is that of an
inverse Laplace transform:
\begin{equation}\label{equ-laplace-op}
B\equiv\left.\lim\limits_{n\rightarrow\infty}\lim
\limits_{Q^2\rightarrow\infty}\right|_{\frac{Q^2}{n^2}=M^2}
\frac{(Q^2)^{(n+1)}}{n!}\left(-\frac{d}{dQ^2}\right)^n
\end{equation}
Applying this operator on any negative power of $(s+Q^2)$, we get
\begin{equation}
B(s+Q^2)^{-k}=\frac{1}{(k-1)!}\left(\frac{1}{M^2}
\right)^{k-1}\!\!e^{-\frac{s}{M^2}},\qquad\,k\geq1.
\end{equation}
After the Borel transformation, for sufficiently low values of the
squared Borel mass $M^2$ the integral over the imaginary part of
the correlator (\ref{dr-ql}) becomes finite and also the
contribution of higher states in (\ref{hadronicside}) gets
suppressed. The truncated power series of the contribution of the
condensates becomes an expansion in negative powers of $M^2$ times
an exponential function, which breaks down for $M^2\rightarrow0$.
This gives a lower bound on $M^2$. On the other hand, the higher
the Borel mass is chosen, the more the uncertainties in the
hadronic spectral density are emphasized. Thus, in the typical sum
rule applications a \textit{working window} of the squared Borel
mass $M^2$ is defined. The boundaries are taken such, that the
contributions from higher states and from the condensate
expansion, respectively, do not exceed a certain percentage value
of the whole expression. One hopes that inside this window the
corresponding hadronic parameter the sum rule is set up for, is
close to reality. The issue of which value of the Borel parameter
should be taken as reading point, will be discussed several times
throughout this thesis.

\subsubsection{The final Sum Rule for the B-meson Decay Constant}

Borel tranforming (\ref{hadronicside}) and the expansion in local
operators (\ref{QCD-part}), we get after equating both sides:
\begin{eqnarray}\label{fb-fastfertig}
\frac{m_B^4f_B^2}{m_b^2}e^{-\frac{m_B^2}{M^2}}\!\!&=&\!\!
\int\limits_{m_b^2}^{\infty}\!ds\rho^{QCD}(s)e^{-\frac{s}{M^2}}
\!-\!\int\limits_{s_0^h}^{\infty}\!ds\rho^h(s)e^{-\frac{s}{M^2}}
+\nonumber\\
&&+\left[-m_b\langle\bar{q}q\rangle+\frac{1}{12}\langle
\frac{\alpha_s}{\pi}GG\rangle-\frac{m_b^2}{2M^2}
\left(1-\frac{m_b^2}{2M^2}\right)m_0^2\langle\bar{q}q\rangle-\right.
\nonumber\\
&&\left.\quad\:-\frac{16\pi}{27}\frac{1}{M^2}
\left(1-\frac{m_b^2}{4M^2}-\frac{m_b^4}{12M^4}\right)
\alpha_s\langle\bar{q}q\rangle^2\right]e^{-\frac{m_b^2}{M^2}},
\end{eqnarray}
where $\rho^{QCD}(s)$ is given in (\ref{rhoQCD}). The hadronic
spectral function $\rho^h(s)$ is not known. By the assumption of
quark-hadron duality, it is related to the perturbative part of
the correlator. It is assumed that at large momentum transfer the
two-point function can be described by the free quark and gluon
fields and thus, the integral over the hadronic spectral density
equals the integral over the perturbative result from a certain
threshold on:
\begin{equation}\label{equ-ansatz-spectralfunction}
\rho^h(s)\Theta(s-s_0^h)=\rho^{QCD}(s)\Theta(s-s_0).
\end{equation}
This ansatz will be discussed more intensively in the next
section. Inserting (\ref{equ-ansatz-spectralfunction}) in
(\ref{fb-fastfertig}), we get the final expression for the B-meson
decay constant:
\begin{eqnarray}\label{equ-fb-sum rule}
m_B^4f_B^2e^{-\frac{m_B^2}{M^2}}\!\!&=&\!\!
\frac{3m_b^2}{8\pi^2}\int\limits_{m_b^2}^{s_0}ds\,\frac{(m_b^2-s)^2}{s}
\left(1+\frac{4\alpha_s}{3\pi}c^1(s)\right)e^{-\frac{s}{M^2}}
+m_b^2\left[-m_b\langle\bar{q}q\rangle+\frac{1}{12}\langle
\frac{\alpha_s}{\pi}GG\rangle \right.\nonumber\\
&&\!\!-\left.\frac{m_b^2}{2M^2}\left(1-\frac{m_b^2}{2M^2}\right)
m_0^2\langle\bar{q}q\rangle-
\frac{16\pi}{27}\frac{1}{M^2}\left(1-\frac{m_b^2}{4M^2}
-\frac{m_b^4}{12M^4}\right)
\alpha_s\langle\bar{q}q\rangle^2\right]e^{-\frac{m_b^2}{M^2}}.\nonumber\\
\end{eqnarray}
This sum rule was the starting point of many analyses through the
last twenty years - see
\cite{Aliev:1983ra,Narison:1987qc,Jamin:2001fw} as examples.
However, the results obtained differ significantly - the estimates
range from $f_B=130\;MeV$ to $f_B=270\;MeV$. The reason for this
discrepancy lies mainly in the sensitivity of the sum rule
(\ref{equ-fb-sum rule}) to the chosen quark mass $m_b$. If one
takes the pole mass $m_{b,pole}$ and varies it from
$m_{b,pole}=4.6\;GeV$ to $m_{b,pole}=4.8\;GeV$, the result changes
from $f_B=210\;MeV$ to $f_B=150\;MeV$. Therefore, an accurate
determination of the heavy quark mass is
necessary\footnote{However, the definition of the pole mass
suffers from an ambiguity of order $\Lambda_{QCD}$ and one cannot
give a more precise value. This will be explained in section
\ref{comparison-jamin-lange}}. The value for the b-quark mass is
usually taken from lattice calculations or bottonium sum rules. In
\cite{Jamin:2001fw}, Jamin and Lange use the $\overline{MS}$ mass
instead of the pole mass. This seems to improve the convergence of
higher order corrections in the strong coupling $\alpha_s$. The
difference of the two analyses of Khodjamirian
\cite{Khodjamirian:1998ji} and Jamin \cite{Jamin:2001fw} will be
discussed in section \ref{comparison-jamin-lange}. We will give
arguments in favor of choosing the $\overline{MS}$ mass and adopt
the sum rule from Jamin and Lange as the starting point of our
analysis.

\section{Quark-Hadron Duality}\label{duality}

The uncertainties in the parameters $m_b,s_0,M^2$ and the
appropriate scale, where the sum rules should be analyzed, are not
the only limitations to the accuracy of the sum rule results. Sum
rules have an intrinsic uncertainty. On one side, they are limited
by the truncation of the operator product expansion. At higher
dimensions of the expansion in $1/Q^2$ \textit{small size
instantons} come into play and OPE cannot take these hard
non-perturbative corrections into account. Furthermore, higher
corrections in the strong coupling to the coefficients of the
condensates and the perturbative calculation are scarcely known.
On the other side, the hadronic spectral function, since it is not
known experimentally to all energies, is approximated by the
arguments of quark-hadron duality. While we do not refer to the
former limitations further, we will discuss the duality
approximation more intense and give arguments for two
modifications of the usual sum rule analysis, related to duality.
\newline

At large negative momentum transfer, the two-point correlator is
believed to be described by the free quark and gluon fields. We
can safely neglect the power expansion on the QCD-side and again
equate the hadronic side (\ref{hadronicside}) with the
perturbative side (\ref{QCD-part}) at large $Q^2$. We get
\begin{equation}\label{global-duality}
Q^4\!\!\int\limits_{s_0^h}^\infty\!\!\frac{\rho^h(s)}{s^2(s+Q^2)}\,ds=
Q^4\!\!\int\limits_{m_b^2}^\infty\frac{\frac{1}{\pi}{Im
\Pi}^p(s)}{s^2(s+Q^2)}\,ds,
\end{equation}
in the limit $Q^2\rightarrow \infty$. We took account for the
subtraction terms by using $\Pi(Q^2)/Q^4$ rather than $\Pi(Q^2)$
as the relevant correlator. Equation (\ref{global-duality}) is
known as \textit{global duality} relation. The integrals over the
hadronic spectrum and the imaginary part of the perturbative
calculation equal in the limit of $Q^2\rightarrow \infty$.
Furthermore, the two integrands should have the same asymptotics
at $s\rightarrow\infty$ to fulfill the equation
(\ref{global-duality}):
\begin{equation}
\rho^h(s)=\frac{1}{\pi}{Im \Pi}^p(s), \quad \mbox{ at } \quad
s\rightarrow\infty.
\end{equation}
Setting the hadronic spectral function point-wise equal to the
perturbative part is known as \textit{local duality}. It is now
argued that one can find a threshold $s_0$, where the integral
over the hadronic function equals the integral over the
perturbative one at finite $Q^2$, leading to the
\textit{semi-local} duality relation:
\begin{equation}\label{semi-local}
Q^4\!\!\int\limits_{s_0^h}^\infty\!\frac{\rho^h(s)}{s^2(s+Q^2)}\,ds\cong
Q^4\!\!\int\limits_{s_0}^\infty\frac{\frac{1}{\pi}{Im
\Pi}^p(s)}{s^2(s+Q^2)}\,ds.
\end{equation}
Thus, the uncertainty of the hadronic spectral function is
replaced by the introduction of one more parameter. This is
certainly only approximative and one has to argue for the right
choice of the threshold-parameter $s_0$. However, Borel
transforming (\ref{semi-local}) introduces an exponential factor,
which suppresses the contribution at higher momenta. After the
transformation, (\ref{semi-local}) is set in (\ref{fb-fastfertig})
and leads to the final sum rule (\ref{equ-fb-sum rule}).
\newline

In section \ref{s(u)} we will introduce a Borel mass dependent
threshold $s_0$. We can find functions $s_0(M^2)$ that will make
$f_B$ constant over the Borel mass $M^2$ in a certain window.
Furthermore, the derivative $M^2\frac{d}{dM^2}\log
\Sigma(M^2,s_0(M^2))$, where $\Sigma(M^2,s_0(M^2))$ is the right
hand side of (\ref{equ-fb-sum rule}), can be set constant over
$M^2$ and equal to the squared meson mass $m_B^2$. The threshold
$s_0(M^2)$ is not a unique function - we can find a continuum of
functions, leading to a continuum of results for $f_B$. However,
we can give arguments for a preferred value of $f_B$.
\newline

We will also apply a second method of analyzing the sum rules,
which is related to the \textit{local duality} approach used
several times by Radyushkin and collaborators
\cite{Nesterenko:1982gc,Szczepaniak:1998sa}. Taking the Borel
parameter $M^2$ to infinity justifies the truncation of the
condensate expansion. However at the same time, the exponential
suppression of the contribution of the hadronic spectral function
vanishes. In these approaches, the threshold $s_0$ is usually
taken to be around the midpoint between the ground state and the
first resonance of the corresponding hadron. In the limit
$M^2\rightarrow\infty$, the sum rule results show a rather high
sensitivity on the chosen threshold $s_0$. In section
\ref{sec-u-tu-infinity}, we suggest setting the daughter sum rule
equal to the meson mass, for which a distinct value for $s_0$ and
therefore for the decay constant $f_B$ can be found.
\newline

In the next section, we will introduce the light-cone sum rules,
which will be used with the above modification in this thesis to
get values for the weak form factor $f_B^+(0)$ and the coupling
$g^{}_{B^*\!B\pi}$.
\section{Light-Cone Sum Rules}\label{sec-lcsr}
\begin{figure}[t]
\begin{center}
\epsfig{file=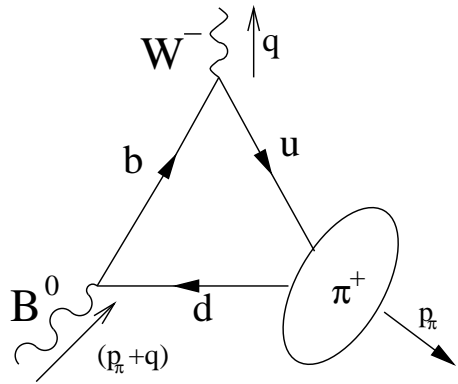,width=6cm}
\end{center}
\vspace{0cm} \caption{\label{fig-wavefunction-diagram}\small
Leading order diagram to the $B\rightarrow \pi$-transition
amplitude. The u- and d-quark combine to form the pion state,
parametrized by its distribution amplitude in the light-cone sum
rule approach.} \vspace{3mm}
\end{figure}

Considering processes of the type $A\rightarrow B+C$, involving
three hadrons or external currents, one needs to modify the
original sum rules. A natural extension of the SVZ sum rules would
be to sandwich three currents, interpolating hadrons by quark
fields, between the physical vacuum,
\begin{equation}
T_{ABC}(p,q)=-\!\int\! d^4\!x\,d^4\!y
\langle0|T\left\{j_A(x)j_B(0)j_C(y)\right\}|0\rangle,
\end{equation}
and apply a similar procedure as for the two-point sum rules.
However, this method, known as three-point sum rules, leads to
difficulties in several applications. In section \ref{sec-Form
Factor}, we analyze the weak form factor $f^+(0)$ for the
semileptonic decay of the B-meson, at maximal recoil ($q^2=0$). It
is defined by the matrix element for the $B\rightarrow\pi$
transition:
\begin{eqnarray}\label{fplus-definition}
\langle\pi(p_\pi)|\bar{u}\gamma_\mu d|B(p_B)\rangle&=&
(p_B+p_\pi)_\mu f^+(q^2)+(p_B-p_\pi)_\mu f^-(q^2)\nonumber\\
&=&2f^+(q^2){p_\pi}_\mu +(f^+(q^2)+f^-(q^2))q_\mu,
\end{eqnarray}
$q=p_B-p_\pi$ being the momentum transfer to the leptons. In the
following, we concentrate only on the form factor $f^+(q^2)$ in
front of the pion-momentum. In the case of large recoil ($q^2=0$),
the b-quark of the meson decays into a u-quark bearing a large
energy of about $E_u=m_b/2$. This process is shown in figure
\ref{fig-wavefunction-diagram}. The u-quark has to recombine with
the d-quark to form the pion. This can be done by exchanging a
hard gluon over a small separation of the Fock state partons (hard
contribution). If such an exchange is not present, the end-point
regions, where one of the partons carries almost all momentum of
the pion are enhanced (soft contribution).
\newline

Expanding the corresponding three-point amplitude in terms of
local condensates immediately leads to problems of convergence.
Replacing the d-quark propagator by the quark-condensate gives
contributions to the correlation function only at the end-point
regions, since the condensates are made out of slowly varying
vacuum fields and do not carry high momenta. This leads to the
fact that the condensates in the OPE are accompanied by increasing
powers of the heavy quark mass $m_b$ \cite{Braun:1997kw}. One has
to take into account a whole chain of condensates of increasing
dimension to make the result numerically stable.
\newline

This is, where the light-cone wave functions come into play
\cite{Braun:1989qv,Ball:1991bs}. Here, the expansion runs in the
transverse distance of the partons instead of the short distance
expansion of the OPE. This leads to a partial resummation of
operators of any dimension into a single wave-function of given
twist, where the twist of an expression is defined by the
difference of its mass dimension and its Lorentz spin.
\newline

Starting point for the light-cone sum rules for the weak form
factor $f^+(q^2)$ is the T-product of a pseudoscalar current,
interpolating the B-meson, and a vector current sandwiched between
the vacuum and the pion state:
\begin{equation}\label{equ-startLCSR}
\Pi_\mu(q,p_\pi)=i\!\int\!\! d^4\!x\,
e^{iqx}\langle\pi(p_\pi)|T\{\bar{u}(x)\gamma_\mu
b(x),\bar{b}(0)i\gamma_5 d(0)\}|0\rangle,
\end{equation}
where $q$ is the momentum carried away by the W-boson. To lowest
order, we contract the b-quark fields and plug in the free
propagator. This process is depicted in figure
\ref{fig-wavefunction-diagram}. We get the expression
\begin{eqnarray}\label{equ-lc-freeprop}
\Pi_\mu(q,p_\pi)\!\!&=&\!\!-i\!\int\!\!
\frac{d^4\!k}{(2\pi)^4}\,d^4\!x\,
e^{i(q-k)x}\frac{1}{k^2-m_b^2}\,\,\big[m_b\langle\pi(p_\pi)|\bar{u}(x)
\gamma_\mu
\gamma_5 d(0)|0\rangle \nonumber \\
&&\hspace{5.2cm}+k^\nu \langle\pi(p_\pi)|\bar{u}(x) \gamma_\mu
\gamma_\nu \gamma_5 d(0)|0\rangle \big].
\end{eqnarray}
The two matrix elements are now expanded around $x^2=0$. The first
term of the first element defines the twist 2 pion light-cone wave
function:
\begin{equation}\label{twist2}
\langle\pi(p_\pi)|\bar{u}(x)\gamma_\mu \gamma_5
d(0)|0\rangle=-i{p_\pi}_\mu f_\pi\!\int\limits_0^1\! du\,
\varphi_\pi(u) e^{iup_\pi x}+\cdots,
\end{equation}
where the ellipses stand for higher terms in the expansion.
$\varphi_\pi(u)$ is the two-particle Fock wave-function of the
pion - its argument $u$ is the momentum fraction carried by one of
the constituents. Instead of the local condensates in the
two-point sum rule approach, the correlation function is now
expanded in non-local matrix elements, which are parametrized by
the wave functions. Replacing in (\ref{equ-lc-freeprop}) the first
matrix element by (\ref{twist2}), we get after integration:
\begin{equation}\label{equ-qcd-twist2}
\Pi_\mu(q,p_\pi)=-{p_\pi}_\mu m_b
f_\pi\!\int\limits_0^1\!\frac{\varphi_\pi(u)du}{(up_\pi+q)^2-m_b^2}
+\cdots.
\end{equation}
Expanding the integrand around $p_\pi=0$ leads to a series in
higher momenta of the wave function $u^n \varphi_\pi(u)$, which
are directly related to higher derivatives of the matrix element
in (\ref{twist2}) and thus to higher dimensional condensates. The
crucial point is that the coefficients of this series are given by
(at $q^2=0$):
\begin{equation}
c_n=\frac{(2qp_\pi)^n}{(m_b^2-q^2)^{n+1}}=\frac{(p_\pi+q)^{2n}}
{{(m_b^2)}^{n+1}}\simeq
1.
\end{equation}
Thus, the expansion in local condensates as one would have
obtained in the 3-point sum rule approach would be useless, since
the coefficients are not suppressed and one had to take the
infinite series of condensates into account. In the LCSR approach
this series is replaced by a single wave function.
\newline

The shape of the pion wave function is not very well known. In the
appendix \ref{app-lc-wf}, we give a definition in terms of an
expansion in Gegenbauer-polynomials. This gives corrections to the
asymptotic form $\varphi_\pi(u)=6u(1-u)$. The values of the
scale-dependent coefficients are frequently discussed in the
literature.
\newline

The hadronic side of the light-cone sum rules for the form factor
is obtained following the same procedure as in the SVZ sum rules
for the weak decay constant. Inserting in (\ref{equ-startLCSR}) a
complete set of states, we get
\begin{equation}
\Pi_\mu(q,p_\pi)=\frac{\langle\pi(p_\pi)|\bar{u}\gamma_\mu
b|B(p_\pi+q)\rangle\langle
B(p_\pi+q)|\bar{b}i\gamma_5d|0\rangle}{m_B^2-(p_\pi+q)^2}+\sum
\!\!\!\!\!\!\!\!\int\limits_n
\frac{\langle\pi(p_\pi)|\bar{u}\gamma_\mu b|n\rangle\langle
n|\bar{b}i\gamma_5d|0\rangle}{m_B^2-(p_\pi+q)^2},
\end{equation}
where $|n\rangle$ are the states of higher resonances and the
continuum contribution. We rewrite these contribution in form of a
dispersion integral over the hadronic spectral density as in
(\ref{hadronicside}) for the decay constant. Defining the relevant
correlator $\Pi^+(q,p_\pi)$ by
\begin{equation}
\Pi_\mu(q,p_\pi)=\Pi^+(q^2,(p_\pi+q)^2){p_\pi}_\mu+\Pi^{\pm}
(q^2,(p_\pi+q)^2)q_\mu,
\end{equation}
and using the definitions for the decay constant
(\ref{fb-definition}) and the form factors
(\ref{fplus-definition}), we get
\begin{equation}
\Pi^+(q^2,(p_\pi+q)^2)=\frac{2m_B^2f_Bf^+(q^2)}{m_b(m_B^2-(p_\pi+q)^2)}
+\int\limits_{s_0^h}^\infty\!\frac{\rho^h(s)ds}{s-(p_\pi+q)^2}.
\end{equation}
Applying the Borel transformation with respect to $(p_\pi+q)^2$,
we finally get for the hadronic side
\begin{equation}
\Pi^+(q^2,M^2)=\frac{2m_B^2}{m_b}f_Bf^+(q^2)e^{-\frac{m_B^2}{M^2}}
+\int\limits_{s_0^h}^\infty \!ds\rho^h(s)e^{-\frac{s}{M^2}}.
\end{equation}
Again, the hadronic spectral function will be replaced by the
perturbative result via the assumption of quark-hadron duality.
Thus, we need to rewrite (\ref{equ-qcd-twist2}) in form of a
dispersion relation\footnote{Here, only a single dispersion
relation is necessary, which is another advantage of the LCSR
approach compared to the ordinary three-point sum rules. In the
case of the B-meson form factor, these make use of double
dispersion relations \cite{Ball:1991ma}, which leads to further
uncertainties of the results.} with respect to $(p_\pi+q)^2$.
Taking the on-shell pion to be massless ($p_\pi^2=0$), we write
the denominator as:
\begin{equation}
(up_\pi+q)^2-m_b^2=u\left((p_\pi+q)^2+\frac{(1-u)q^2}{u}-
\frac{m_b^2}{u}\right)
\end{equation}
By making the substitution $s=(m_b^2-q^2)/u+q^2$, we get
\begin{equation}
\Pi^+(q^2,(q+p_\pi)^2)=m_b
f_\pi\int\limits_0^1\frac{du\,\varphi_\pi(u)}{u\left(
\left(\frac{m_b^2-q^2}{u}+q^2\right)-(p_\pi+q)^2\right)}= m_b
f_\pi\int\limits_{m_b^2}^\infty\frac{ds\,\widetilde{\varphi}_\pi(s)}
{(s-q^2)(s-(p_\pi+q)^2)}.
\end{equation}
Thus, we identify $\widetilde{\varphi}_\pi(s)/(s-q^2)$ with the
imaginary part of $\Pi^+(q^2,(p_\pi+q)^2)$:
\begin{equation}
\frac{1}{\pi}Im\Pi^+(q^2,(p_\pi+q)^2)=m_b
f_\pi\frac{\widetilde{\varphi}_\pi(s)}{(s-q^2)}.
\end{equation}
Taking the simple duality ansatz,
\begin{equation}
\rho^h(s)\,\Theta(s-s_0^h)=\frac{1}{\pi}Im\Pi^+(q^2,(p_\pi+q)^2)\,
\Theta(s-s_0),
\end{equation}
and subtracting the integral over the hadronic spectral density on
both sides, we effectively cut the integral on the right hand side
at $s_0$. After Borel transformation with respect to $(p_\pi+q)^2$
and taking back the substitution, the right hand side of the sum
rule reads:
\begin{equation}
\Pi^+_{rhs}(q^2,M^2)=m_b
f_\pi\int\limits_{m_b^2}^{s_0}\frac{ds\,\widetilde{\varphi}_\pi(s)}
{s-q^2}e^{-\frac{s}{M^2}}
=m_b
f_\pi\int\limits_{\Delta}^{1}\frac{du}{u}\,\varphi_\pi(u)
e^{-\frac{m_b^2-(1-u)q^2}{uM^2}},
\end{equation}
where the lower limit in the integral is given by
\begin{equation}
\Delta=\frac{m_b^2-q^2}{s_0-q^2}.
\end{equation}
Equating the hadronic and the perturbative side we get the LCSR to
lowest order:
\begin{equation}
f_Bf^+(q^2)e^{-\frac{m_B^2}{M^2}}=\frac{m_b^2f_\pi}{2m_B^2}
\int\limits_\Delta^1\frac{du}{u}\varphi_\pi(u)e^{-\frac{m_b^2-(1-u)q^2}
{uM^2}}.
\end{equation}
Further contributions to the QCD-side come from higher terms in
the expansion around the light-cone $x^2=0$ of the matrix element
(\ref{twist2}):
\begin{eqnarray}
\langle\pi(p_\pi)|\bar{u}(x)\gamma_\mu \gamma_5
d(0)|0\rangle&=&-i{p_\pi}_\mu f_\pi\int\limits_0^1 du e^{iup_\pi
x}\left(\varphi(u)+x^2g_1(u)\right)\nonumber\\
&&+f_\pi\left(x_\mu-\frac{x^2{p_\pi}_\mu}{p_\pi
x}\right)\int\limits_0^1due^{iup_\pi x}g_2(u)+\cdots.
\end{eqnarray}
Here, $g_1$ and $g_2$ are two particle twist 4 wave functions.
Their definition is given in the appendix \ref{app-lc-wf}. The
second matrix element in (\ref{equ-lc-freeprop}) can be split into
a symmetric and an antisymmetric part. The corresponding elements
define the two-particle wave functions of twist 3:
\begin{equation}
\langle\pi(p_\pi)|\bar{u}(x)\gamma_5 d(0)|0\rangle=f_\pi
\mu_\pi\int\limits_0^1\!du\,e^{iup_\pi x}\varphi_p(u),
\end{equation}
multiplied by the symmetric tensor $g_{\mu\nu}$. The antisymmetric
matrix element is
\begin{equation}
\langle\pi(p_\pi)|\bar{u}(x)\sigma_{\mu\nu}\gamma_5
d(0)|0\rangle=i({p_\pi}_\mu x_\nu-{p_\pi}_\nu x_\mu)\frac{f_\pi
\mu_\pi}{6}\int\limits_0^1\!du\,e^{iup_\pi x}\varphi_\sigma(u),
\end{equation}
where
\begin{equation}\label{equ-mu-pi}
\mu_\pi=\frac{m^2_\pi}{m_u+m_d}.
\end{equation}
This normalization factor is directly related to the quark
condensate via the Gell-Mann-Oakes-Renner relation, and leads to a
further error source besides the coefficients in the definitions
of the wave functions.
\newline

In appendix \ref{app-lc-wf} we list further contributions to the
LCSR, from taking into account higher orders of the contraction of
the b-fields. This leads effectively to an additional gluon line
going from the b-quark into the pion wave function. The relevant
amplitudes are parametrized by three particle twist 3 and 4 wave
functions. We also list $O(\alpha_s)$-corrections to the twist 2
pion wave function. Again, all these terms have to be integrated
over $x$ and $k$ in (\ref{equ-lc-freeprop}) and then written in
form of a dispersion relation. Here, terms proportional to higher
powers of $1/((up_\pi+q)^2-m_b^2)$ than one have to be partially
integrated, which leads to further contributions to the sum rule
in form of surface terms.
\newline

Up to twist 4 we give the LCSR for the form factor at maximum
recoil ($q^2=0$) explicitly in chapter \ref{sec-Form Factor}. The
expression is taken from Khodjamirian and R\"uckl
\cite{Khodjamirian:1998ji} with first order corrections to the
twist 2 wave function $\varphi_\pi(u)$ taken from Bagan, Ball and
Braun \cite{Bagan:1998bp}.

\chapter{Decay Constant $f_B$}\label{decay-constant}

The leptonic decay constant $f_B$ of the B-meson is a hadronic
parameter of high interest in today's physics. It appears in most
calculations concerning the B-meson. Its squared value enters in
the calculation of the $B-\overline{B}$-mixing amplitude, which is
an important tool to measure CKM elements and investigate
CP-violation. Therefore, a precise knowledge of $f_B$ is strongly
desired. However, there are no experimental measurements of the
decay constant and one has to rely on theoretical predictions.
There exist numerous sum rule analyses of the decay constant. In
the following section, we will reanalyze the SVZ- sum rules for
the decay constant to NLO-accuracy in the perturbative expansion,
following Jamin and Lange\cite{Jamin:2001fw}. We compare the
results with those of Khodjamirian and
R\"uckl\cite{Khodjamirian:1998ji}. The two analyses mainly differ
in the the use of the heavy quark mass - the running mass and the
pole mass, respectively.
\newline

We then proceed with our own analysis using the sum rule from
\cite{Jamin:2001fw} up to first order corrections in the strong
coupling and neglecting the light quark masses.

\section{Discussion of Sum Rule Analyses from the Literature}
\label{comparison-jamin-lange}
The Borel sum rules in the form of equation (\ref{equ-fb-sum
rule}) are known for over twenty years. Unfortunately, the sum
rule is very sensitive to the input parameters like the quark mass
and the values of the threshold $s_0$ and the Borel mass $M^2$.
Therefore, the results of the various calculations vary in a
rather wide range. In 1983, Aliev and Eletskii \cite{Aliev:1983ra}
used this sum rule to obtain $f_B=130\;MeV$. This rather low
value, compared to today's results, is due to the use of a quite
high value of the pole mass ($m_b=4.8\;GeV$) and low value of the
QCD-scale ($\Lambda=100\;MeV$). Using the same sum rule
(\ref{equ-fb-sum rule}), Narison \cite{Narison:1987qc} extracted a
typical value of about $f_B=185\;MeV$ with a very high threshold
$s_0=50\;GeV^2$ and a low Borel parameter $M^2\approx 2.8\;GeV^2$.
\newline

In the late 90's analyses were applied using the running mass of
the heavy quark in the $\overline{MS}$-scheme instead of the pole
mass. Their definitions are given in appendix \ref{app-rg}.
Although the pole mass is independent of the renormalization
scheme, it is sensitive to long distance
dynamics\cite{Beneke:1998ui}. This can be observed by considering
the self energy diagram. Replacing the gluon propagator by the
full propagator with fermion loops, leads to a IR singularity of
the integral. Therefore, an ambiguity, which is of order
$\Lambda_{QCD}$, enters the definition of the pole mass and one
cannot assign a value to it, which is more precise. In the case of
the B-meson decay constant, using the pole mass in
(\ref{equ-fb-sum rule}), the contributions from the first order
and second order corrections are of the same size as the one loop
perturbative results, whereas the expansion is under good control,
if the running mass is used - see \cite{Jamin:2001fw} and also
table \ref{tab-f2cont}.
\newline

We will now briefly demonstrate the analysis of the sum rule for
$f_B$, following Jamin and Lange \cite{Jamin:2001fw}. Second order
corrections to the perturbative result were found by Chetyrkin and
Steinhauser \cite{Chetyrkin:2001je} and are available as
mathematica file \textit{rvs.m}. Jamin and Lange also gave an
expansion in the light quark mass up to $O(m^4)$. These
corrections are numerically negligible, however, from terms of
order $O(m^3)$ on, mass logarithms of the form $\log(\mu^2/m^2)$
appear. These can be cancelled by rewriting the normal ordered
condensates into non-normal ordered \cite{Jamin:1993se}. With this
at hand, Jamin and Lange found corrections to the quark-condensate
to first order in the strong coupling $\alpha_s$:
\begin{equation}
c^{(1)}_{\overline{q}q}(M^2,s_0,\mu)\!=\! 2\!\int
\limits_{\frac{m_b^2}{M^2}}^{\frac{s_0}{M^2}}\!dt\, e^{-t}t^{-1}
-2 \left[1+\left(1-\frac{m_b^2}{M^2}\right)
\left(\log\frac{\mu^2}{m_b^2}+\frac{4}{3}\right)\right]
e^{-\frac{m_b^2}{M^2}}.
\end{equation}
This expression is already Borel transformed. The dependence on
the threshold $s_0$ in the upper limit of the integral appears,
when the initial expression is written as a dispersion relation
and the continuum contribution from the hadronic side is
subtracted.
\newline

The calculation of the simple quark loop involved the replacement
of the quark propagators with delta functions, thus setting the
quarks on-shell. It follows that the heavy quark mass $m_b$,
appearing in the perturbative part of the sum rule
(\ref{equ-fb-sum rule}), is per definition the pole mass. To
express the decay constant $f_B$ in terms of the running mass
$\overline{m}(\mu)$, one has to apply equation
(\ref{equ-pole-running-mass}). Effectively, this replacement only
changes the size of the different loop-contributions - the total
contribution should stay the same. However, the zero order
perturbative contribution to the decay constant rises
significantly after insertion of the running mass, while the first
and second order show a convergent behavior. Thus, with the change
to the running mass one can truncate the expansion even after the
$O(\alpha_s)$-corrections. Keeping the pole mass leads to first-
and second-order corrections of about the same size as the zeroth
order. The truncation of the expansion in the strong coupling
after the second order terms might lead to further errors, when
contributions of higher order terms are not taken into account.
The different contributions to the final results in the two
schemes are shown in table (\ref{tab-f2cont}).
\newline

The quark masses appearing in the leading order coefficients of
the condensates are not defined to be the pole- or the running
mass, either. We use the running mass without imposing
$O(\alpha_s)$-corrections to the coefficients.
\newline

The final sum rule for the decay constant up to $O(\alpha_s^2)$
and without corrections in the light quark mass reads:
\begin{equation}\label{equ-fb-sum rule-sigma}
m_B^4f_B^2e^{-\frac{m_B^2}{M^2}}=\Sigma(M^2,s_0),
\end{equation}
where the right hand side is given by
\begin{eqnarray}\label{equ-fb-sum rule-jl}
\Sigma(M^2,s_0)\!\!&=&\!\!
\frac{3m_b^2}{8\pi^2}\!\!\int\limits_{m_{b,pole}^2}^{s_0}\!\!\!\!ds
\left[\frac{(m_b^2-s)^2}{s}
\left(1+\frac{4\alpha_s}{3\pi}\left(c^1(s)+\Delta^{(1)}\right)\right)+
\frac{\alpha_s^2}{3\pi^2}\left(c^{(2)}_{CS}(s)
+\Delta^{(2)}\right)\right]e^{-\frac{s}{M^2}}\nonumber\\
&&+m_b^2\left[\left(-1+\frac{\alpha_s}{\pi}
c^{(1)}_{\overline{q}q}\right)m_b\langle\bar{q}q\rangle+\frac{1}{12}
\langle
\frac{\alpha_s}{\pi}GG\rangle-\frac{m_b^2}{2M^2}\left(1-\frac{m_b^2}
{2M^2}\right)m_0^2\langle\bar{q}q\rangle\right.\nonumber\\
&&\hspace{1cm}-\left.
\frac{16\pi}{27}\frac{1}{M^2}\left(1-\frac{m_b^2}{4M^2}-\frac{m_b^4}
{12M^4}\right)
\alpha_s\langle\bar{q}q\rangle^2\right]e^{-\frac{m_b^2}{M^2}}.
\end{eqnarray}
Here, $\Delta^{(1)}$ and $\Delta^{(2)}$ are the additional first
and second order corrections coming from the replacement of the
pole masses by use of equation (\ref{equ-pole-running-mass}). In
the expression (\ref{equ-fb-sum rule-jl}), the mass $m_b$ is now
understood as being the scale dependent running mass
$\overline{m}_b(\mu)$. The quark mass in the lower limit of the
integral is kept to the pole mass. It indicates the start of the
branch cut on the positive $q^2$-axis. Replacing it by the running
mass would lead to corrections starting at order $O(\alpha_s^3)$.
The numerical results, if the running mass is used instead of the
pole mass differ about $5\%$. A similar sum rule, except for the
$O(\alpha_s)$- corrections to the coefficient of the quark
condensate, was also used by Narison, who was further able to
extract the running mass $\overline{m}_b(\overline{m}_b)$ at the
same time \cite{Narison:1998ui,Narison:2001pu}\footnote{Narison
adjusted the parameters $s_0$,\;$M^2$ and the quark mass $m_b$
such that the sum rule is in the most stable region. He extracted
$\overline{m}_b(\overline{m}_b)=(4.05\pm0.06)\;GeV$, which is
rather small compared to typical values.}.
\newline

Besides the QCD-parameters $\alpha_s, \overline{m}_b(\mu)$ and the
phenomenological condensates, the function $\Sigma(M^2,s_0)$ also
depends on the artificial parameters $s_0$, coming from the crude
duality ansatz (\ref{equ-ansatz-spectralfunction}), and the Borel
mass $M^2$. Ideally, if one knew the hadronic spectral density to
all energies and the whole perturbative expansion and
non-perturbative corrections to it on the right hand side of the
sum rules, the resulting function for the decay constant would not
depend on the momentum transfer $q^2$ and thus not on the Borel
mass. However, in the real world one has to give arguments for the
values of these two parameters, where the decay constant is read
off. After the Borel transformation, the contribution of the
hadronic continuum is exponentially suppressed and the Borel
parameter $M^2$ should be small enough to ensure that errors
coming from the approximation (\ref{equ-ansatz-spectralfunction})
do not contribute to a large extend. On the other side, the Borel
mass should be high enough, such that the expansion in the
condensates converges sufficiently fast and and the truncation of
the series is justified. In our analysis of the sum rule
($\ref{equ-fb-sum rule-sigma}$), we will take the window $4
\;GeV^2<M^2<8\;GeV^2$, where on the left boundary the contribution
from the mixed quark gluon condensate contributes to less than
$8\%$ and the four quark contribution is negligible. The continuum
contribution of the hadronic side is rather large in the whole
window and rises to about the same size as the final sum rule on
the right boundary. Therefore, the threshold parameter $s_0$ has
to be chosen carefully. It is considered to be roughly in the
range of the first exited state of the hadron under investigation
and taken such, that the decay constant does not vary strongly in
the Borel window.
\newline

\begin{figure}[t]
\begin{picture}(140,100)
\put(4,57){
\begin{minipage}[t]{8cm}
\epsfig{file=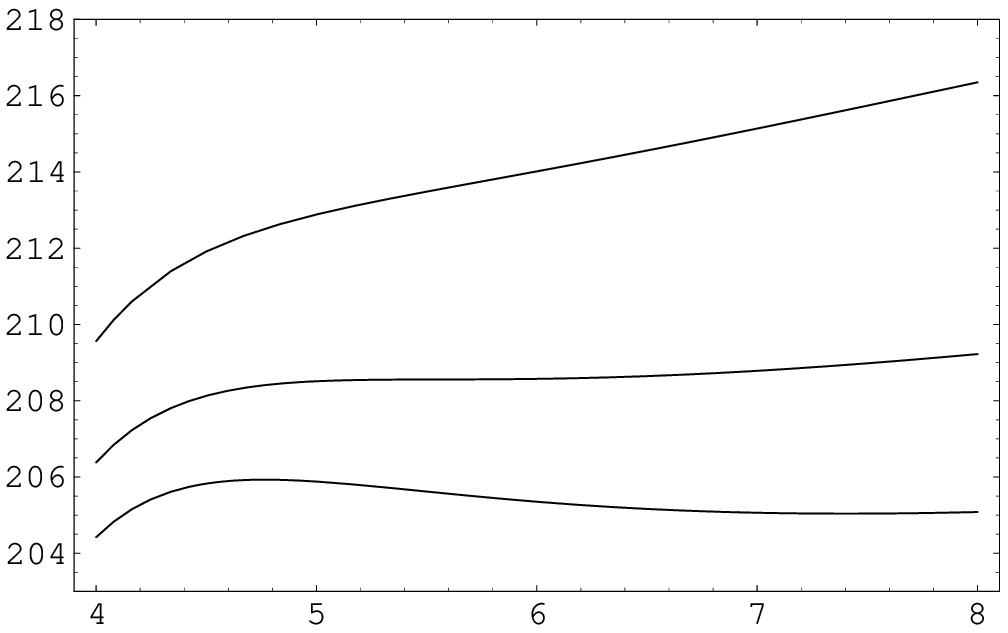,width=7.8cm}
\end{minipage}}
\Text(76,55.5)[c]{\scalebox{.8}{$M^2[GeV^2]$}}
\Text(11,107)[l]{\scalebox{.8}{$f_B[MeV]$}}
\Text(75,101)[r]{\scalebox{.7}{a)}}
\Text(25,93)[l]{\scalebox{.6}{$s_0=34.9\;GeV^2$}}
\Text(60,80)[l]{\scalebox{.6}{$s_0=34.1\;GeV^2$}}
\Text(35,70)[l]{\scalebox{.6}{$s_0=33.7\;GeV^2$}}
\put(25.75,67){\line(0,1){4}} \put(39.2,75){\line(0,1){4}}
\put(71,64.5){\line(0,1){4}} \put(2,2){
\begin{minipage}[t]{8cm}
\epsfig{file=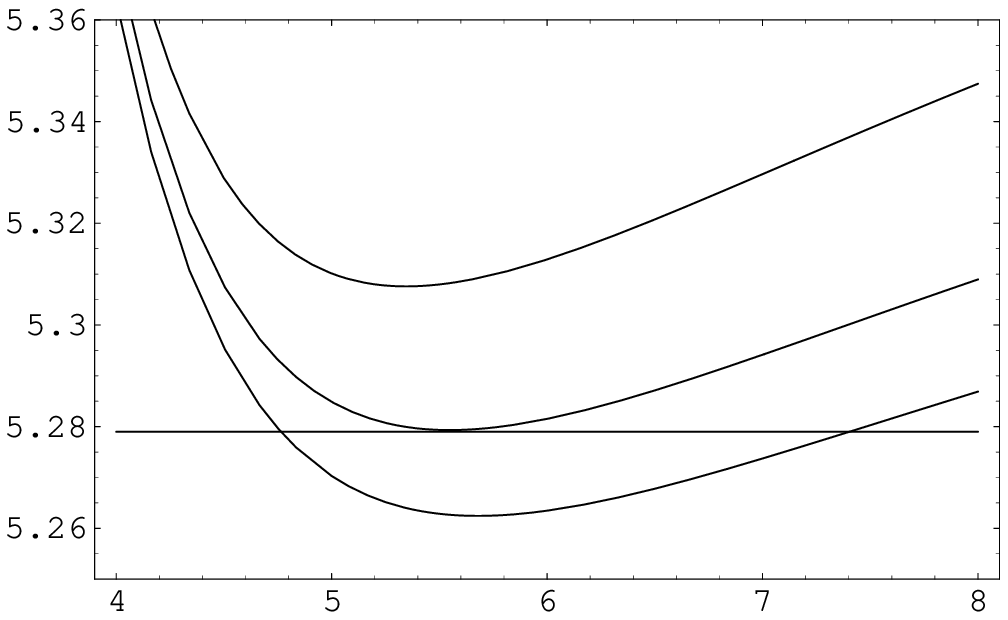,width=8cm}
\end{minipage}}
\Text(76,0.3)[c]{\scalebox{.8}{$M^2[GeV^2]$}}
\Text(11,52)[l]{\scalebox{.8}{$\Re^{1/2}[GeV]$}}
\Text(75,47)[r]{\scalebox{.7}{b)}}
\Text(35,33)[l]{\scalebox{.6}{$s_0=34.9\;GeV^2$}}
\Text(55,27)[l]{\scalebox{.6}{$s_0=34.1\;GeV^2$}}
\Text(13,13)[l]{\scalebox{.6}{$s_0=33.7\;GeV^2$}}
\put(25.75,15.25){\line(0,1){4}} \put(39.2,15.25){\line(0,1){4}}
\put(71,15.25){\line(0,1){4}} \put(95,107){
\begin{minipage}[t]{6cm}
 \caption{\small Decay constant $f_B$ (graph a) and the
Ratio $\Re^{1/2}$ (b), defined in equation
(\ref{equ-first-derivative}), as functions of the Borel parameter
$M^2$, plotted for three different values of the threshold $s_0$.
The intersections of $\Re^{1/2}$ with the experimentally measured
meson mass $m_B=5.279 \: GeV$ correspond to extremal points of the
decay constant. In the case of $s_0=34.1 \:GeV$, also the second
derivative of the corresponding function of the decay constant
vanishes.}\label{fig-mass}
\end{minipage}}
\end{picture}
\vspace{3mm}
\end{figure}

An often suggested procedure to get the reading point
($\hat{M}_0^2$,$\hat{s}_0$) of $f_B$ is to look at the daughter
sum rule of (\ref{equ-fb-sum rule-jl}) \cite{Dominguez:1993sj},
defined by
\begin{equation}\label{equ-daughter-sumrule}
\Re(M^2,s_0)=M^4\frac{d}{dM^2}\log\Sigma(M^2,s_0).
\end{equation}
Applying the derivative on the left hand side of equation
(\ref{equ-fb-sum rule-sigma}) it is then argued that the resulting
function is independent of $f_B$ and gives the squared
pseudoscalar mass as a function of $M^2$. Setting this function
equal to the experimental mass $m_B^2$, gives reading points for
the threshold and the Borel mass. However, in our opinion this
method sometimes is overrated in the literature. The crucial point
is that the sum rule (\ref{equ-fb-sum rule-sigma}) leads to a
rather complicated function for the decay constant $f_B$, which is
by no means independent of $M^2$. The parameter that is constant
in the sum rule (\ref{equ-fb-sum rule-sigma}) is the mass $m_B$ of
the pseudoscalar meson. Its input value in the final sum rule is
the experimental mass, independent of $M^2$. Taking the derivative
of the logarithm of both sides in (\ref{equ-fb-sum rule-sigma})
yields:
\begin{equation}\label{equ-first-derivative}
m_B^2+\frac{M^4 {f_B^2}'(M^2)}{f_B^2(M^2)}=\Re(M^2,s_0).
\end{equation}
Thus, setting the ratio $\Re(\hat{M}^2,\hat{s}_0)$ equal to the
squared mass $m_B^2$ at a point $M_0^2$ is just a rather
complicated procedure to find an extremum of the initial function
$f_B(M^2)$. The argument for taking this distinct value for the
Borel mass $\hat{M}^2$ is then not the fact that it represents the
\textit{right} value for the meson mass (this is always the case),
but rather that the function for the decay constant is in a stable
region around this point. In figure \ref{fig-mass}, we plotted the
decay constant $f_B(M^2)$ for several values of the threshold
$s_0$ and the corresponding ratios $\Re^{1/2}(M^2,s_0)$, defined
in (\ref{equ-daughter-sumrule}) as functions of $M^2$. The
intersections with the experimentally measured mass
$m_B=5.279\;GeV$ give the values of the squared Borel parameter
$M^2$, where $f_B$ reaches an extremum.
\newline

There is a distinct pair of the threshold and the Borel parameter,
where the ratio $\Re^{1/2}(M^2,s_0)$ reaches a minimum at the
meson mass. Taking one more derivative with respect to $M^2$ in
(\ref{equ-first-derivative}), gives
\begin{equation}
2\frac{M^2{f_B^2}'(M^2)}{f_B^2(M^2)}-\frac{M^4{{f_B^2}'}^2(M^2)}
{f_B^4(M^2)}
+\frac{M^4{f_B^2}''(M^2)}{f_B^2(M^2)}=\Re'(M^2,s_0).
\end{equation}
Setting the minimum of $\Re(M^2,s_0)$ equal to the mass $m_B^2$
yields the condition
\begin{equation}
f_B''(\hat{M}^2,\hat{s}_0)=0.
\end{equation}
Therefore, the pair ($\hat{M}^2$,$\hat{s}_0$) found in this way
gives a stable region, where the first two derivatives of the
decay constant vanish, and one can extract a value for $f_B$.
However, there is no physical argument except the stability of the
sum rule around this point that this value resembles the true
decay constant at best.
\newline

Using the sum rule (\ref{equ-fb-sum rule-jl}) with additional
corrections in the light quark mass up to order $O(m_q^4)$, Jamin
and Lange extracted for the B-meson decay constant
\begin{equation}\label{equ-fb-jl}
f_B=(210 \pm 19)\; MeV.
\end{equation}
Here, the main uncertainties come from the value of the heavy
quark mass $m_b=4.21\pm 0.05$, the scale $\mu=(3-6)\: GeV$ and the
value of the quark condensate
$\langle\overline{q}q\rangle(2\;GeV)=-(267\pm 17)^3 \;MeV^3$.
\begin{table}[b]\label{tab-f2cont}
\begin{tabular}{|c||c|c|c|c|c|}
\hline
$f^2_{B,JL}$&$f_1^{(0)}$&$f_1^{(1)}$&$f^{(0)}_{\langle\overline{q}
q\rangle}$
&$f^{(1)}_{\langle\overline{q}q\rangle}$&$f_{\langle-\rangle}$\\
\hline\hline 43468&33064&1855&13201&-4203&-449\\
 \hline
\end{tabular}
\hspace{.6cm}
\begin{tabular}{|c||c|c|c|c|c|}
\hline
$f^2_{B,KR}$&$f_1^{(0)}$&$f_1^{(1)}$&$f^{(0)}_{\langle\overline{q}
q\rangle}$&$f_{\langle-\rangle}$\\
\hline\hline 32314&12045&11170&11388& -2290\\
 \hline
\end{tabular}
 \caption{\small Contributions of the different terms to the
 squared decay constant $f_B$ for the two analyses of
 Jamin and Lange (left table) and Khodjamirian and R\"uckl
 (right table). $f_{\langle-\rangle}$ indicates the contribution
 from condensates with mass dimension four or higher.}
\end{table}
\newline

We repeated the analysis using the procedure described above,
however, only including first order corrections. Replacing the
pole mass by the running quark mass leads to corrections to the
$O(\alpha_s)$-terms of:
\begin{equation}
\Delta^{(1)}=\frac{s-3\overline{m}_b^2}{s-\overline{m}_b^2}
\left(2+\frac{3}{2}\log\frac{\mu^2}{\overline{m}_b^2}\right).
\end{equation}
\newline
For the running b-quark mass, we take the value from El-Khadra
\cite{El-Khadra:2002wp}, $m_b=4.24\pm0.11 \;GeV$, which is
obtained as an average from a large collection of different
analyses, including lattice simulations and sum rule methods for
several systems containing b-quarks. The corresponding pole mass
is $m_{b,pole}=4.69\:GeV$, using the two-loop relation
(\ref{equ-pole-running-mass}). The values for the other input
parameters we used in this analysis and the following ones are
listed in table \ref{tab-parameter}. Taking the mean values we
obtained
\begin{equation}\label{equ-result-jl}
f_B=208 \;MeV \hspace{1cm} s_0=34.1 \;GeV^2 \hspace{1cm}
M^2=5.5\;GeV^2.
\end{equation}
Although we only included first order corrections and used a
different quark running mass and corresponding pole mass, this
result is very close to (\ref{equ-fb-jl}). We do not give an
estimate of the error, since we are only interested in the
relative size of the different contributions to the decay
constant.
\newline

The contributions to $f_B^2$ from the several terms of the sum
rule are shown in table \ref{tab-f2cont}. This table shows also
our result for the sum rule analysis following Khodjamirian and
R\"uckl \cite{Khodjamirian:1998ji}. By using the suggested values
of $m_{b,pole}=4.7\; GeV$ for the b-quark pole mass and $s_0=35
\;GeV^2$ for the threshold, we could exactly reproduce the value
$f_B=180\;MeV$. Note that the first order corrections in the
strong coupling are of about the same size than the zeroth order
perturbative result and contribute to about $35\%$ of the final
result. By switching to the running mass the main contribution of
the perturbative calculation is shifted to the zeroth-order term,
whereas the $O(\alpha_s)$-corrections are about $4\%$ of the final
result. Including the second order corrections from Chetyrkin and
Steinhauser \cite{Chetyrkin:2001je} in the pole mass scheme, the
contribution is again of the same size than the zeroth and first
order corrections. In the analysis of Jamin and Lange the
$O(\alpha_s^2)$-corrections are negligible.
\newline

Since the expansion in the strong coupling constant is under good
control compared to the sum rule using the pole mass for the heavy
quark, we will use the sum rule in the form of equation
(\ref{equ-fb-sum rule-sigma}) up to $O(\alpha_s)$ as the starting
point of our own analysis.

\section{The Limit $M^2\rightarrow\infty$}\label{sec-u-tu-infinity}

As argued above, the method of setting the daughter sum rule
(\ref{equ-daughter-sumrule}) equal to the meson mass is equivalent
to search for a threshold parameter $s_0$, for which the sum rule
for the decay constant is in a stable region. Another method was
suggested several times by Radyushkin
\cite{Nesterenko:1982gc,Szczepaniak:1998sa} in sum rule analyses
of the pion form factor. By taking the limit $M^2\rightarrow
\infty$, one can get rid off higher dimensional condensates and
the sum rule is again in a stable region. In this limit, the
uncertainties coming from the truncation of the power series
vanish, whereas higher resonances and the continuum contribution
are not suppressed exponentially anymore. Thus, the duality
approximation is the crucial assumption in this approach and the
source of the main error. In the limit $M^2\rightarrow\infty$, the
sum rule reads:
\begin{eqnarray}\label{equ-fb-sum rule-limit}
m_B^4f_B^2\!&=&\!
\frac{3m_b^2}{8\pi^2}\!\!\int\limits_{m_{b,pole}^2}^{s_0}\!\!\!ds
\frac{(m_b^2-s)^2}{s}
\left(1+\frac{4\alpha_s}{3\pi}\left(c^1(s)+\Delta^{(1)}\right)\right)
\nonumber\\
&&+m_b^2\left[\left(-1-2\frac{\alpha_s(\mu)}{\pi}
\left(\frac{7}{3}+\log\frac{\mu^2}{m_b^2}\right)\right)
m_b\langle\bar{q}q\rangle+\frac{1}{12}\langle
\frac{\alpha_s}{\pi}GG\rangle\right].
\end{eqnarray}
Apart from the constant contributions of the two condensates, the
decay constant is now described as the average of the transition
of a free quark pair of invariant mass $s$ in the duality interval
$m^2_{b,pole}<s<s_0$. In the analysis of the pion form factor
\cite{Nesterenko:1982gc}, the threshold is estimated to be the
midpoint between the first ($\pi$) and second ($A_1$) resonance.
Applied to the B-meson sum rule, the threshold should then be
around $s_0\simeq 31\;GeV^2$, assuming that the mass of the first
excited state $B'$ is about $0.5 \;GeV$ higher than the ground
state.
\newline

Instead of setting the threshold to the above value and varying it
within a certain interval, we propose to take a distinct value of
$s_0$. In the limit $M^2\rightarrow\infty$, the sum rule for the
decay constant is only a function of the threshold. Its first
derivative scales like ${f_B^2}'\propto 1/M^4$. Although $f_B$ is
asymptotic for any value of the threshold, we suggest taking the
value $s_0^*$, that satisfies
\begin{equation}\label{equ-asmypt}
M^4{f_B^2}'(M^2,s_0^*)\stackrel{M^2\rightarrow\infty}{\longrightarrow}0,
\end{equation}
for which the sum rule reaches its asymptotic value the fastest.
Using (\ref{equ-asmypt}) in equation (\ref{equ-first-derivative}),
$s_0^*$ can be found by setting the daughter sum rule of
(\ref{equ-fb-sum rule-jl}) to the meson mass in the limit
$M^2\rightarrow\infty$:
\begin{figure}[t]
\begin{picture}(150,50)
\put(2,2){
\begin{minipage}[t]{7cm}
\epsfig{file=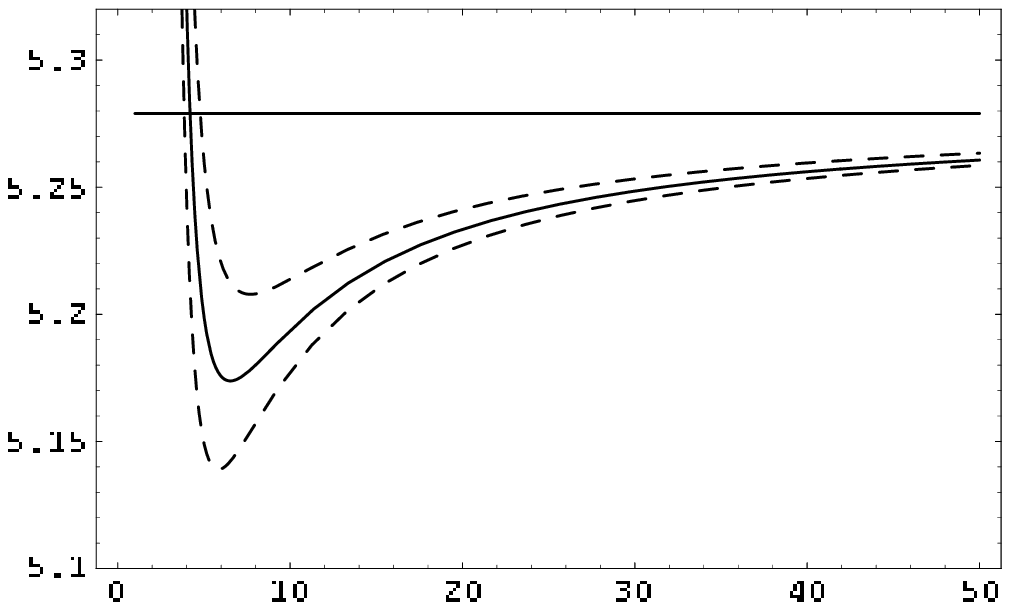,width=7cm}
\end{minipage}}
\Text(66,0)[c]{\scalebox{.8}{$M^2[GeV^2]$}}
\Text(24,46)[r]{\scalebox{.8}{$\Re^{1/2}[GeV]$}}
\Text(70,40)[r]{\scalebox{.7}{a)}}
\Text(50,25)[r]{\scalebox{.6}{$\overline{m}_b=4.35\;GeV$}}
\Text(47,21)[r]{\scalebox{.6}{$\overline{m}_b=4.24\;GeV$}}
\Text(44,17)[r]{\scalebox{.6}{$\overline{m}_b=4.13\;GeV$}}
\Text(50,38)[r]{\scalebox{.6}{$m_B=5.279\;GeV$}}
\put(35,25){\line(-4,1){7}} \put(32,21){\line(-5,1){7}}
\put(29,17){\line(-1,0){6}} \put(80,2){
\begin{minipage}[t]{7cm}
\epsfig{file=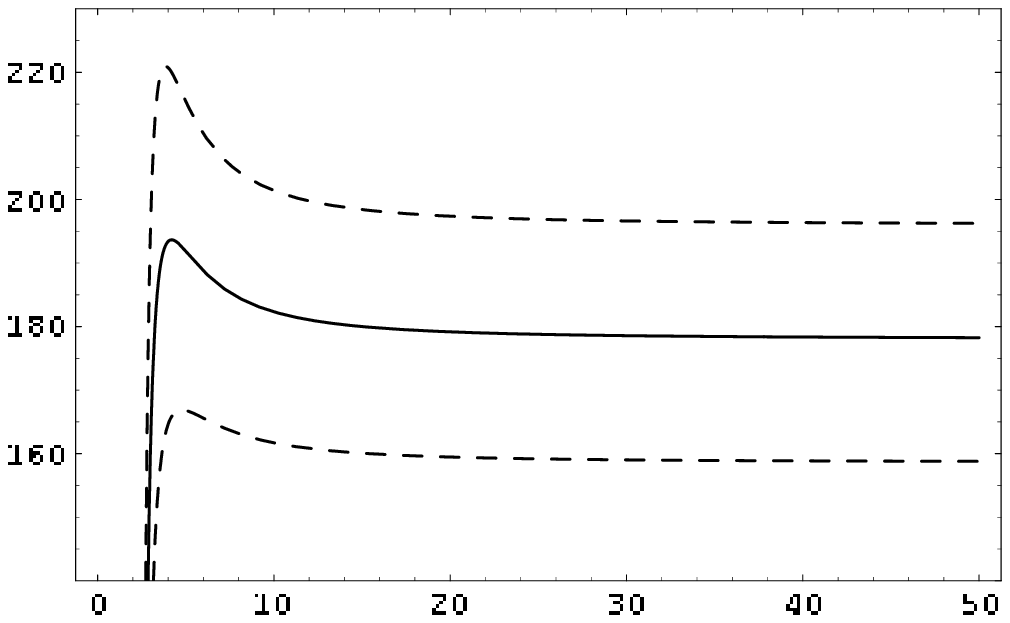,width=7cm,height=3.8cm}
\end{minipage}}
\Text(144,0)[c]{\scalebox{.8}{$M^2[GeV^2]$}}
\Text(99,46)[r]{\scalebox{.8}{$f_B[MeV]$}}
\Text(148,40)[r]{\scalebox{.7}{b)}}
\Text(132,31)[r]{\scalebox{.6}{$\overline{m}_b=4.13\;GeV$}}
\Text(132,23)[r]{\scalebox{.6}{$\overline{m}_b=4.24\;GeV$}}
\Text(132,14)[r]{\scalebox{.6}{$\overline{m}_b=4.35\;GeV$}}
\end{picture}
\caption{\small a: Matching of the daughter sum rule
(\ref{equ-daughter-sumrule}) to the meson mass $m_B=5.279 \;GeV$
in the limit $M^2\rightarrow\infty$ for different values of the
running quark mass $\overline{m}_b$. - b: The corresponding sum
rules for the decay constant, using the threshold $s_0^*$, found
by the matching procedure in the left
figure.}\label{fig-u-infinity} \vspace{3mm}
\end{figure}
\begin{equation}\label{equ-limit-dsr}
\Re(M^2,s_0^*)\stackrel{M^2\rightarrow\infty}{\longrightarrow}m_B^2.
\end{equation}
In figure \ref{fig-u-infinity}, we demonstrate the matching of
$\Re(M^2,s_0)$ to the meson mass for different values of the
running quark mass and show the functions of the corresponding sum
rules for the decay constant $f_B$. Using the values of table
\ref{tab-parameter}, we find a mean threshold of
$s_0^*=31.7\;GeV$. The corresponding decay constant is:
\begin{table}[b]\label{tab-parameter}
\centering
\begin{tabular}{|r||r|l|c|}
\hline
Parameter&Value&$\Delta f_B$&Taken from\\
\hline\hline
$\overline{m}_b(\overline{m}_b)$&$(4.24\pm0.11)\;GeV$&$\pm18$&
\cite{El-Khadra:2002wp}\\
$\langle\overline{q}q\rangle$&$-(267\pm16)^3\;MeV$&$\pm5$&
\cite{Jamin:2002ev}\\
$\mu$&$(3.5-6)\;GeV$&$\pm14$&-\\
$\langle\frac{\alpha_s}{\pi}GG\rangle$&$(0.012\pm0.012)
\;GeV^4$&$\pm1$&\cite{Khodjamirian:1998ji}\\
$m_0^2$&$(0.8\pm0.2)\;GeV^2$&$\pm0$&\cite{Jamin:2001fw}\\
$\alpha_s\langle\overline{q}q\rangle^2$&$(8\pm8)^{-5}\;GeV^6$&$
\pm0$&\cite{Khodjamirian:1998ji}\\
\hline
\end{tabular}
\caption{\small Values of the parameters used for the analysis
$M^2\rightarrow\infty$ and the corresponding errors they inflict
on the final value for the decay constant $f_B=178\;MeV$.}
\end{table}
\begin{equation}\label{equ-fb-u-infinity}
f_B=(178\pm18\pm20)\;MeV.
\end{equation}
Here, the first error is inflicted by the uncertainties in the
quark mass. We separate this error from the remaining ones, since
when the sum rule for the form factor in section \ref{sec-ff-uinf}
is divided by $f_B$ and the quark mass is varied simultaneously,
the uncertainties coming from the mass will almost cancel in the
final result for $f^+(0)$.
\newline

The uncertainties of the parameters used and the corresponding
errors to the final result are listed in table
\ref{tab-parameter}. The decay constant is very sensitive to the
running quark mass and the scale. At lower values of $\mu$, the
perturbative expansion becomes quite important. The inclusion of
second order corrections in the strong coupling would give a
better control of the scale dependence. However, we do not intend
to give a too accurate result, for the accuracy of this method
itself is limited intrinsically. Since the exponential damping of
the higher excitations and the continuum contribution is taken
away by the limiting procedure $M^2\rightarrow\infty$, errors
coming from the crude duality ansatz are rather emphasized. We
therefore sum up linearly all errors inflicted by the several
input parameters except the quark mass, leading to the relatively
large overall error in (\ref{equ-fb-u-infinity}).
\newline

Although the decay constant $f_B$ is very sensitive to the values
of the input parameters, the corresponding threshold $s_0^*$ is
not. Varying the parameters within the errors indicated in table
\ref{tab-parameter} and adjusting the threshold such that the
daughter sum rule matches the meson mass $m_B$ in the limit
$M^2\rightarrow\infty$ (\ref{equ-limit-dsr}), $s_0^*$ always stays
in the rather small interval
\begin{equation}
31.5\;GeV^2<s_0^*<32.0\;GeV^2.
\end{equation}
Vice versa, the value of the decay constant is quite sensitive to
the threshold $s_0$. Thus, taking the limit $M^2\rightarrow
\infty$ and guessing the threshold to be in a certain region,
instead of fixing it by the method described above, leads to large
uncertainties of the final result.
\newline

In section \ref{sec-fb-relation}, we will relate the value
(\ref{equ-fb-u-infinity}) with the result of the analysis of the
next section and compare our results with other analyses from the
literature.

\section{Borel Mass dependent Threshold $s_0(M^2)$}\label{s(u)}

In this section, we propose a further approach to analyze the sum
rule (\ref{equ-fb-sum rule}). As argued above and illustrated in
figure \ref{fig-mass}, the method of setting the logarithmic
derivative (\ref{equ-daughter-sumrule}) equal to the meson mass is
a rather sophisticated way to search for a value of the threshold
$s_0$, for that the sum rule of the decay constant is in a stable
region. This works nicely for the B-meson decay constant - the
resulting function of $f_B$ is almost constant over the Borel
window. However, there are sum rules, as will be seen later, in
which the resulting function for the hadronic parameter has a
strong dependence on $M^2$. Here, one cannot find a threshold
$s_0$ in a physical accessible region for which the first two
derivatives of the sum rule vanish and sometimes not even the
first one.
\newline

Ideally, the decay constant is not a function of the Borel
parameter. From a certain value of $M^2$ on, the expansion in the
vacuum condensates in equation (\ref{equ-fb-sum rule-jl}) is under
good control and its truncation does not induce a significant
error to the final result. Assuming that all errors to the sum
rule come from the duality approximation (\ref{semi-local}), we
introduce a Borel mass dependent threshold $s_0(M^2)$ that
compensates this error. This means, we search for functions
$s_0(M^2)$, for which the decay constant $f_B$ is not a function
of the Borel mass.
\newline

At first sight, this modification is quite similar to the initial
approach of adjusting the constant threshold such, that the sum
rule is in stable region. Instead of having a decay constant that
is dependent on $M^2$, this dependence is now shifted to the
threshold $s_0(M^2)$. Furthermore one does not find a unique value
of the decay constant. For each value of $f_B$, one should be able
to find a function $s_0(M^2)$ . However, this approach has an
intuitive interpretation. It is rather clear that the hadronic
spectral density can only be represented approximately by the
perturbative free quark spectral function. Introducing in
(\ref{semi-local}) a $q^2$-dependent threshold can compensate for
this error. For each distinct value of the momentum transfer, one
can than impose a value $s(q^2)$ rendering the approximation
exact. After Borel transformation, the resulting sum rule for the
decay constant should then be independent of $M^2$ provided that
the errors induced from the truncation of the power series and the
expansion in the strong coupling are negligible\footnote{In the
analysis below, we introduce the functional dependence of the
threshold after the Borel transformation, since the limiting
procedure of equation (\ref{equ-laplace-op}) would be non-trivial
if one included the dependence on the momentum transfer $q^2$
before the transformation. Nevertheless, the physical picture,
that shifting the $q^2$($M^2$)-dependence to the threshold is
compensating errors coming from the duality approximation, stays
the same}.
\newline

It still remains the question of where to read off the value of
the decay constant, since one can find a continuum of functions
$s_0(M^2)$, for which the sum rule is constant over $M^2$. At this
point, the duality assumption reenters the analysis. Assuming
(\ref{semi-local}) is a good approximation in the sense that one
can find a constant threshold for which the two sides equal, we
suggest looking for the most stable function $s_0(M^2)$ to reflect
this duality ansatz. Thus, whereas in most other applications the
threshold is a constant and its value enters by physical reasoning
or is found by searching for the most stable function $f(M^2)$,
the proposed modification features constant functions $f_B$ in a
given Borel window and we search for the most stable functions
$s_0(M^2)$.
\begin{figure}[t]
\begin{picture}(145,90)
\put(2.75,51){
\begin{minipage}[t]{7cm}
 \epsfig{file=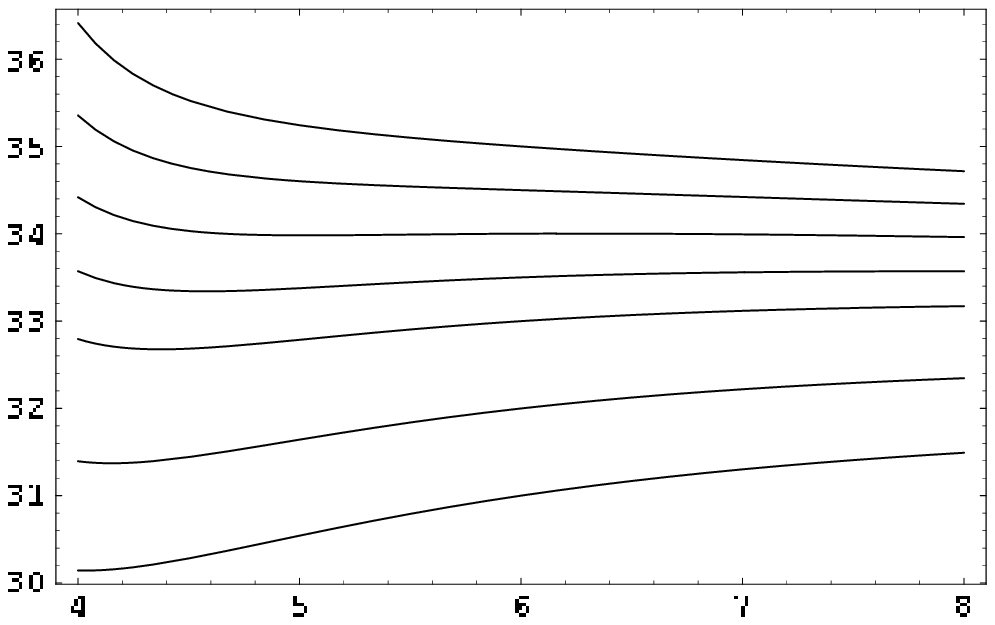,width=6.8cm,height=3.9cm}
\end{minipage}}
\Text(64,49.5)[c]{\scalebox{.8}{$M^2[GeV^2]$}}
\Text(9,95.5)[l]{\scalebox{.8}{$s_0(M^2)[GeV^2]$}}
\Text(68,90)[r]{\scalebox{.7}{a)}} \put(0,2){
\begin{minipage}[t]{7cm}
 \epsfig{file=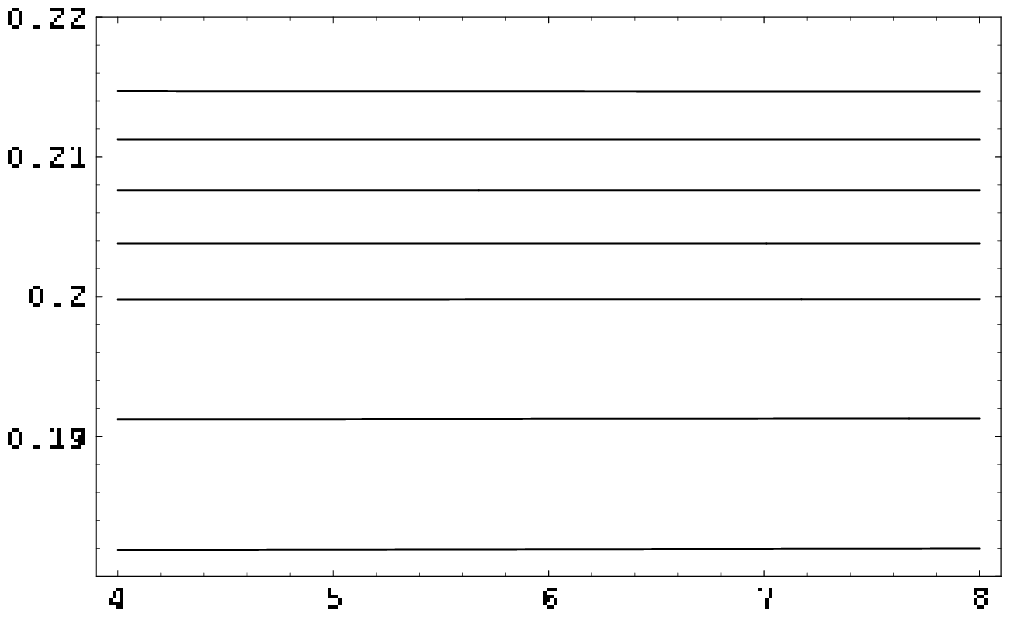,width=7cm}
\end{minipage}}
\Text(64,0.3)[c]{\scalebox{.8}{$M^2[GeV^2]$}}
\Text(9,46)[l]{\scalebox{.8}{$f_B[GeV]$}}
\Text(68,41)[r]{\scalebox{.7}{b)}}
\Text(87,46)[l]{\scalebox{.8}{$f_B[GeV]$}} \put(78,51){
\begin{minipage}[t]{8cm}
 \epsfig{file=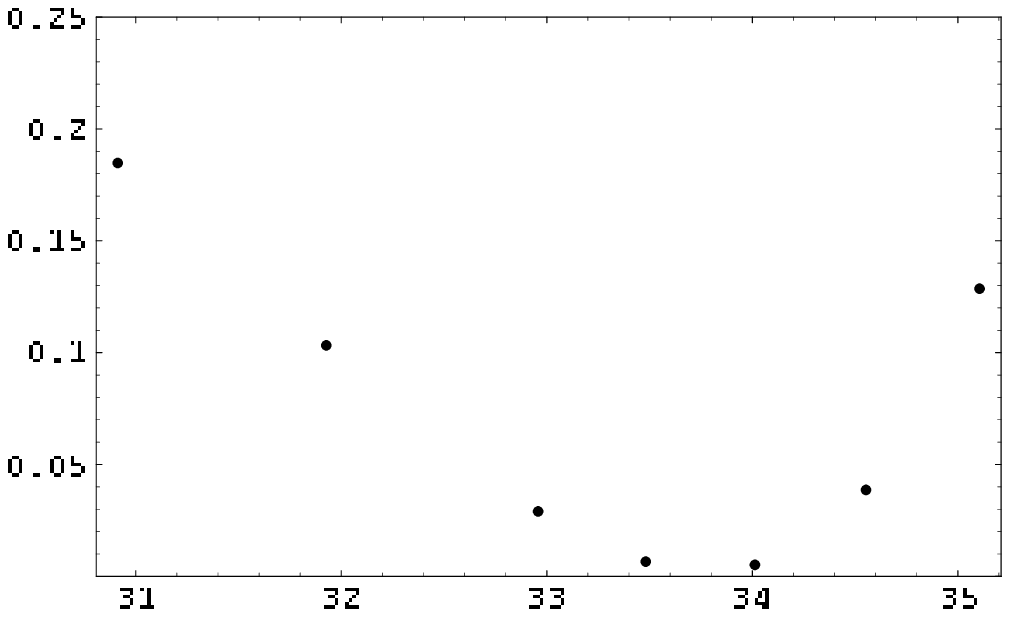,width=7cm}
\end{minipage}}
\Text(141,49.5)[c]{\scalebox{.8}{$\overline{s}_0[GeV^2]$}}
\Text(89,94.5)[c]{\scalebox{.8}{$s_{var}$}}
\Text(144,90)[r]{\scalebox{.7}{c)}} \put(78,2.2){
\begin{minipage}[t]{8cm}
 \epsfig{file=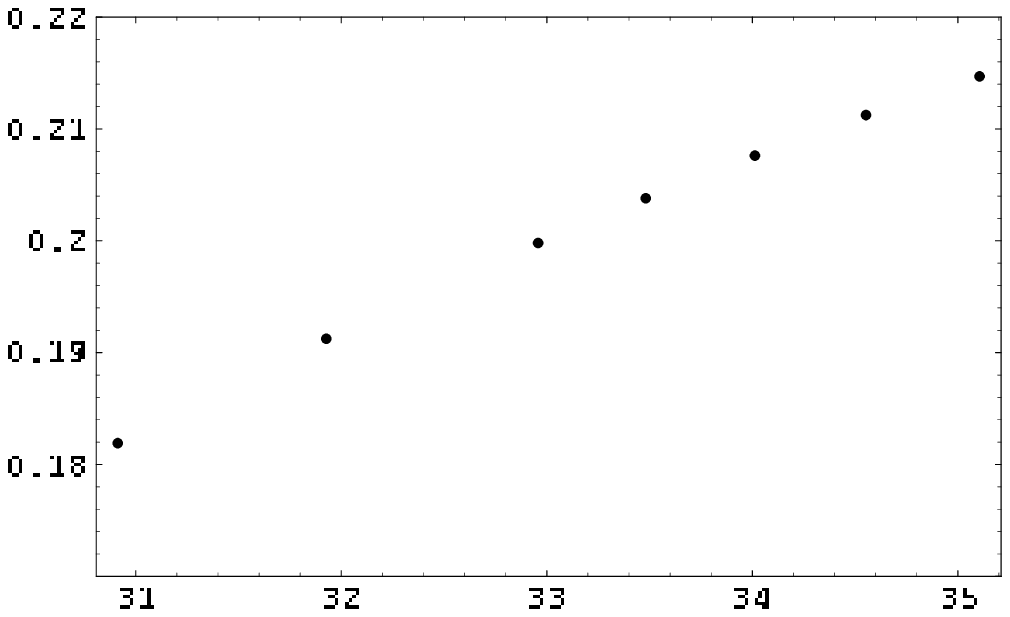,width=7cm,height=4.28cm}
\end{minipage}}
\Text(142,0.3)[c]{\scalebox{.8}{$\overline{s}_0[GeV^2]$}}
\Text(144,41)[r]{\scalebox{.7}{d)}}
\end{picture}
\caption{\small Sum rule analysis for Borel mass dependent
thresholds. Graph a) shows various functions for $s_0(M^2)$, which
are solutions to the differential equation (\ref{equ-diff-equ}).
Below, the corresponding functions for the decay constant are
plotted (b). Graph c) shows the variation of the threshold
functions $s_0(M^2)$ plotted in a), as a function of their average
value in the given Borel window. Diagram d) shows the decay
constant plotted over the average values of the
thresholds.}\label{fig-sofu} \vspace{3mm}
\end{figure}
\newline

Since the sum rule for $f_B(M^2)$, which results from setting the
minimum of the daughter sum rule equal to the meson mass is
already quite constant in the window $4\:GeV\leq M^2 \leq 8\:GeV$
(see figure \ref{fig-mass}), it is not surprising, that the
following analysis will lead to the same result.
\newline

Taking the derivative $M^4 d/dM^2$ of the logarithm of both sides
of equation (\ref{equ-fb-sum rule-sigma}) yields:
\begin{equation}\label{equ-fb-log-abl}
m_B^2+\frac{M^4
{f_B^2}'(M^2)}{f_B^2(M^2)}=\frac{M^4\frac{\partial}{\partial
M^2}\Sigma(M^2,s_0(M^2))+M^4 s_0'(M^2)\frac{\partial}{\partial
s_0}\Sigma(M^2,s_0(M^2))}{\Sigma(M^2,s_0(M^2))}.
\end{equation}
Setting the right hand side equal to the squared meson mass
$m_B^2$, we get a differential equation for the function
$s_0(M^2)$:
\begin{equation}\label{equ-diff-equ}
s_0'(M^2)=\frac{m_B^2\Sigma(M^2,s_0(M^2))-M^4\frac{\partial}{\partial
M^2}\Sigma(M^2,s_0(M^2))}{M^4\frac{\partial}{\partial
s_0}\Sigma(M^2,s_0(M^2))}.
\end{equation}
We do not write the explicit form of the right hand side, since it
is a rather complicated function of $s_0(M^2)$ and its derivation
is straightforward. Setting the solutions of equation
(\ref{equ-diff-equ}) into the sum rule for the decay constant will
yield constant functions of $f_B$.
\newline

This equation was solved numerically. We started at certain values
for the threshold $s_0$ at $M^2=6\;GeV^2$ and the functions were
then evaluated with mathematica, using a numerical procedure with
adaptive step size.
\newline

Figure \ref{fig-sofu}$\,$a shows some solutions of $s_0(M^2)$
we obtained for the mean values of the input parameters listed in
table \ref{tab-parametersofu}. The corresponding decay constants
are given below (b). As argued above, we read off the decay
constant at the most stable function of $s_0(M^2)$. This could be
done by simply looking at the graphs of the threshold $s_0(M^2)$,
since the method itself contains implicit errors and one should
not seek for a too accurate result. However, to demonstrate the
equivalence to the result (\ref{equ-result-jl}) obtained by
following the analysis of Jamin and Lange \cite{Jamin:2001fw}, we
introduce a variation of the threshold functions:
\begin{figure}[t]
\begin{picture}(145,90)
\put(3.5,51){
\begin{minipage}[t]{7cm}
 \epsfig{file=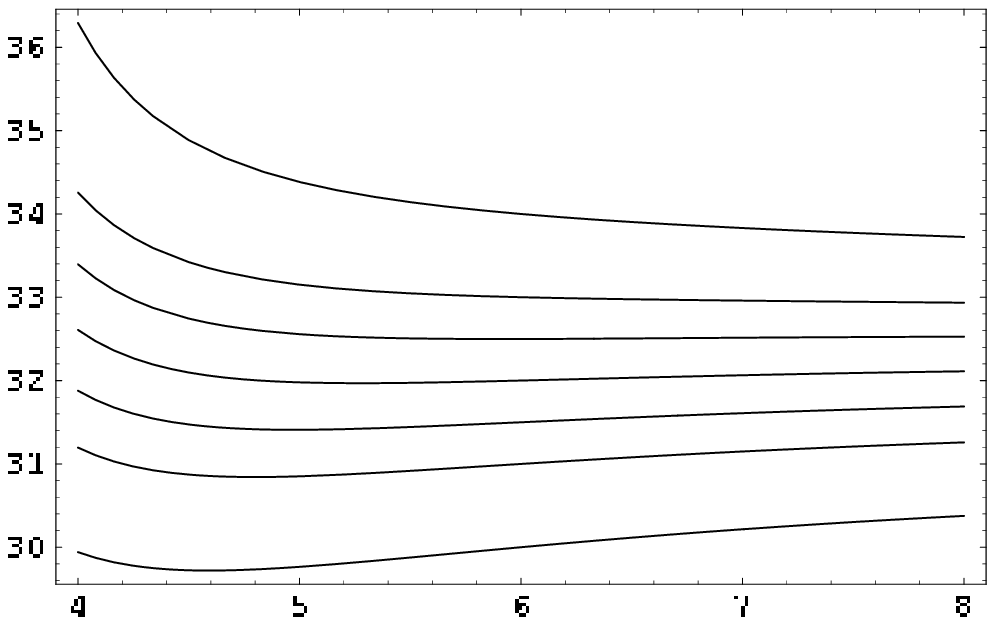,width=7cm,height=4cm}
\end{minipage}}
\Text(64,49.5)[c]{\scalebox{.8}{$M^2[GeV^2]$}}
\Text(10,95.5)[l]{\scalebox{.8}{$s_0(M^2)[GeV^2]$}}
\Text(68,90)[r]{\scalebox{.7}{a)}} \put(1,2){
\begin{minipage}[t]{7cm}
 \epsfig{file=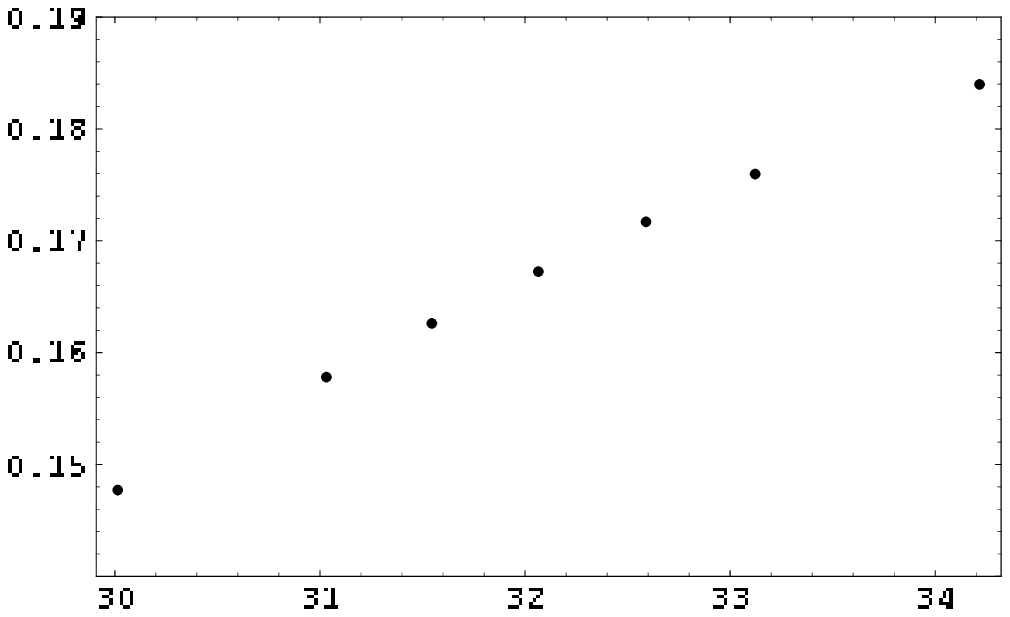,width=7cm,height=4.2cm}
\end{minipage}}
\Text(64,0.3)[c]{\scalebox{.8}{$\overline{s}_0[GeV^2]$}}
\Text(10,46)[l]{\scalebox{.8}{$f_B[GeV]$}}
\Text(68,41)[r]{\scalebox{.7}{b)}}
\Text(88,46)[l]{\scalebox{.8}{$f_B[GeV]$}} \put(80.5,51){
\begin{minipage}[t]{8cm}
 \epsfig{file=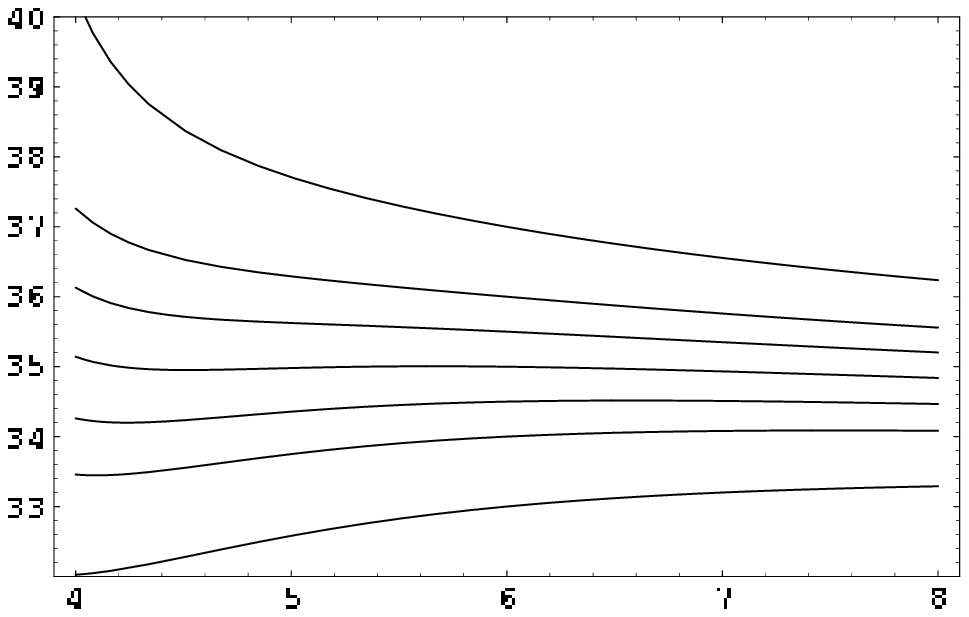,width=7cm,height=4.3cm}
\end{minipage}}
\Text(142,49.5)[c]{\scalebox{.8}{$M^2[GeV^2]$}}
\Text(89,95.5)[l]{\scalebox{.8}{$s_0(M^2)[GeV^2]$}}
\Text(120,95.5)[l]{\scalebox{.8}{$m_b=4.13\;GeV$}}
\Text(144,90)[r]{\scalebox{.7}{c)}} \put(79,2){
\begin{minipage}[t]{8cm}
 \epsfig{file=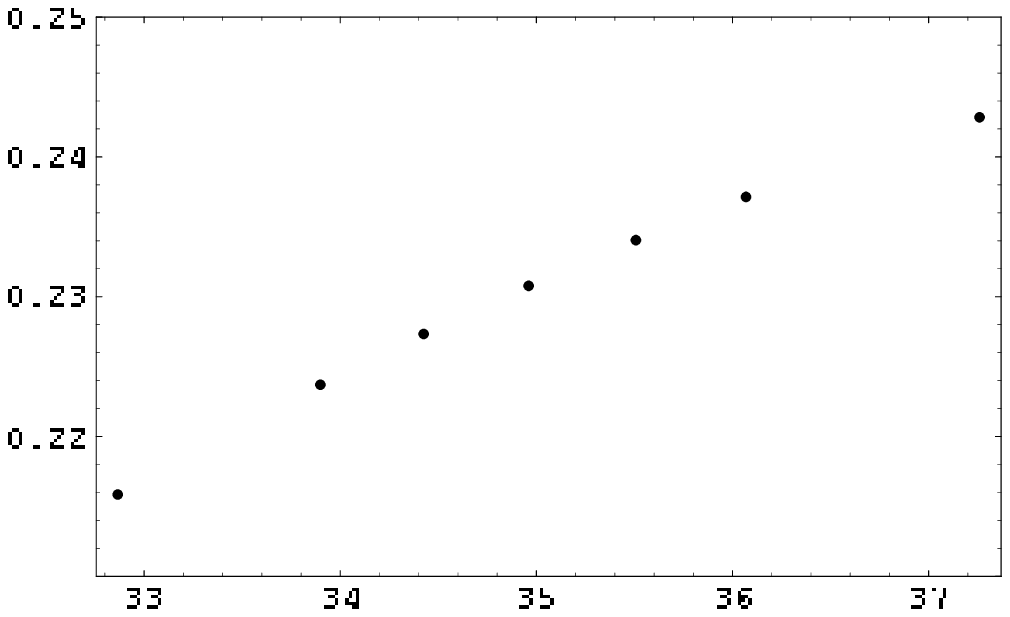,width=7cm,height=4.3cm}
\end{minipage}}
\Text(143,0.5)[c]{\scalebox{.8}{$\overline{s}_0[GeV^2]$}}
\Text(40,95.5)[l]{\scalebox{.8}{$m_b=4.35\;GeV$}}
\Text(144,41)[r]{\scalebox{.7}{d)}}
\end{picture}
\caption{\small Borel mass dependent functions for the threshold
$s_0(M^2)$ and corresponding decay constants $f_B$, plotted over
the average value of the thresholds. On the left side
$m_b=4.35\;GeV$ was used as input value for the running quark mass
- on the right side $m_b=4.13\;GeV$.}\label{fig-sofu-mass-errors}
\vspace{3mm}
\end{figure}
\begin{equation}
{s^n_{var}}=\frac{1}{4}\int\limits_4^8(s^n_0(M^2)-
\overline{s^n_0})^2dM^2.
\end{equation}
In figure \ref{fig-sofu}$\,$c, we listed the variations of the
functions $s_0(M^2)$ over their average value. Graph
\ref{fig-sofu}$\,$d shows the corresponding decay constants, again
plotted as a function of the mean value of the threshold. Reading
off the decay constant at the point with the least variation gives
\begin{equation}
f_B=208 \:MeV.
\end{equation}
Thus, we reproduced the result (\ref{equ-result-jl}) from the
analysis in section \ref{comparison-jamin-lange}, following Jamin
and Lange.
\newline

In table \ref{tab-parametersofu}, we listed the value of the input
parameter we used for this analysis and the corresponding errors
they inflict on the final result. In figure
\ref{fig-sofu-mass-errors} we plotted the results of our analysis
for the b-quark masses $m_b=4.13\;GeV$ and $m_b=4.35\;GeV$. Again,
$f_B$ is very sensitive to the running mass and the scale. We
again separate the error coming from the quark mass from the
remaining errors, added in quadrature:
\begin{table}[b]\label{tab-parametersofu}
\centering
\begin{tabular}{|r||r|c|c|c|}
\hline
Parameter&Value&$-\Delta f_B$&+$\Delta f_B$&Taken from\\
\hline\hline
$\overline{m}_b(\overline{m}_b)$&$(4.24\pm0.11)\;GeV$&$231$&$167$&
\cite{El-Khadra:2002wp}\\
$\langle\overline{q}q\rangle$&$-(267\pm16)^3\;MeV$&$200$&$212$&
\cite{Jamin:2002ev}\\
$\mu$&$(3.5-6)\;GeV$&$192$&$221$&-\\
$\langle\frac{\alpha_s}{\pi}GG\rangle$&$(0.012\pm0.012)
\;GeV^4$&$207$&$208$&\cite{Khodjamirian:1998ji}\\
$m_0^2$&$(0.8\pm0.2)\;GeV^2$&$208$&$203$&\cite{Jamin:2001fw}\\
$\alpha_s\langle\overline{q}q\rangle^2$&$(8\pm8)^{-5}
\;GeV^6$&$208$&$208$&\cite{Khodjamirian:1998ji}\\
\hline
\end{tabular}
\caption{\small Input parameters used for the analysis with a
Borel mass dependent threshold. The column $-\Delta f_B$($+\Delta
f_B$) shows the deviations from the central result $f_B=208\;MeV$,
when the lower(higher) limit of the input parameter is used.}
\end{table}
\begin{equation}\label{equ-decay-constant-s(u)}
f_B=(208^{+23\;\;+19}_{-41\;\;-14})\:MeV
\end{equation}
The total error of $f_B$ is rather large. This is mainly due to
the large uncertainties, which we assigned to the quark mass. In
the following two chapters, we will analyze the sum rules for the
B-meson semi-leptonic form factor $f^+(0)$ and the coupling
$g^{}_{B^*\!B\pi}$. These sum rules are proportional to the decay
constant. When the two sum rule results are divided by $f_B$ and
the quark masses are varied simultaneously, the final errors due
to the uncertainties in the quark mass decrease drastically.
\newline

We also checked that changes in the boundaries of the Borel window
have a minor effect on the corresponding variations of the
functions $s_0(M^2)$ and the point of extraction of the decay
constant.

\section{Summary of the Results and Relation to the Literature}
\label{sec-fb-relation}

The two results (\ref{equ-fb-u-infinity}) and
(\ref{equ-decay-constant-s(u)}) differ about $15\%$. In the sum
rules we analyze in this thesis, the method of taking the limit
$M^2\rightarrow\infty$ will always lead to significantly lower
results than the method of imposing a Borel mass dependent
threshold $s_0(M^2)$. Considering only the LO-perturbative term of
the daughter sum rule
\begin{table}[b]
\centering
\begin{tabular}{|l|c|l|}
\hline $f_B[MeV]$&Ref.&Method \\
\hline \hline
$178\pm27$&This Thesis&\small SVZ-SR $O(\alpha_s)$; running mass;
$\lim M^2\rightarrow\infty$ \\
$208^{+30}_{-43}$&This Thesis&\small SVZ-SR $O(\alpha_s)$; running
mass; $s_0(M^2)$\\
$180\pm30$&\small Khodjamirian '98\cite{Khodjamirian:1998ji}&
\small SVZ-SR $O(\alpha_s)$; pole mass\\
$190\pm9$&\small Narison '98\cite{Narison:1998ui}&
\small SVZ-SR $O(\alpha_s)$; running mass\\
$210\pm19$&\small Jamin '01\cite{Jamin:2001fw}&
\small SVZ-SR $O(\alpha_s^2)$; running mass\\
$205\pm23$&\small Narison '01\cite{Narison:2001pu}&
\small SVZ-SR $O(\alpha_s^2)$; running mass\\
$173\pm13^{+34}_{-1}$&\small Abada APE'99\cite{Abada:1999xd}&
\small Lattice\\
$218\pm5^{+5}_{-41}$&\small Bowler UKQCD'00\cite{Bowler:2000xw}&
\small Lattice\\
$191\pm23^{+0}_{-19}$&\small Wittig '03\cite{Wittig:2003cf}&
\small
Lattice-summary of unquenched results\\
\hline
\end{tabular}
\caption{\label{tab-literature-fb}\small Comparison of our results
with a collection of values from other analyses in the
literature.}
\end{table}
\begin{equation}\label{equ-lo-pert}
\int\limits_{m_b^2}^{s_0}ds \,s f(s)e^{-\frac{s}{M^2}},
\end{equation}
it is clear from the positivity of the integrand $f(s)$ and the
vanishing of the exponential damping in the limit
$M^2\rightarrow\infty$, that the relevant value of the integral
needed in the fitting procedure is reached at a rather small
threshold $s_0$. In the second method, we are looking for
functions $s_0(M^2)$ that do not depend strongly on the Borel
mass. Thus, the contribution from the term proportional to
$s'_0(M^2)$ in (\ref{equ-fb-log-abl}) is small compared to
(\ref{equ-lo-pert}) and therefore, the upper limit $s_0(M^2)$ of
the integral has to be higher than in the former case. The
difference in the size of the two thresholds translates into the
discrepancy of the final results for the decay constant. This can
be inferred from the figures \ref{fig-u-infinity} and
\ref{fig-mass}. Since the limit $\Re(M^2,s_0^*)\rightarrow m_B^2$
is reached from below, the corresponding sum rule for the decay
constant is monotonously decreasing and in the Borel window of
figure \ref{fig-mass} it would lie below the plotted curves.
\newline

In table \ref{tab-literature-fb}, we listed our results for the
decay constant and several results from the literature. The latest
sum rule analyses \cite{Jamin:2001fw} and \cite{Narison:2001pu},
including three-loop corrections to the perturbative calculations
and using the running quark mass, coincide with our result from
using a Borel mass dependent threshold. The result from Wittig
\cite{Wittig:2003cf} is a global estimate of recent unquenched
lattice calculations. It is about $10\%$ lower than the sum rule
results. Nevertheless, all the quoted values are in agreement,
when the error bars are taken into account.
\newline

Before we proceed with the analysis of the B-meson weak form
factor $f^+(0)$ and the strong coupling $g^{}_{B^*\!B\pi}$, we
present briefly the analysis of the two point sum rule for the
vector meson decay constant $f_{B^*}$.

\section{Weak Decay Constant $f_{B^*}$}\label{sec-fstar}

The decay constant $f_{B^*}$ is needed in the analysis for the
strong coupling $g^{}_{B^*\!B\pi}$, since the sum rule we use to
extract this coupling is proportional to the decay constants of
the $B$- and the $B^*$-meson. We will analyze the light cone sum
rule for the coupling by the use of the two new methods introduced
in this chapter. To get a consistent value for $g^{}_{B^*\!B\pi}$,
we will divide the sum rule by the corresponding decay constants
$f_B$ and $f_{B^*}$, obtained within the same method. We therefore
summarize in this section our analysis of the $B^*$ decay
constant.
\begin{table}[b]
\centering
\begin{tabular}{|r||r|c|c|c|}
\hline
Parameter&Value&$-\Delta f_{B^*}$&$+\Delta f_{B^*}$&Taken from\\
\hline\hline
$\overline{m}_b(\overline{m}_b)$&$(4.24\pm0.11)\;GeV$&$221$&$174$&
\cite{El-Khadra:2002wp}\\
$\langle\overline{q}q\rangle$&$-(267\pm16)^3\;MeV$&$192$&$203$&
\cite{Jamin:2002ev}\\
$\mu$&$(3.5-5)\;GeV$&$182$&$208$&-\\
$\langle\frac{\alpha_s}{\pi}GG\rangle$&$(0.012\pm0.012)
\;GeV^4$&$197$&$198$&\cite{Khodjamirian:1998ji}\\
$m_0^2$&$(0.8\pm0.2)\;GeV^2$&$197$&$197$&\cite{Jamin:2001fw}\\
\hline
\end{tabular}
\caption{\label{tab-fbstar-errors-uinf}\small Values of the input
parameters used in the sum rule analysis of $f_{B^*}$ in the limit
$M^2\rightarrow\infty$ and the corresponding errors they inflict
on the final result. The central result is $f_{B^*}=197 \;MeV$.}
\end{table}
\begin{table}[t]
\centering
\begin{tabular}{|r||r|c|c|c|}
\hline
Parameter&Value&$-\Delta f_{B^*}$&+$\Delta f_{B^*}$&Taken from\\
\hline\hline
$\overline{m}_b(\overline{m}_b)$&$(4.24\pm0.11)\;GeV$&$281$&$213$&
\cite{El-Khadra:2002wp}\\
$\langle\overline{q}q\rangle$&$-(267\pm16)^3\;MeV$&$233$&$257$&
\cite{Jamin:2002ev}\\
$\mu$&$(3.5-5)\;GeV$&$226$&$258$&-\\
$\langle\frac{\alpha_s}{\pi}GG\rangle$&$(0.012\pm0.012)
\;GeV^4$&$245$&$246$&\cite{Khodjamirian:1998ji}\\
$m_0^2$&$(0.8\pm0.2)\;GeV^2$&$252$&$243$&\cite{Jamin:2001fw}\\
\hline
\end{tabular}
\caption{\label{tab-fbstar-errors-s(u)}\small Values of the input
parameters used in the sum rule analysis of $f_{B^*}$ with a Borel
mass dependent threshold $s_0(M^2)$ and the corresponding errors
they inflict on the final result. The central result of this
analysis is $f_{B^*}=245 \;MeV$.} \vspace{3mm}
\end{table}
\newline

We collected the sum rule for the heavy-light vector meson
constant from Dominguez\cite{Dominguez:1990nz} and
Reinders\cite{Reinders:1985sr}. Neglecting the numerically
insignificant contribution of the four quark condensate, we get up
to mass dimension 5 in the condensates and first order in the
perturbative expansion:
\begin{eqnarray}
f_{B^*}^2m_{B^*}^2e^{-\frac{m_{B^*}^2}{M^2}}&=&\frac{1}{8\pi}
\int\limits_{m_b^2}^{s_0}ds\frac{(s-m_b^2)^2}{s}
\left(2+\frac{m_b^2}{s}\right)\left[1+\frac{4\alpha_s}{3\pi}
\left(f^{(1)}(s)+\Delta^{(1)}(s)\right)\right]e^{-\frac{s}{M^2}}
\nonumber\\
&&+\left[-m_b\langle\overline{q}q\rangle
\left(1-\frac{m_0^2m_b^2}{4M^4}\right)+
\frac{1}{12}\langle\frac{\alpha_s}{\pi}GG\rangle\right]
e^{-\frac{m_b^2}{M^2}},
\end{eqnarray}
where the first order correction in the strong coupling is given
by:
\begin{eqnarray}
f^{(1)}(s)&=&\frac{13}{4}+2Li_2\left(\frac{m_b^2}{s}\right)+
\log\left(\frac{m_b^2}{s}\right)\log\left(1-\frac{m_b^2}{s}\right)
+\frac{3}{2}\frac{m_b^2}{2s+m_b^2}\log\left(\frac{m_b^2}{s-m_b^2}
\right)\nonumber\\
&&-\log\left(1-\frac{m_b^2}{s}\right)-
\frac{(4s^2-m_b^2s-m_b^4)}{(s-m_b^2)^2(2s+m_b^2)}
m_b^2\log\left(\frac{m_b^2}{s}\right)-
\frac{5s^2-m_b^2s-2m_b^4}{(s-m_b^2)(2s+m_b^2)}.\nonumber\\
\end{eqnarray}
We again replaced the pole mass in the integrand by the running
$\overline{MS}$-mass. This lead to the additional term:
\begin{equation}
\Delta^{(1)}(s)=\frac{6m_b^2(s+m_b^2)}{m_b^4+m_b^2s-2s^2}
\left(1+\frac{3}{4}\log\frac{\mu^2}{m_b^2}\right)
\end{equation}
As before, with the pole mass replaced by the running mass, the
sum rule shows better convergence. The first order correction
amounts to about $20\%$ of the perturbative part, whereas before
this contribution was about $60\%$. The use of the running mass
increases the overall perturbative contribution about $30\%$.
\newline

We analyzed this sum rule using the two methods introduced in the
preceding sections. The values for the decay constant we extracted
are:
\begin{equation}\label{equ-fstar-uinf}
f_{B^*}=(197 \pm 24\pm 16)\;MeV,
\end{equation}
when the limit $M^2\rightarrow\infty$ is taken and
\begin{equation}\label{equ-fstar-s(u)}
f_{B^*}=(245\pm 35\pm 23)\;MeV
\end{equation}
in the case of a Borel mass dependent threshold $s_0(M^2)$. Again,
the first errors come from the uncertain running quark mass
$\overline{m}_b(\overline{m}_b)=(4.24\pm0.11)\;GeV$. The second
error summarizes further uncertainties. The individual
contributions of the uncertainties of the input parameters and the
dependence on the scale $\mu$ are collected in the tables
\ref{tab-fbstar-errors-uinf} and \ref{tab-fbstar-errors-s(u)}.
\newline

The result (\ref{equ-fstar-uinf}) is close to a NLO-result from
Khodjamirian\cite{Khodjamirian:1999hb}, $f_{B^*}=(195\pm35)\;MeV$,
whereas (\ref{equ-fstar-s(u)}) is far off that value. This can be
traced back to the strong dependence of the sum rule on the Borel
parameter if a constant threshold $s_0$ is used. To stabilize the
sum rule by introduction of a Borel mass dependent threshold will
require rather high average values of $s_0(M^2)$. Nevertheless, we
will use the two decay constants from this analysis in the
extraction of the coupling $g^{}_{B^*\!B\pi}$.

\chapter{B-Form Factor $f^+(0)$}\label{sec-Form Factor}

In this chapter, we analyze the weak semileptonic form factor
$f_B^+(0)$ of the B-meson at maximal recoil. In section
\ref{sec-lcsr} we already gave the sum rule for the form factor of
the twist two pion wave function. Up to twist 4 we adopt the sum
rule from Khodjamirian \cite{Khodjamirian:1998ji}. For $q^2=0$
this is:
\begin{equation}\label{equ-sr-ff}
f_Bf^+(0)e^{-\frac{m_B^2}{M^2}}=\frac{m_b^2f_\pi}{2m_B^2}\Phi(s_0,M^2),
\end{equation}
where $\Phi(s_0,M^2)$ is given by:
\begin{eqnarray}\label{equ-form-factor}
\Phi(s_0,M^2)&=&\int\limits_\Delta^1\frac{du}{u}e^{-\frac{m_b^2}{uM^2}}
\Bigg[\varphi_\pi(u)
+\frac{\mu_\pi}{m_b}\left(u\varphi_p(u)+\frac{\varphi_\sigma(u)}{3}
\left(1+\frac{m_b^2}{2uM^2}\right)\right)-\frac{4m_b^2}{u^2M^4}g_1(u)
\nonumber\\
&&\hspace{2.3cm}-\frac{4m_b^2}{u^2M^4}g_1(u)+
\frac{2}{uM^2}\int\limits_0^u dt
g_2(t)\left(1+\frac{m_b^2}{uM^2}\right)\Bigg]\nonumber\\
&&\hspace{2.1cm}+A^+(s_0,M^2)+f^+_{3P}(s_0,M^2)-
\frac{2\alpha_s}{3\pi}T^1(s_0,M^2).
\end{eqnarray}
Here, the lower limit in the integral is now given by
$\Delta=m_b^2/s_0$. $A^+(s_0,M^2)$ is the surface term, coming
from the partial integration necessary to build up a dispersion
relation. The explicit form of the wave functions as well as the
three particle contributions and the first order corrections to
the twist 2 wave function is given in appendix \ref{app-lc-wf}.
\newline

The right hand side of equation (\ref{equ-form-factor}) is
effectively an expansion in powers of $1/uM^2$ rather than just
$M^2$. Taking the Borel window of the sum rule for the decay
constant $f_B$, $4\;GeV^2<M^2<8\;GeV^2$, and dividing the
boundaries by an average value of the integration variable $u$, we
obtain a new window that takes the change from the parameter $M^2$
to $uM^2$ into account. Evaluating the integral in
(\ref{equ-form-factor}) in the Borel window for the decay
constant, we find an average value of $\overline u$ varying
between $0.5$ and $0.7$, for which the integrand is equal to the
evaluated integral. Thus, we use for the analysis of the form
factor a window with boundaries of about the doubled values as for
the decay constant:
\begin{figure}[t] \centering
\begin{picture}(110,75)
\put(0,2) {\begin{minipage}[t]{11cm}
\epsfig{file=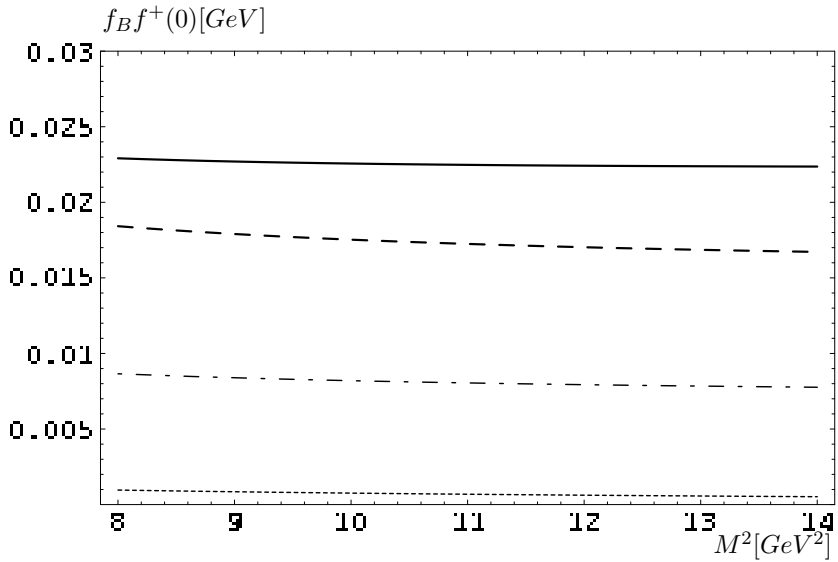,width=11cm}
\end{minipage}}
\Text(110,0)[r]{\scalebox{.9}{$M^2[GeV^2]$}}
\Text(13,70)[l]{\scalebox{.9}{$f_Bf^+(0)[GeV]$}}
\end{picture}
\caption{\small Contributions to the final sum rule for
$f_Bf^+(0)$: twist 2 (full line); 2P - twist 3 (dashed);
$O(\alpha_s)$-twist 2 (dash-dotted); 3P - twist 3 and twist 4
(dotted). We used the central values $m_b=4.7\;GeV$ and
$s_0=35\;GeV^2$ as input parameters. The sum of all contributions
is plotted in figure
\ref{fig-Form-Factor-Khod}.}\label{fig-anteile} \vspace{3mm}
\end{figure}
\begin{equation}
8\;GeV^2<M^2<14\;GeV^2.
\end{equation}
We checked that the hadronic contributions of higher states and
the continuum, which are subtracted from the QCD side by the
duality assumption are of about $25\%$ of the final result on the
right edge of the Borel window. Higher twist contributions are
under good control and the contributions of the three particle
wave functions are almost negligible. For central input
parameters, we depicted the contributions from the single terms to
the total sum rule in figure \ref{fig-anteile}. The dotted line
collects the two particle twist 4 and all the three particle twist
3 and 4 contributions. The sum of these terms is only about $2\%$.
Therefore, the truncation of the expansion after these terms is
justified.
\newline

\begin{figure}[t]
\begin{picture}(150,50)
\put(2,2.5){
\begin{minipage}[t]{7cm}
 \epsfig{file=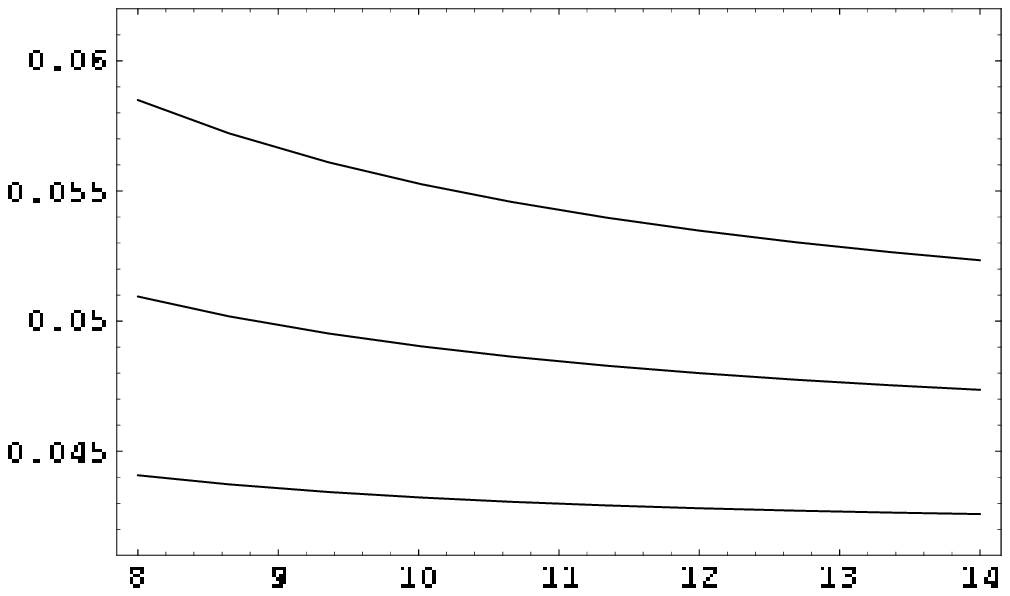,width=7cm}
\end{minipage}}
\Text(66,0.2)[c]{\scalebox{.8}{$M^2[GeV^2]$}}
\Text(31,45)[r]{\scalebox{.8}{$f_Bf^+(0)[GeV]$}}
\Text(70,40)[r]{\scalebox{.7}{a)}}
\Text(50,32)[r]{\scalebox{.6}{$m_b=4.6\;GeV$}}
\Text(47,21)[r]{\scalebox{.6}{$m_b=4.7\;GeV$}}
\Text(44,11)[r]{\scalebox{.6}{$m_b=4.8\;GeV$}} \put(80,2.1){
\begin{minipage}[t]{7cm}
\epsfig{file=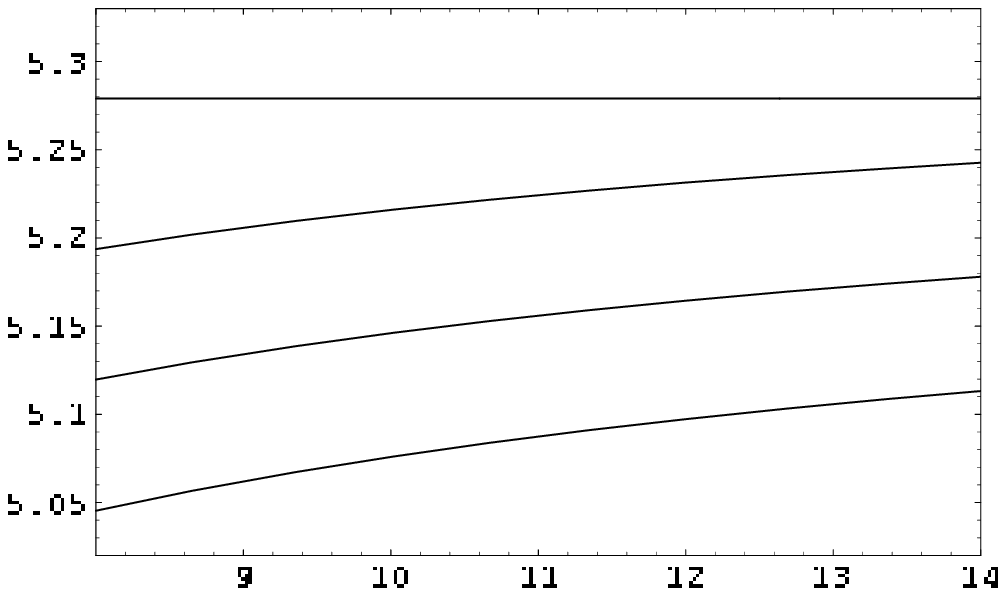,width=7cm,height=3.8cm}
\end{minipage}}
\Text(142,0.2)[c]{\scalebox{.8}{$M^2[GeV^2]$}}
\Text(102,45)[r]{\scalebox{.8}{$\Re^{1/2}[GeV]$}}
\Text(147,40)[r]{\scalebox{.7}{b)}}
\Text(115,38)[r]{\scalebox{.6}{$m_B=5.279\;GeV$}}
\Text(125,32)[r]{\scalebox{.6}{$m_b=4.8\;GeV$}}
\Text(125,23)[r]{\scalebox{.6}{$m_b=4.7\;GeV$}}
\Text(125,15)[r]{\scalebox{.6}{$m_b=4.6\;GeV$}}
\end{picture}
\caption{\small a: Sum rule for the product $f_Bf^+(0)$ for
$s_0=35\;GeV$ and quark masses $m_b=(4.6,\,4.7,\,4.8)\;GeV$. Graph
b) shows the corresponding daughter sum rule. The curves do not
intersect with the meson mass, and therefore $f_Bf^+(0)$ decreases
monotonously and does not reach a minimum in the Borel
window.}\label{fig-Form-Factor-Khod} \vspace{3mm}
\end{figure}

We will evaluate the sum rule at the scale
\begin{equation}\label{equ-scale-mub}
\mu_b=\sqrt{m_B^2-m_b^2}=2.4 \; GeV.
\end{equation}
This scale, which is often used in analyses of hadronic properties
of the B-meson, reflects the average virtuality of the b-quark in
the meson. A further ambiguity lies in the use of the quark mass.
Nevertheless, taking the running mass
$\overline{m}_b(\overline{m}_b)=4.24\pm0.11\;GeV$ and running it
down to the scale (\ref{equ-scale-mub}), we get
$\overline{m}_b(\mu_b)=4.69 \;GeV$, which is of about the same
size as the pole mass obtained from the two-loop relation
(\ref{equ-pole-running-mass}). Therefore, we will take the
typically used value of the pole mass:
\begin{equation}
m_b=4.7\pm0.1\;GeV.
\end{equation}
The errors are also of the same size as for the running mass. In
figure \ref{fig-Form-Factor-Khod}a, we plotted $f_Bf^+(0)$ for the
three different masses $m_b=\{4.6\;GeV,\,4.7\;GeV,\,4.8\;GeV\}$
and $s_0=35 \;GeV^2$. Other input parameters and the values of the
coefficients of the wave functions can be found in appendix
\ref{app-lc-wf}.
\newline

Taking the mean values $M^2=11 \;GeV^2$ and $s_0=35\;GeV^2$, we
read off $f_Bf^+(0)=0.049\; GeV$, consistent with the analysis of
Khodjamirian and R\"uckl, who extracted $f^+(0)=0.27$ with a
rather low central value for the decay constant, $f_B=180\;MeV$.
However, the sum rule is quite unstable in the given Borel window
as can be seen in figure \ref{fig-Form-Factor-Khod}. It decreases
monotonously in $M^2$ about $10\%$. In figure
\ref{fig-Form-Factor-Khod}$\,$b, we plotted the corresponding
daughter sum rules. They are clearly lower than the experimentally
measured meson mass $m_B=5.279 \;GeV$. Intersections with the mass
of the B-meson will only occur at higher values of the Borel
parameter or the threshold $s_0$. A unique pair
$(\hat{M}^2,\hat{s}_0)$ that would correspond to a vanishing
second derivative of $f_Bf^+(0)(M^2)$ is never reached within
finite values of the Borel mass and the threshold. The choice of
the read off pair $(\hat{M}^2,\hat{s}_0)$ is therefore rather
ambiguous. This motivates the application of the two modified
analyses introduced before.

\section{The Limit $M^2\rightarrow\infty$}\label{sec-ff-uinf}

In this section we will apply the method of local duality,
introduced in section \ref{sec-u-tu-infinity}, to the sum rule for
$f_Bf^+(0)$. Taking the limit $M^2\rightarrow\infty$ justifies the
truncation of the power expansion in $1/M^2$ and minimizes
uncertainties coming from higher twist wave functions. On the
other side, uncertainties in the hadronic spectral function are
emphasized, since they are not exponentially damped anymore.
\newline

\begin{figure}[t]
\begin{picture}(150,50)
\put(2,2.5){
\begin{minipage}[t]{7cm}
 \epsfig{file=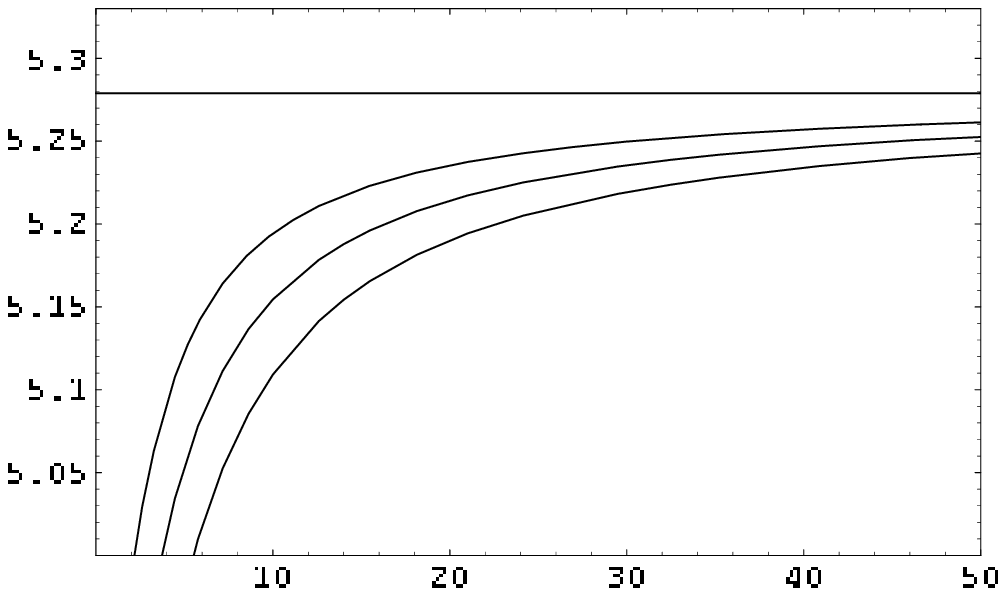,width=7cm}
\end{minipage}}
\Text(64,0.3)[c]{\scalebox{.8}{$M^2[GeV^2]$}}
\Text(25,45.5)[r]{\scalebox{.8}{$\Re^{1/2}[GeV]$}}
\Text(70,40)[r]{\scalebox{.7}{a)}}
\Text(35,39)[r]{\scalebox{.6}{$m_B=5.279\;GeV$}}
\Text(29,33)[r]{\scalebox{.6}{$m_b=4.8\;GeV$}}
\Text(47,21)[r]{\scalebox{.6}{$m_b=4.7\;GeV$}}
\Text(38,16)[r]{\scalebox{.6}{$m_b=4.6\;GeV$}}
\put(33,32){\line(-3,1){4}} \put(32,21){\line(-4,1){9}}
\put(23,16){\line(-1,0){2}}
\put(80,2){
\begin{minipage}[t]{7cm}
\epsfig{file=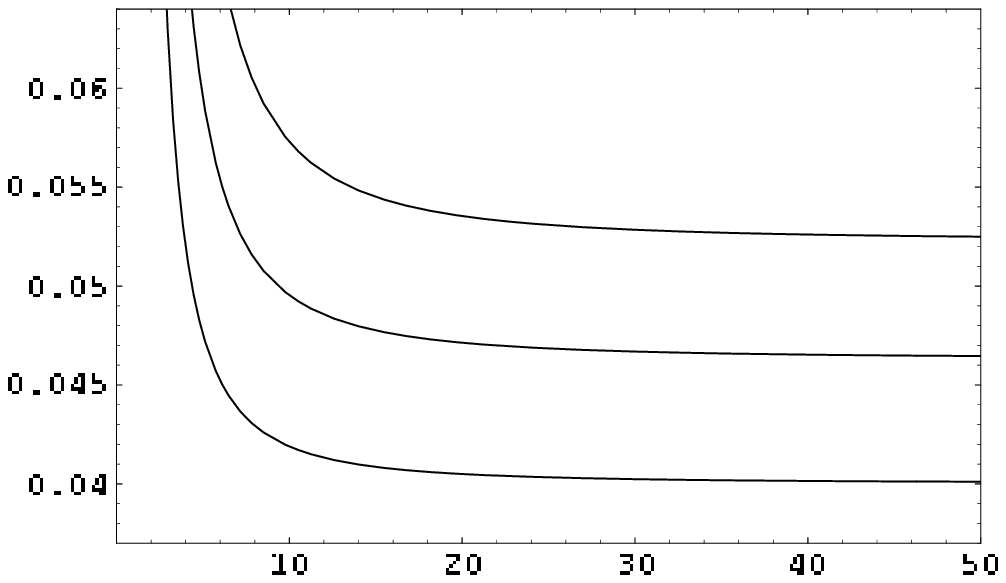,width=7cm,height=3.6cm}
\end{minipage}}
\Text(142,0.3)[c]{\scalebox{.8}{$M^2[GeV^2]$}}
\Text(110,45.5)[r]{\scalebox{.8}{$f_Bf^+(0)[GeV]$}}
\Text(147,40)[r]{\scalebox{.7}{b)}}
\Text(135,29)[r]{\scalebox{.6}{$m_b=4.6\;GeV$}}
\Text(135,20)[r]{\scalebox{.6}{$m_b=4.7\;GeV$}}
\Text(135,11)[r]{\scalebox{.6}{$m_b=4.8\;GeV$}}
\end{picture}
\caption{\label{fig-Form-Factor-uinf}\small Matching of the
daughter sum rule to the experimental meson mass $m_B=5.279\;GeV$
in the limit $M^2\rightarrow\infty$ (a). The values for the
thresholds obtained by the matching are used to determine the
values for the product $f_Bf^+(0)$ in the same limit (b).}
\vspace{3mm}
\end{figure}

We again take the derivative $M^4\;d/dM^2$ of the logarithm of the
right hand side in equation (\ref{equ-form-factor}). We then
adjust the threshold $s_0$ such, that the square root of this
daughter sum rule takes on the value of the meson mass
$m_B=5.279\;GeV$ in the limit $M^2\rightarrow \infty$. Using the
threshold obtained in this way in the sum rule (\ref{equ-sr-ff}),
the product $f_Bf^+(M^2)$ reaches its asymptotic value very fast.
In figure \ref{fig-Form-Factor-uinf}, we plotted the matching of
the daughter sum rules for the center value of the quark mass and
its largest errors $m_b=(4.7\pm0.1)\;GeV$, as well as the
corresponding functions for $f_Bf^+(0)$ of $M^2$. The central
value is
\begin{equation}
\lim\limits_{M^2\rightarrow\infty}f_Bf^+(M^2)=0.0463 \;GeV,
\end{equation}
and the error inflicted by the uncertainties of the quark mass is
about $12\%$.
\newline

Since the decay constant obtained with this method was rather
sensitive to the scale $\mu$, at which the sum rule was evaluated,
we also calculated the values of $f_Bf^+(0)$ at the scales
$\mu=2\; GeV$ and $\mu=3\;GeV$. There are many $\mu$-dependent
parameters in (\ref{equ-form-factor}) and we will give their
scaling laws in the following.
\newline

The scaling behavior of $\mu_\pi$ can be obtained from the
Gell-Mann-Oakes-Renner relation (\ref{equ-gell-mann}) and is
therefore directly related to the running of quark masses. The
scaling law of the coefficients of the twist two pion wave
function $\varphi_\pi(u)$ is given in (\ref{equ-coeff-running}).
The scale dependence of the parameters appearing in the twist 3
and 4 wave functions are taken from Braun and
Filianov\cite{Braun:1990iv}:
\begin{eqnarray}
f_{3\pi}(\mu_2^2)&=&\left(\frac{\alpha_s(\mu_2^2)}
{\alpha_s(\mu_1^2)}\right)^{\frac{11}{15}}f_{3\pi}(\mu_1^2)\\
f_{3\pi}\omega_{1,0}(\mu_2^2)&=&\left(\frac{\alpha_s(\mu_2^2)}
{\alpha_s(\mu_1^2)}\right)^{\frac{104}{75}}f_{3\pi}\omega_{1,0}
(\mu_1^2)\\
f_{3\pi}\left(\begin{array}{c}
\omega_{1,1} \\
  4\omega_{2,0} \\
\end{array}
\right)(\mu_2^2)&=&\left(1-\frac{\alpha_s}{4\pi}
\log\frac{\mu_2^2}{\mu_1^2}\left(\begin{array}{cc}
\frac{122}{9}&\frac{5}{3}\\
\frac{21}{5}&\frac{511}{45}
\end{array}\right)\right)f_{3\pi}\left(\begin{array}{c}
\omega_{1,1} \\
  4\omega_{2,0} \\
\end{array}
\right),
\end{eqnarray}
for the parameters of the two and three particle wave functions of
twist 3 and
\begin{eqnarray}
\delta^2(\mu_2^2)=\left(\frac{\alpha_s(\mu_2^2)}{\alpha_s(\mu_1^2)}
\right)^{\frac{32}{75}}\delta^2(\mu_1^2)\\
\delta^2\varepsilon(\mu_2^2)=\left(\frac{\alpha_s(\mu_2^2)}
{\alpha_s(\mu_1^2)}\right)^{\frac{6}{5}}\delta^2\varepsilon(\mu_1^2)
\end{eqnarray}
for the twist 4 wave functions. Using these scaling laws, we
evaluated the sum rule (\ref{equ-sr-ff}) at the scales
$\mu=2\;GeV$ and $\mu=3\;GeV$. The two functions of the daughter
sum rule are almost similar and the matching leads to two
thresholds, which are also very close to each other:
\begin{equation}
s_0(2\;GeV)=35.46 \;GeV,\hspace{1cm}s_0(3\;GeV)=35.45\;GeV.
\end{equation}
Setting these two thresholds in the corresponding functions for
$f_Bf^+(M^2)$, we get a small scale dependence of about $\pm4\%$,
when varying the scale $\mu=2\;GeV$ to $\mu=3\;GeV$. This is shown
in figure \ref{fig-Form-Factor-uinf-scale-ass}$\,$a. We finally
obtain:
\begin{figure}[t]
\begin{picture}(150,50)
\put(2,2){
\begin{minipage}[t]{7cm}
\epsfig{file=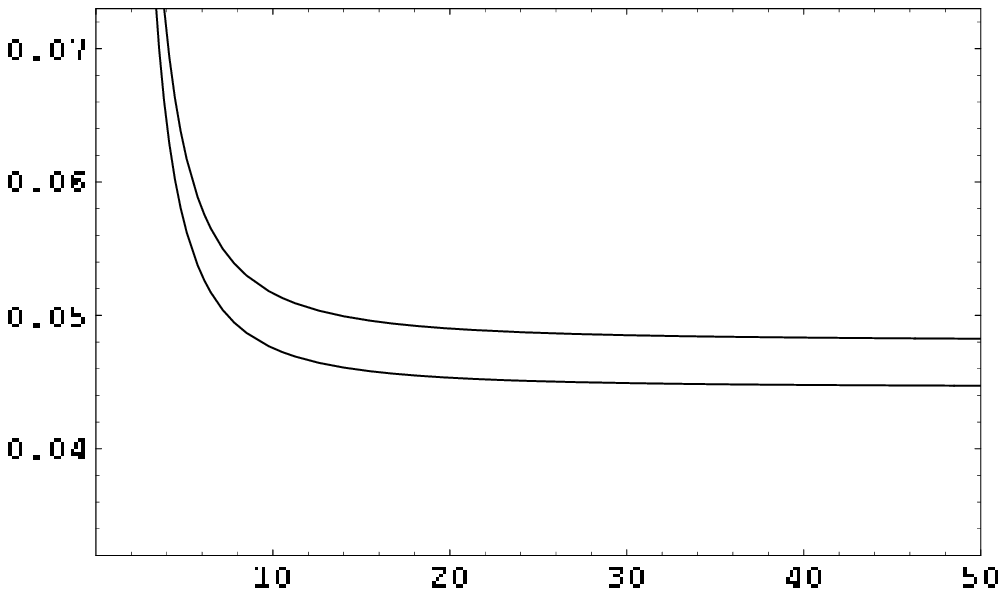,width=7cm}
\end{minipage}}
\Text(64,0.3)[c]{\scalebox{.8}{$M^2[GeV^2]$}}
\Text(10,45)[l]{\scalebox{.8}{$f_Bf^+(0)[GeV]$}}
\Text(69,39)[r]{\scalebox{.7}{a)}}
\Text(56,22)[r]{\scalebox{.6}{$\mu=2\;GeV$}}
\Text(56,15)[r]{\scalebox{.6}{$\mu=3\;GeV$}} \put(80,2){
\begin{minipage}[t]{7cm}
 \epsfig{file=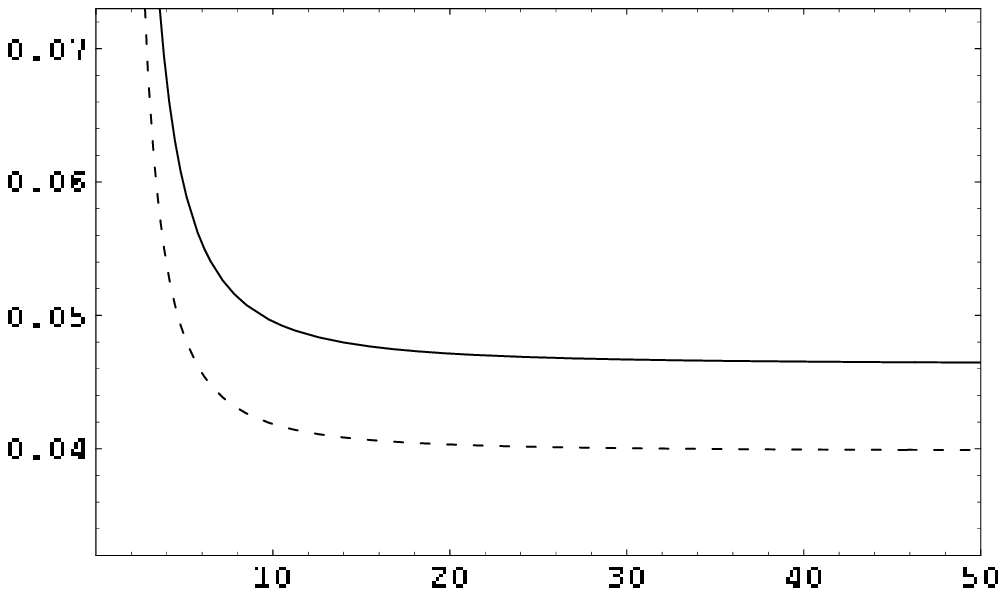,width=7cm,height=3.9cm}
\end{minipage}}
\Text(142,0.3)[c]{\scalebox{.8}{$M^2[GeV^2]$}}
\Text(88,45)[l]{\scalebox{.8}{$f_Bf^+(0)[GeV]$}}
\Text(147,39)[r]{\scalebox{.7}{b)}}
\end{picture}
\caption{\small a: Sum rule for $f_Bf^+(0)$ evaluated at the
scales $\mu=2\;GeV$ and $\mu=3\;GeV$. b: $f_Bf^+(0)$, when the
asymptotic pion wave function is used (dashed line), compared to
the central result (solid
line).}\label{fig-Form-Factor-uinf-scale-ass} \vspace{3mm}
\end{figure}
\begin{equation}\label{equ-fbfpl-uinf-final}
\lim\limits_{M^2\rightarrow\infty}f_Bf^+(M^2)=
(0.046\pm0.006\pm0.002)\;GeV,
\end{equation}
where the first error is due to uncertainties in the quark mass
and the second comes from the scale at which the sum rule
(\ref{equ-form-factor}) is evaluated. There are further errors
coming from the uncertainties of the input parameters $\mu_\pi$
and $f_{3\pi}$, however, they are small compared to the values
above and we do not take them into account in this analysis.
\newline

A rather large error is inflicted by the uncertainties of the
coefficients of the light cone wave functions, which are not known
to a high accuracy. We therefore also analyzed the sum rule for
the asymptotic twist 2 pion wave function. In figure
\ref{fig-Form-Factor-uinf-scale-ass}b, we plotted the sum rule for
the asymptotic wave function, compared to our central result. The
value for the product $f_Bf^+(0)$ decreases about $13\%$. This
number can be interpreted as a rather conservative error estimate
on the result (\ref{equ-fbfpl-uinf-final}).
\newline

For consistency, we divide (\ref{equ-fbfpl-uinf-final}) by the
decay constant (\ref{equ-fb-u-infinity}), obtained with the same
method in section \ref{sec-u-tu-infinity}. In the ratio, the two
errors from the quark masses almost cancel, when they are varied
simultaneously. Thus, although both sum rules show a high
sensitivity to the chosen quark mass, this dependence becomes very
weak if the two results are combined. We obtain:
\begin{equation}\label{equ-fpl-uinf}
f^+(0)=0.26\pm0.05.
\end{equation}
The error due to the unknown quark mass amounts to less than
$4\%$. We included the error estimate of $13\%$ from the
uncertainties of the coefficients of the wave functions. Added to
the other error sources, the final error estimate is about $20\%$.
In section \ref{sec-ff-relation}, we relate our results to values
found in the literature.
\newline

The two individual analyses of the decay constant $f_B$ in section
\ref{sec-u-tu-infinity} and the product $f_Bf^+(0)$ in this
section show rather low values compared to other methods. However,
since the ratio of the two values is taken, some errors cancel as
shown above and the error of about $20\%$ of the final result is
reasonable. It is hoped that the intrinsic error coming from the
method itself also shrinks when the ratio of the two results is
taken. This will be highly supported in the next section.

\section{Borel Mass dependent Threshold $s_0(M^2)$}

The sum rule for the form factor (\ref{equ-sr-ff}) leads to
functions that have a rather strong dependence on $M^2$ for
typical values of the input parameters, as can be seen from figure
\ref{fig-Form-Factor-Khod}. With higher values of the thresholds
the functions develop a minimum in the Borel window, however, the
second derivative does not vanish as was the case in the analysis
of the decay constant. Thus, one cannot find a unique pair
($\hat{M}^2,\hat{s}_0$) for which the daughter sum rule is equal
to the meson mass at its extremal point.
\begin{figure}[t]
\begin{picture}(145,90)
\put(4,52){
\begin{minipage}[t]{7cm}
 \epsfig{file=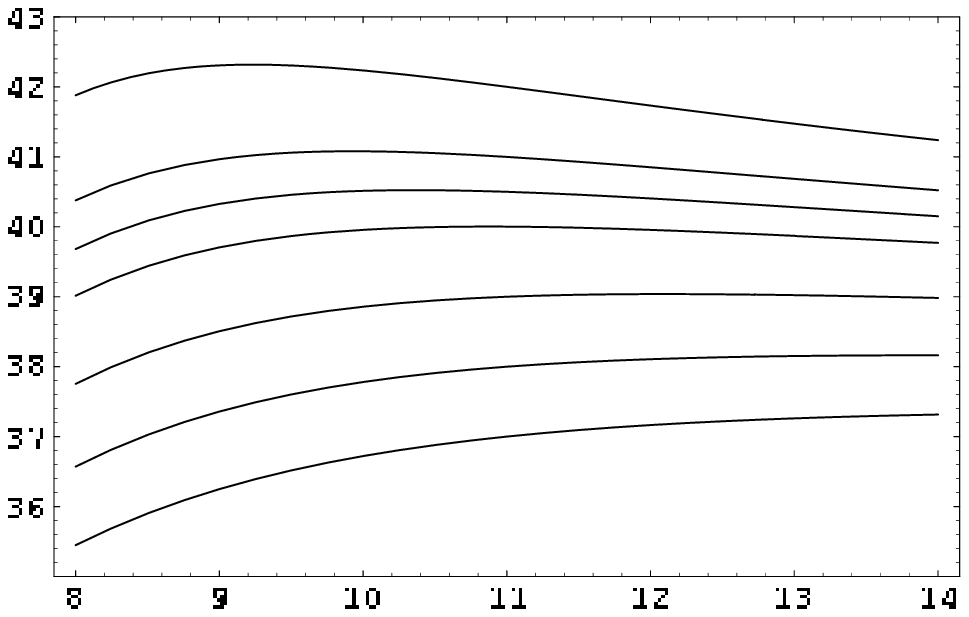,width=6.8cm,height=4cm}
\end{minipage}}
\Text(64,49.5)[c]{\scalebox{.8}{$M^2[GeV^2]$}}
\Text(11,93.5)[l]{\scalebox{.8}{$s_0(M^2)[GeV^2]$}}
\Text(68,89)[r]{\scalebox{.7}{a)}} \put(1,2.5){
\begin{minipage}[t]{7cm}
 \epsfig{file=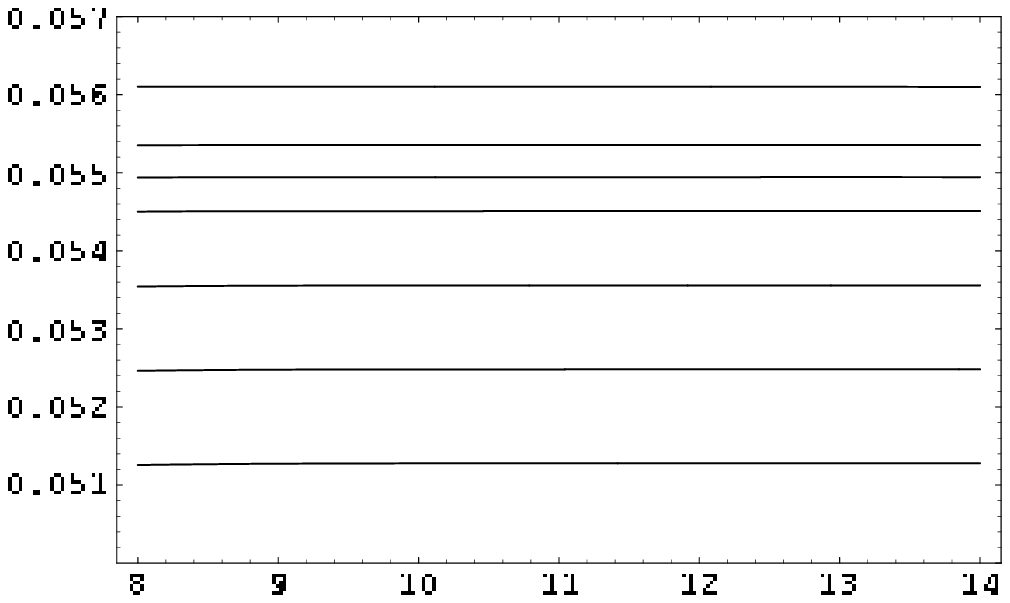,width=7cm}
\end{minipage}}
\Text(64,0.3)[c]{\scalebox{.8}{$M^2[GeV^2]$}}
\Text(11,46)[l]{\scalebox{.8}{$f_Bf^+(0)[GeV]$}}
\Text(68,41)[r]{\scalebox{.7}{b)}}
\Text(88,46)[l]{\scalebox{.8}{$f_Bf^+(0)[GeV]$}} \put(79.5,51){
\begin{minipage}[t]{8cm}
 \epsfig{file=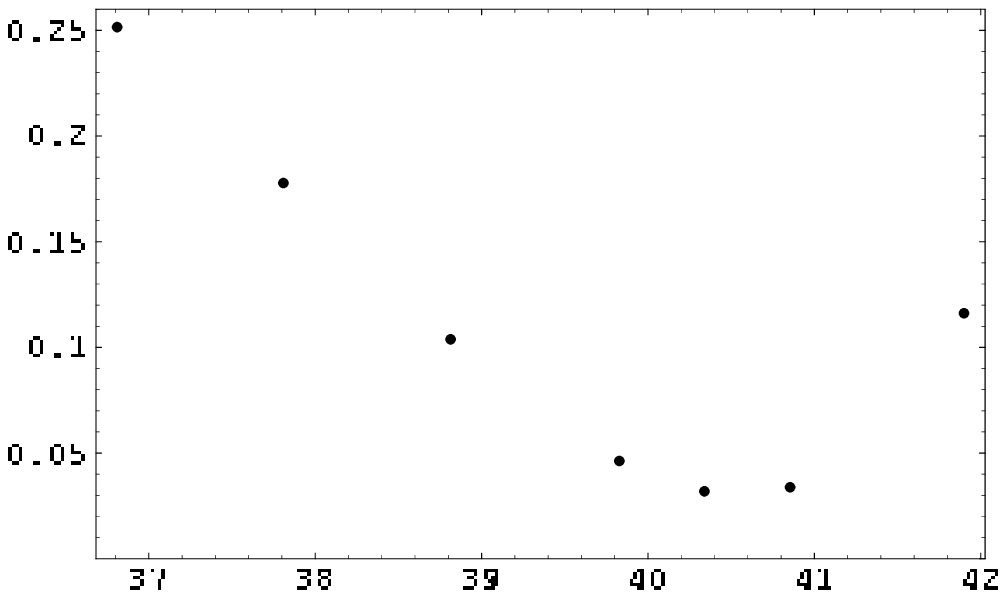,width=6.9cm}
\end{minipage}}
\Text(141,49.5)[c]{\scalebox{.8}{$\overline{s}_0[GeV^2]$}}
\Text(90,93)[c]{\scalebox{.8}{$s_{var}$}}
\Text(144,89)[r]{\scalebox{.7}{c)}} \put(78,2.5){
\begin{minipage}[t]{8cm}
 \epsfig{file=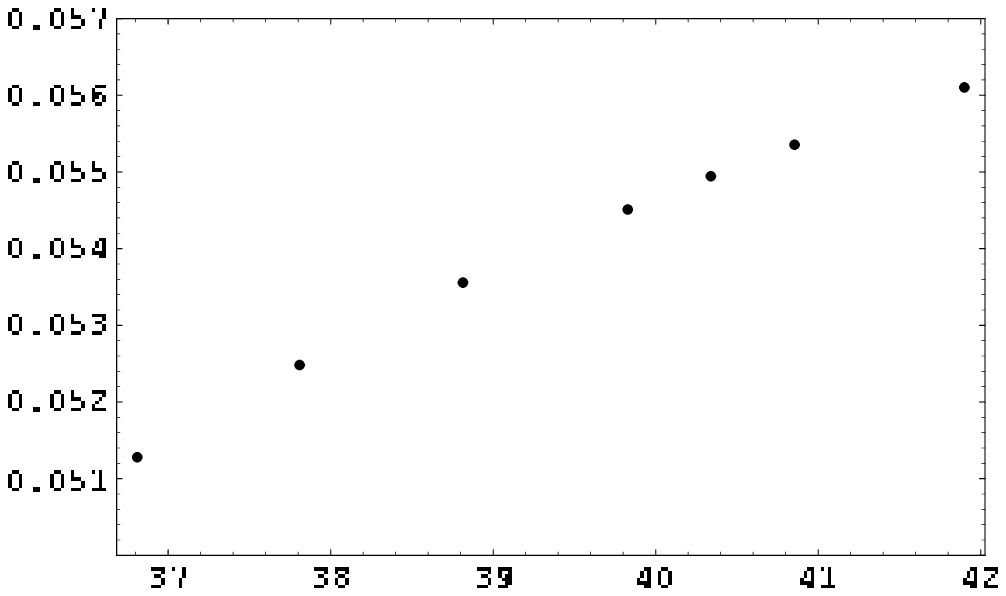,width=7cm,height=4.2cm}
\end{minipage}}
\Text(142,0.3)[c]{\scalebox{.8}{$\overline{s}_0[GeV^2]$}}
\Text(144,41)[r]{\scalebox{.7}{d)}}
\end{picture}
\caption{\small Sum rule analysis for $f_Bf^+(0)$ with a Borel
mass dependent threshold $s_0(M^2)$. Diagram a) shows several
solutions of the differential equation (\ref{equ-ff-diff-equ}).
Plot d) shows the corresponding values of $f_Bf^+(0)$ over the
mean values of the thresholds. Graph c) shows the variations of
the functions of the threshold to give the region where the value
of $f_Bf^+(0)$ should be read off. Diagram b) shows the functions
$f_Bf^+(0)$ when the solutions $s_0(M^2)$ are inserted and gives a
crosscheck of the numerical analysis.}\label{fig-ff-sofu-47}
\vspace{3mm}
\end{figure}
\newline

In this section we apply the method outlined in section \ref{s(u)}
to the sum rule of the form factor. Thus, we shift the functional
dependence of the form factor to the threshold $s_0$ by solving
the differential equation
\begin{equation}\label{equ-ff-diff-equ}
s_0'(M^2)=\frac{m_B^2\Phi(M^2,s_0(M^2))-M^4\frac{\partial}{\partial
M^2}\Phi(M^2,s_0(M^2))}{M^4\frac{\partial}{\partial
s_0}\Phi(M^2,s_0(M^2))},
\end{equation}
where $\Phi(M^2,s_0(M^2))$ is given in (\ref{equ-form-factor}).
For central values of the input parameters, we plotted several
solutions in figure \ref{fig-ff-sofu-47}. We again plotted the
variations of the curves to read off the value for the product
$f_Bf^+(0)$ at the most stable threshold. This threshold is rather
high ($\overline{s}_0\approx 40.4 \;GeV^2$) compared to the
earlier analyses. As can be seen in figure
\ref{fig-Form-Factor-Khod}, the dependence of the sum rule on the
Borel parameter is rather strong for smaller values of the
threshold. Thus, one needs to impose a function $s_0(M^2)$ that is
also strongly varying with the Borel mass. To higher thresholds,
this dependence of the sum rules on $M^2$ becomes weaker, which is
then also the case for $s_0(M^2)$.
\begin{figure}[p]
\begin{picture}(145,90)
\put(3.75,52){
\begin{minipage}[t]{7cm}
 \epsfig{file=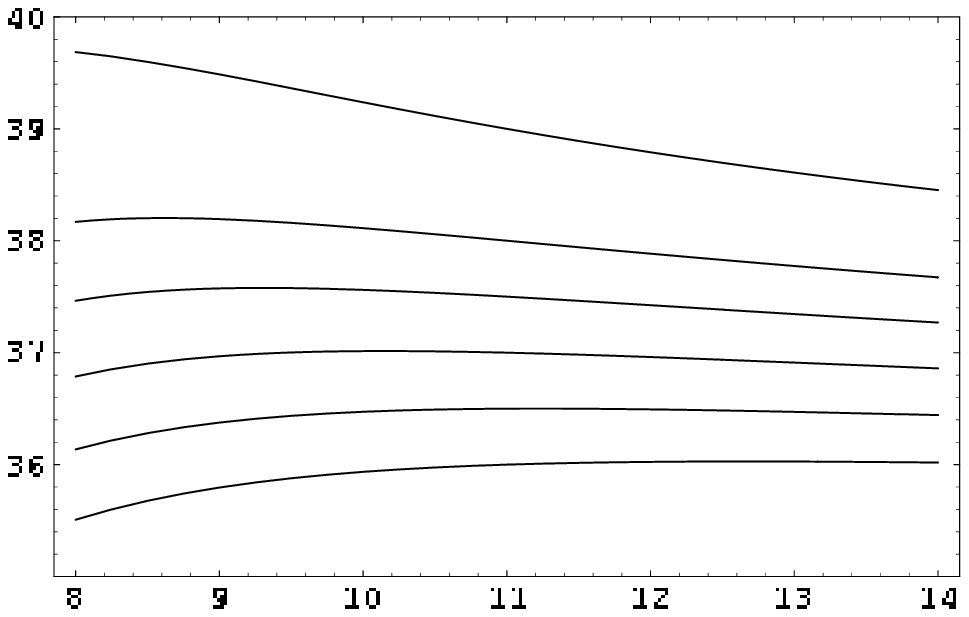,width=6.875cm,height=4.2cm}
\end{minipage}}
\Text(64,49.5)[c]{\scalebox{.8}{$M^2[GeV^2]$}}
\Text(10,95)[l]{\scalebox{.8}{$s_0(M^2)[GeV^2]$}}
\Text(68,90)[r]{\scalebox{.7}{a)}} \put(1,2.5){
\begin{minipage}[t]{7cm}
 \epsfig{file=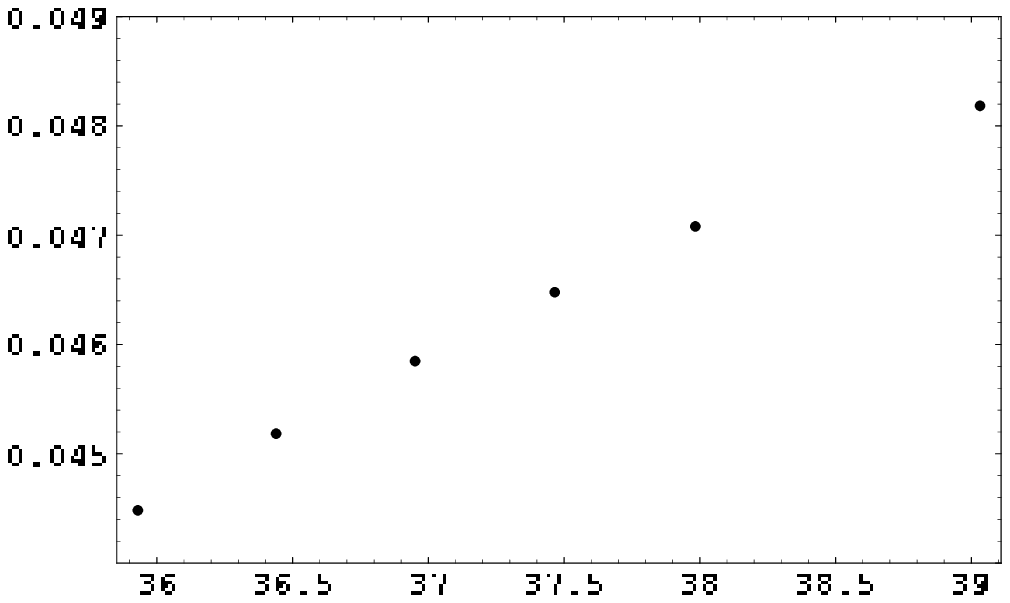,width=7cm}
\end{minipage}}
\Text(64,0.3)[c]{\scalebox{.8}{$\overline{s}_0[GeV^2]$}}
\Text(11,46)[l]{\scalebox{.8}{$f_Bf^+(0)[GeV]$}}
\Text(68,41)[r]{\scalebox{.7}{b)}}
\Text(88,46)[l]{\scalebox{.8}{$f_Bf^+(0)[GeV]$}} \put(81.5,51.5){
\begin{minipage}[t]{8cm}
 \epsfig{file=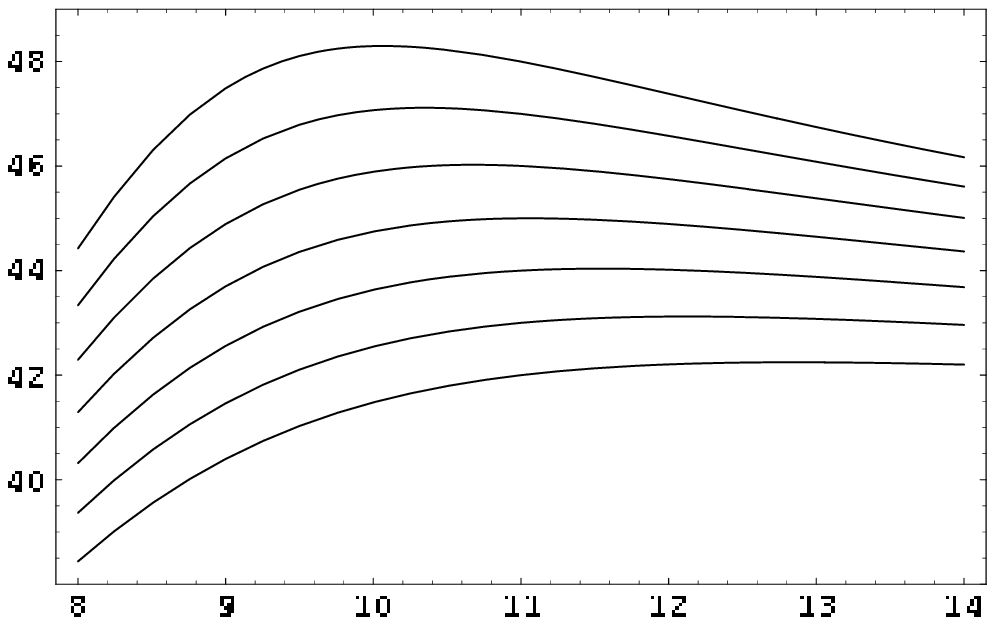,width=6.9cm,height=3.9cm}
\end{minipage}}
\Text(142,49.5)[c]{\scalebox{.8}{$M^2[GeV^2]$}}
\Text(89,95.5)[l]{\scalebox{.8}{$s_0(M^2)[GeV^2]$}}
\Text(120,95)[l]{\scalebox{.8}{$m_b=4.6\;GeV$}}
\Text(144,90)[r]{\scalebox{.7}{c)}} \put(76,2){
\begin{minipage}[t]{8cm}
 \epsfig{file=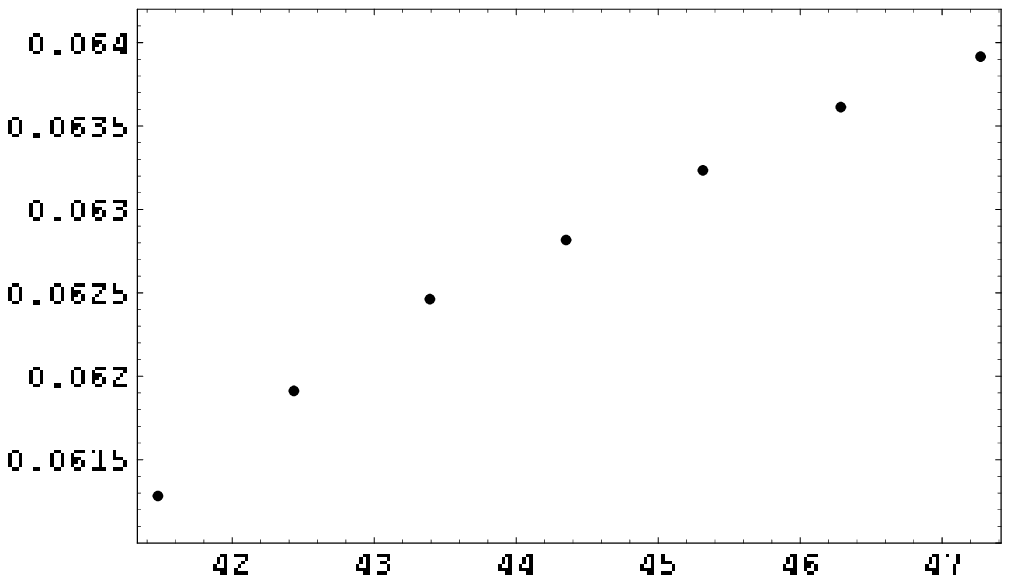,width=7.3cm,height=4cm}
\end{minipage}}
\Text(143,0.5)[c]{\scalebox{.8}{$\overline{s}_0[GeV^2]$}}
\Text(40,95)[l]{\scalebox{.8}{$m_b=4.8\;GeV$}}
\Text(144,41)[r]{\scalebox{.7}{d)}}
\end{picture}
\caption{\small Analyses for the different values of the quark
mass: $m_b=4.8\;GeV$(left side) and $m_b=4.6\;GeV$(right side).
The upper diagrams show the solutions for the thresholds
$s_0(M^2)$ - the lower diagrams show the product $f_Bf^+(0)$,
plotted over the average values of the
thresholds.}\label{fig-ff-sofu-48}
\end{figure}
\begin{figure}[p]
\begin{picture}(145,90)
\put(5.25,51.5){
\begin{minipage}[t]{7cm}
\epsfig{file=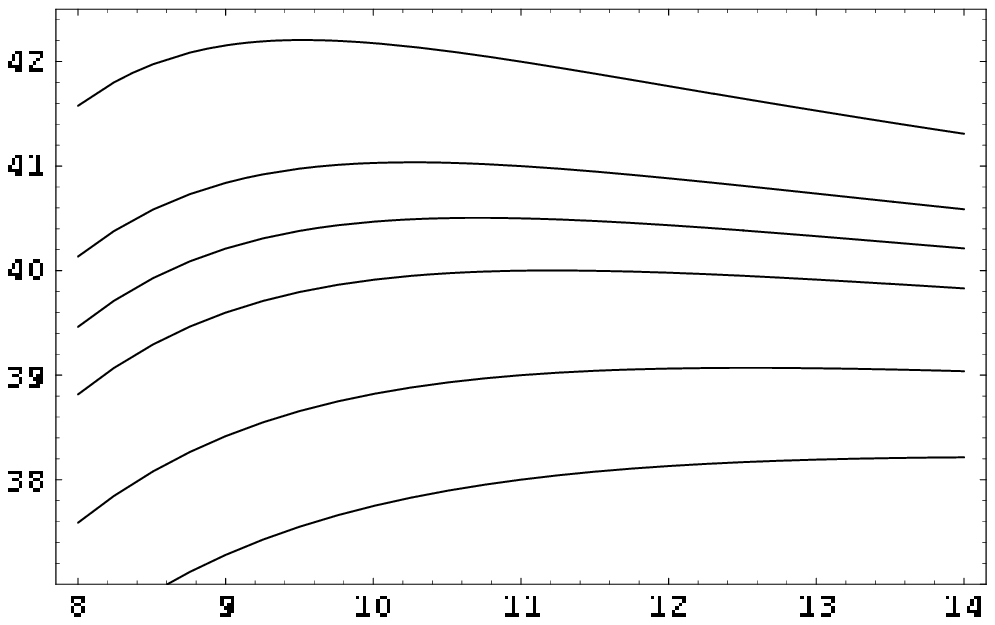,width=6.65cm,height=3.9cm}
\end{minipage}}
\Text(64,49.5)[c]{\scalebox{.8}{$M^2[GeV^2]$}}
\Text(11,95)[l]{\scalebox{.8}{$s_0(M^2)[GeV^2]$}}
\Text(68,90)[r]{\scalebox{.7}{a)}} \put(1,2){
\begin{minipage}[t]{7cm}
 \epsfig{file=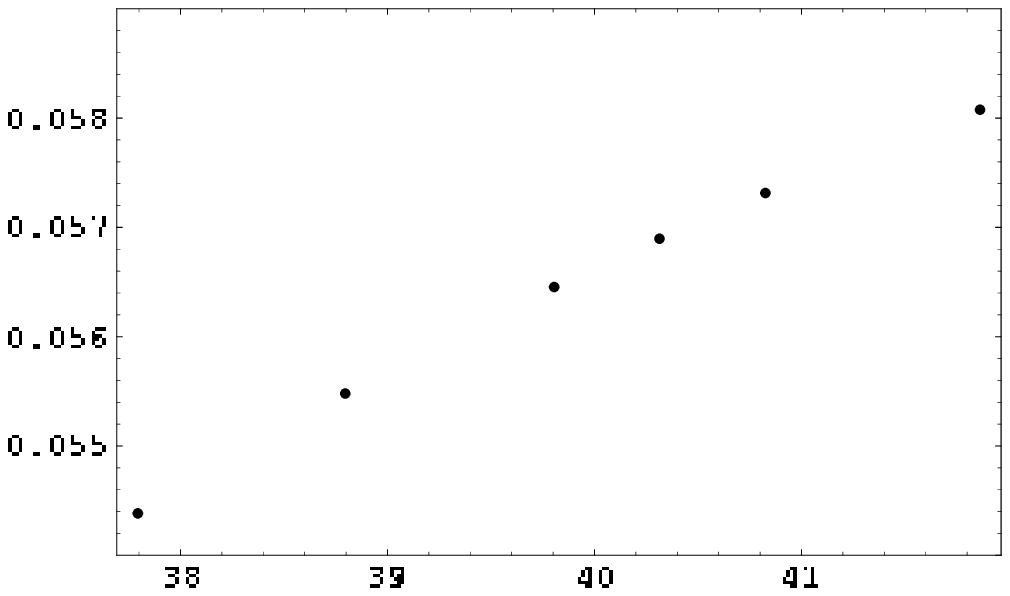,width=5.8cm}
\end{minipage}}
\Text(64,0.3)[c]{\scalebox{.8}{$\overline{s}_0[GeV^2]$}}
\Text(11,45)[l]{\scalebox{.8}{$f_Bf^+(0)[GeV]$}}
\Text(68,41)[r]{\scalebox{.7}{b)}}
\Text(88,45)[l]{\scalebox{.8}{$f_Bf^+(0)[GeV]$}} \put(80.5,51.5){
\begin{minipage}[t]{8cm}
 \epsfig{file=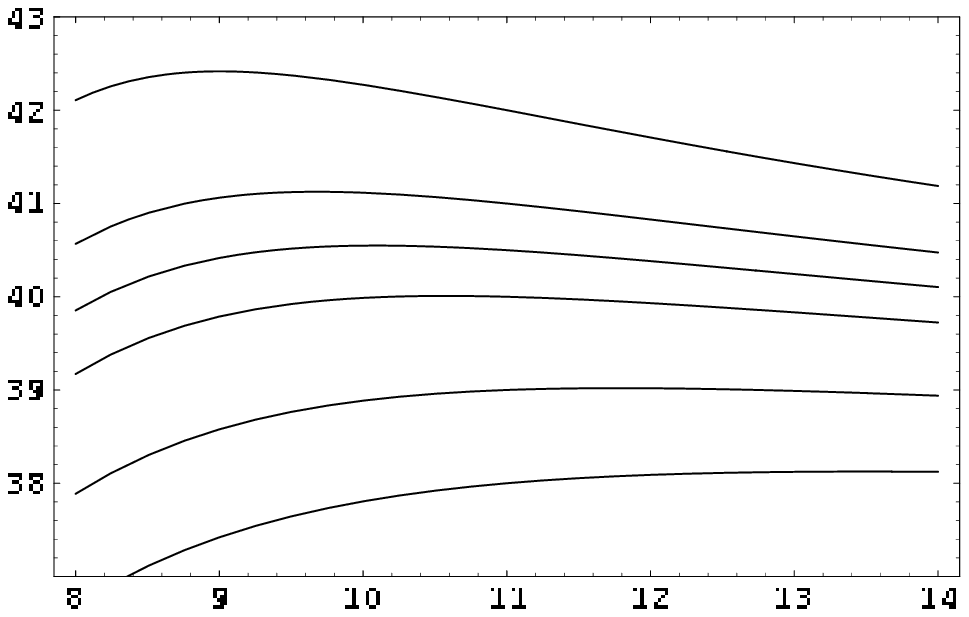,width=6.8cm}
\end{minipage}}
\Text(142,49.5)[c]{\scalebox{.8}{$M^2[GeV^2]$}}
\Text(89,95.5)[l]{\scalebox{.8}{$s_0(M^2)[GeV^2]$}}
\Text(120,95)[l]{\scalebox{.8}{$\mu=3\;GeV$}}
\Text(144,90)[r]{\scalebox{.7}{c)}} \put(77,2){
\begin{minipage}[t]{8cm}
 \epsfig{file=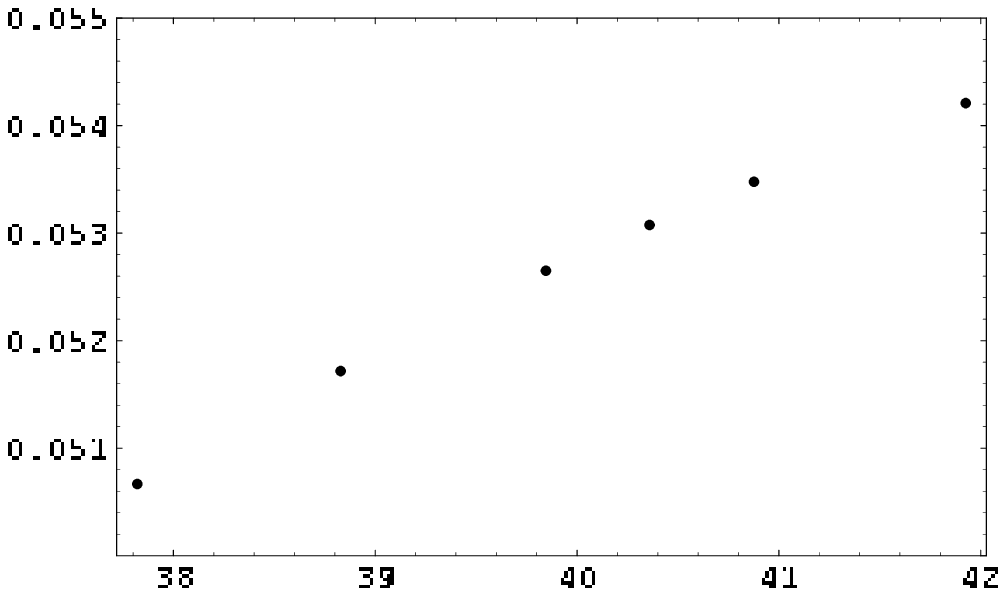,width=7.3cm,height=4.15cm}
\end{minipage}}
\Text(143,0.5)[c]{\scalebox{.8}{$\overline{s}_0[GeV^2]$}}
\Text(40,95)[l]{\scalebox{.8}{$\mu=2\;GeV$}}
\Text(144,41)[r]{\scalebox{.7}{d)}}
\end{picture}
\caption{\small Scale dependence of the sum rule for $f_Bf^+(0)$.
On the left side, the results are shown for the scale $\mu=2\;GeV$
and on the right side, the scale $\mu=3\;GeV$ was
used.}\label{fig-ff-scale}
\end{figure}
\newline

In figure \ref{fig-ff-sofu-48}, we plotted the results from using
$m_b=4.6\;GeV$ and $m_b=4.8\;GeV$ as the values of the quark mass.
As before, the error to the product $f_Bf^+(0)$ coming from the
variation of the b-quark mass is large. It is about $15\%$ of the
final result.
\begin{figure}[t]
\begin{picture}(150,50)
\put(2,2.5){
\begin{minipage}[t]{7cm}
 \epsfig{file=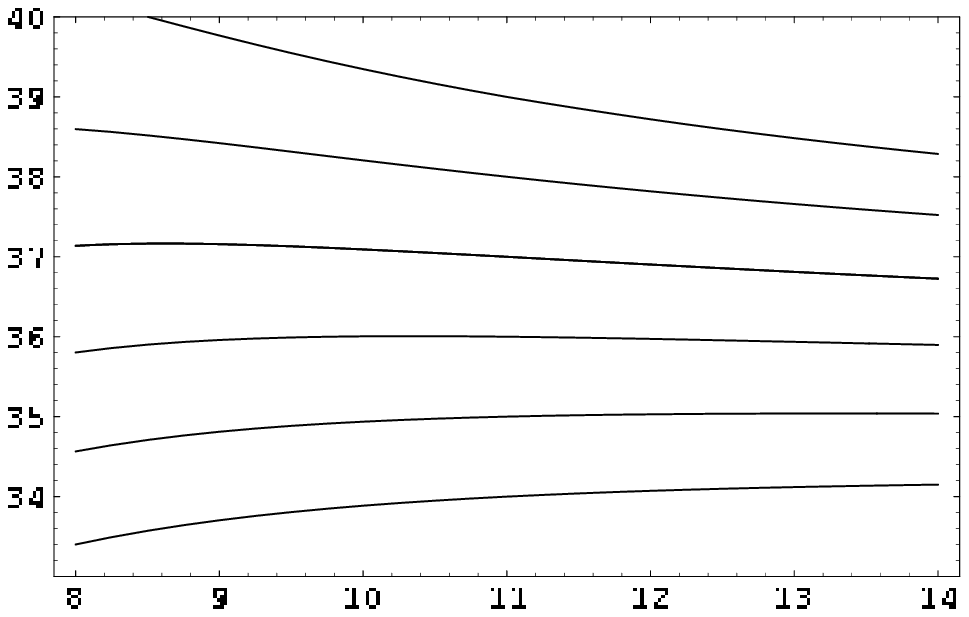,width=6.7cm}
\end{minipage}}
\Text(64,0.3)[c]{\scalebox{.8}{$M^2[GeV^2]$}}
\Text(10,46)[l]{\scalebox{.8}{$s_0(M^2)[GeV^2]$}}
\Text(69,39)[r]{\scalebox{.7}{a)}} \put(80,1.5){
\begin{minipage}[t]{7cm}
\epsfig{file=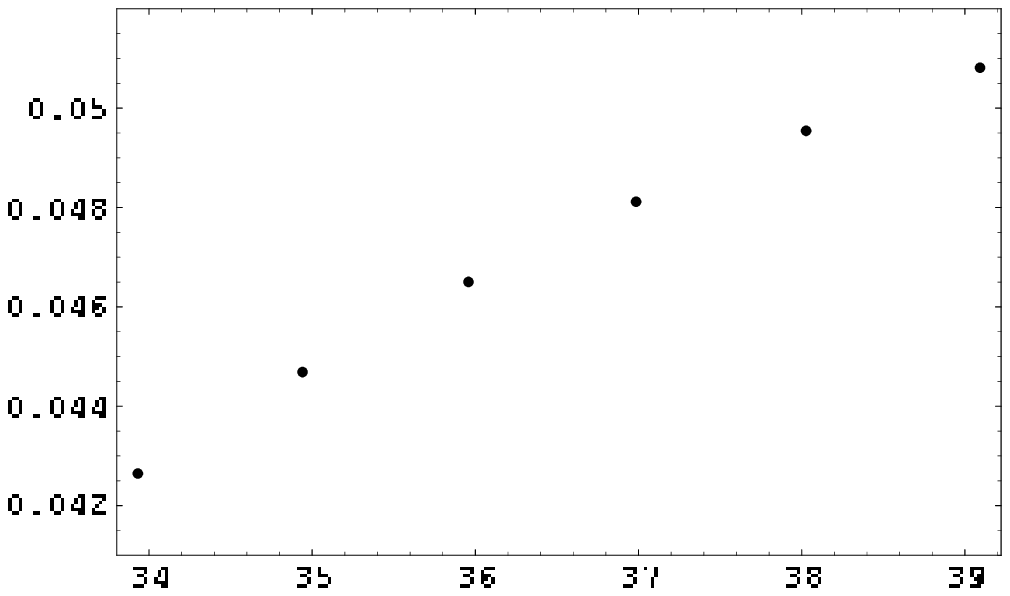,width=7cm,height=3.6cm}
\end{minipage}}
\Text(142,0.3)[c]{\scalebox{.8}{$\overline{s}_0[GeV^2]$}}
\Text(90,46)[l]{\scalebox{.8}{$f_Bf^+(0)[GeV]$}}
\Text(145,40)[r]{\scalebox{.7}{b)}}
\end{picture}
\caption{\small Sum rule analysis for the asymptotic pion wave
function. The product $f_Bf^+(0)$ decreases significantly about
$15\%$.}\label{fig-ff-ass} \vspace{3mm}
\end{figure}
\newline

We also examined the scale dependence of $f_Bf^+(0)$ extracted
from this analysis. Like before, we evaluated the sum rule at
$\mu=2\;GeV$ and $\mu=3\;GeV$. The results are shown in figure
\ref{fig-ff-scale}. The values of $f_Bf^+(0)$ are quite stable to
variations of the scale $\mu$. The error is less than $4\%$. We
again analyzed the sum rule using the asymptotic pion wave
function to take the error from the uncertainties of the
coefficients of the wave functions into account. The resulting
thresholds $s_0(M^2)$ and values for $f_Bf^+(0)$ can be seen in
figure \ref{fig-ff-ass}.
\newline

Neglecting further error sources, we extract for the product
$f_Bf^+(0)$:
\begin{equation}\label{equ-result-ff-s(u)}
f_Bf^+(0)=(0.055\pm0.009\pm0.008)\;GeV,
\end{equation}
where the first error is the one coming from the variations of the
quark mass and the second is due to the scale dependence and
uncertainties of the coefficients of the wave functions. If we
divide this result by the decay constant
(\ref{equ-decay-constant-s(u)}), that was analyzed using the same
method, we obtain:
\begin{equation}\label{equ-fpl-s(u)}
f^+(0)=0.26\pm0.05.
\end{equation}
The change of the quark mass to the extremal values in the two sum
rule results (\ref{equ-result-ff-s(u)}) and
(\ref{equ-decay-constant-s(u)}) leads to slightly higher ratios
(about $3\%$) than the central value in either case. Again, the
main error comes from the conservative error estimate of the
uncertainties in the coefficients of the wave functions.

\section{Summary of the Results and Relation to the
Literature}\label{sec-ff-relation}

It is amazing that the two results (\ref{equ-fpl-uinf}) and
(\ref{equ-fpl-s(u)}) are so close to each other. The central
results differ by less than $2\%$. The two methods have very
different physical meanings. Whereas the first method of taking
the limit $M^2\rightarrow\infty$ emphasizes errors coming from the
duality approximation, the second method tries to suppress these
uncertainties. Therefore, the discrepancy of the individual
results for the decay constant and the product $f_Bf^+$ was rather
natural. It was hoped that when the ratio of the two sum rule
results was taken, the intrinsic errors from the methods would be
lowered. However, we did not expect that the two results for the
semi-leptonic form factor are that close to each other. In the
next chapter we will analyze the sum rule for the strong coupling
$g^{}_{B^*\!B\pi}$. The sum rule is proportional to $f_B$ and
$f_{B^*}$ and will be divided by these constants, analyzed within
the same method. However, this time, the extracted couplings
differ about $13\%$.
\begin{table}[t]
\centering
\begin{tabular}{|l|c|p{8cm}|}
\hline $f^+(0)$&Ref.&Method \\
\hline \hline
$0.26\pm 0.05$&\small This Thesis&\small LCSR;
$O(\alpha_s)$-twist 2 corrections; $\lim M^2\rightarrow\infty$\\
$0.26\pm 0.05$&\small This Thesis&\small LCSR;
$O(\alpha_s)$-twist 2 corrections; $s_0(M^2)$\\
$0.24 \pm 0.025$&\small Ball '91\cite{Ball:1991ma}&\small 3-pt SR\\
$0.27 \pm 0.05$&\small Khodjamirian '97\cite{Khodjamirian:1997ub}&
\small \small LCSR; $O(\alpha_s)$-twist 2 corrections;\\
$0.28\pm 0.05$&\small Bagan '97\cite{Bagan:1998bp}&\small LCSR;
$O(\alpha_s)$-twist 2 corrections;
average of full and asymptotic twist 2 pion wave function\\
$0.26 \pm 0.06 \pm 0.05$&\small Ball '01\cite{Ball:2001fp}&
\small LCSR $O(\alpha_s)$-twist 2 and 3 corrections\\
$0.28 \pm 0.04$&\small Abada APE'99\cite{Abada:1999xd}&
\small Lattice\\
\hline
\end{tabular}
\caption{\label{tab-literature-ff}\small Comparison of our results
to a collection of values from the literature} \vspace{3mm}
\end{table}
\newline

We did not take $O(\alpha_s)$-corrections to the two particle
twist 3 wave functions into account. These were recently obtained
in \cite{Ball:2001fp}. The $O(\alpha_s)$-corrections to the twist
2 wave function raised this contribution about $30\%$. Assuming
that the new corrections are also about $30\%$ of the twist 3 wave
functions, we expect the final result of $f^+(0)$ to raise about
$10\%$.
\newline

In table \ref{tab-literature-ff} we collected some values for the
form factor $f^+(0)$ from the literature. Our results seem to fit
quite well in this table, especially when the large error bars are
taken into account. Furthermore, the results from LCSR analyses
and 3-point sum rules coincide with lattice calculations. This
will be very different in case of the strong coupling
$g^{}_{B^*\!B\pi}$ analyzed in the next section.

\chapter{Coupling $g^{}_{B^*\!B\pi}$}\label{coupling}

In this chapter, we will apply our two methods on the sum rule for
the strong coupling constant $g^{}_{B^*\!B\pi}$. The process
$B^*\rightarrow B\pi$ is kinematically forbidden, thus one has to
rely on theoretical predictions. This coupling is closely related
to the $B \rightarrow \pi$ form factor. It is assumed that the
form factor is dominated by the vector meson at zero recoil. In
the single pole model, the form factor at large $q^2\approx
m_{B^*}^2$ is parametrized by
\begin{equation}\label{equ-pole-model}
f^+(q^2)=\frac{f_{B^*}g^{}_{B^*\!B\pi}}{2m_{B^*}\left(1-
\frac{q^2}{m_{B^*}^2}\right)}.
\end{equation}
Thus, the coupling is part of the residue of the pole of the form
factor. It was observed in the literature \cite{Ball:1991ma} that
the vector meson dominance is also valid to smaller values of the
transferred momentum. However, at $q^2=0$, the two approaches, the
pole model and the LCSR for the form factor, differ in their
scaling behavior in the heavy quark limit $m_b\rightarrow\infty$
(see \cite{Belyaev:1995zk}). We therefore do not compare our
results for the form factor at maximum recoil with the sum rule
results for the coupling $g^{}_{B^*\!B\pi}$.
\newline

In the next section, we will briefly sketch the derivation of the
sum rule, following \cite{Belyaev:1995zk} and then apply the
modified analyses to it.

\section{Derivation of the Sum Rule}

We take the following definition for the $B^*\!B\pi$-coupling:
\begin{equation}\label{equ-def-coupling}
\langle{\overline{B}^*}^0(q)\pi^-(p_\pi)|B^-(p_\pi+q)\rangle=-
g^{}_{B^*\!B\pi} \varepsilon_\mu p_\pi^\mu,
\end{equation}
where $\varepsilon_\mu$ is the polarization vector of the
$B^*$-meson. Matrix elements of states with different charges are
related to the definition of the coupling by isospin symmetry.
\newline

Starting point is again the correlation function
(\ref{equ-startLCSR}) of two pseudoscalar currents, sandwiched
between a pion state and the vacuum. This function can be split in
two parts
\begin{equation}
\Pi_\mu(p_\pi,q)=\Pi(q^2,(p_\pi+q)^2)
p_{\pi\mu}+\widetilde{\Pi}(q^2,(p_\pi+q)^2)q_\mu,
\end{equation}
and again, we are only interested in the function in front of the
pion momentum.
\newline

On the hadronic side, we first insert a set of intermediate states
carrying B-meson quantum numbers. This leads to the product of two
matrix elements:
\begin{equation}
\langle\pi|\overline{u}\gamma_\mu b|B\rangle\langle
B|\overline{b}i\gamma_5 d|0\rangle
\end{equation}
The second element is proportional to the B-meson decay constant.
Using crossing symmetry and inserting another set of intermediate
states, this time carrying $B^*$-meson quantum numbers, we obtain
the matrix element of the upper definition
(\ref{equ-def-coupling}) and a second element, which defines the
vector-meson decay constant:
\begin{equation}
\langle0|\overline{q}\gamma_\mu
b|B^*\rangle=m_{B^*}f_{B^*}\varepsilon_\mu.
\end{equation}
Using this definition, we analyzed the sum rule for $f_{B^*}$. The
results were summarized in section \ref{sec-fstar}.
\newline

The hadronic side of the correlation function is:
\begin{equation}\label{equ-hadronic-side-coupling}
\Pi(q^2,(p_\pi+q)^2)=\frac{m_B^2m_{B^*}f_B
f_{B^*}g^{}_{B^*\!B\pi}}{m_b(q^2-m_{B^*}^2)((p_\pi+q)^2-m_B^2)}+
\int\limits_{\Sigma_h}\frac{\rho^h(s_1,s_2)ds_1
ds_2}{(s_1-q^2)(s_2-(p_\pi+q)^2)}+\cdots.
\end{equation}
The ellipses stand for subtraction terms, which will be removed
when the Borel transformation is applied. The limit $\Sigma_h$ of
the double dispersion integral is now a region in the
$(s_1,s_2)$-plane.
\newline

When the expressions (\ref{equ-hadronic-side-coupling}) and the
corresponding QCD-side are Borel transformed with respect to $q^2$
and $(p_\pi+q)^2$, the two sides are equaled and the hadronic
continuum is subtracted by the assumption of quark hadron duality:
\begin{figure}[t]
\begin{picture}(150,50)
\put(2,2){
\begin{minipage}[t]{7cm}
\epsfig{file=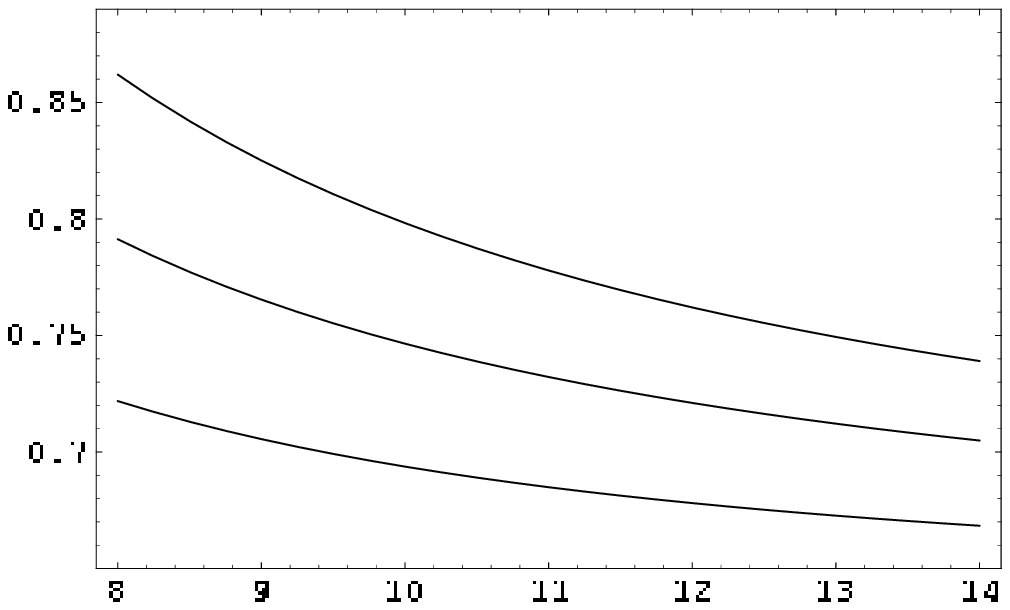,width=7cm}
\end{minipage}}
\Text(64,0.3)[c]{\scalebox{.8}{$M^2[GeV^2]$}}
\Text(37,46)[r]{\scalebox{.8}{$f_Bf_{B^*}g^{}_{B^*\!B\pi}[GeV^2]$}}
\Text(70,40)[r]{\scalebox{.7}{a)}}
\Text(40,33)[r]{\scalebox{.6}{$m_b=4.6\;GeV$}}
\Text(47,21)[r]{\scalebox{.6}{$m_b=4.7\;GeV$}}
\Text(28,11)[r]{\scalebox{.6}{$m_b=4.8\;GeV$}} \put(80,2.3){
\begin{minipage}[t]{7cm}
\epsfig{file=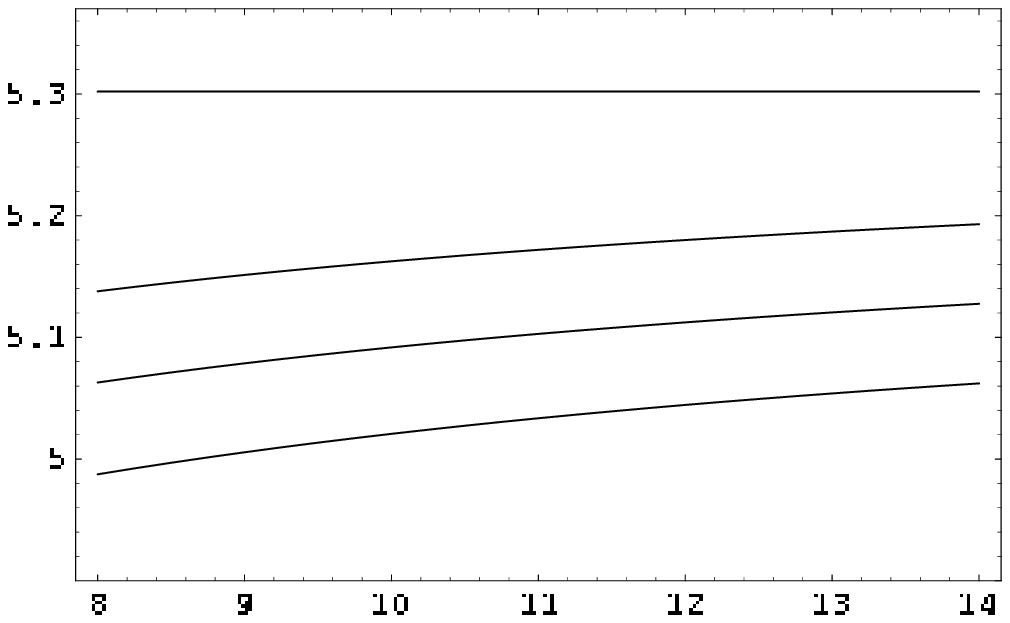,width=7cm,height=3.6cm}
\end{minipage}}
\Text(142,0.3)[c]{\scalebox{.8}{$M^2[GeV^2]$}}
\Text(102,46)[r]{\scalebox{.8}{$\Re^{1/2}[GeV]$}}
\Text(148,40)[r]{\scalebox{.7}{b)}}
\Text(115,40)[r]{\scalebox{.6}{$\hat{m}=5.302\;GeV$}}
\Text(135,30)[r]{\scalebox{.6}{$m_b=4.8\;GeV$}}
\Text(135,24)[r]{\scalebox{.6}{$m_b=4.7\;GeV$}}
\Text(135,18)[r]{\scalebox{.6}{$m_b=4.5\;GeV$}}
\end{picture}
\caption{\small Sum rule for $f_Bf_{B^*}g^{}_{B^*\!B\pi}$ and
corresponding daughter sum rule. The threshold is set to
$s_0=35\;GeV$ and the graphs are plotted for $m_b=(4.6, 4.7,
4.8)\;GeV$. The sum rule decreases about $10\%$ in the Borel
window. }\label{fig-Coupling-usual} \vspace{3mm}
\end{figure}
\begin{equation}
f_Bf_{B^*}g^{}_{B^*\!B\pi}e^{-\frac{m_{B^*}^2}{M_1^2}}
e^{-\frac{m_B^2}{M_2^2}}
=\frac{m_b}{m_B^2m_{B^*}}\int\limits^{\Sigma_0}\!\rho^{QCD}
e^{-\frac{s_1}{M_1^2}}e^{-\frac{s_2}{M_2^2}}ds_1 ds_2.
\end{equation}
The derivation of the QCD spectral function is exactly the same as
in section \ref{sec-lcsr}, since we used the same initial
correlation function. The explicit subtraction procedure is rather
complicated and we refer to the appendix in \cite{Belyaev:1995zk}.
The two Borel parameters are believed to be of the same magnitude.
The choice $M_1^2=M_2^2\equiv 2M^2$ simplifies the derivation of
the final expression for the sum rule:
\begin{equation}\label{equ-cpl-sr}
f_Bf_{B^*}g^{}_{B^*\!B\pi}e^{-\frac{m_B^2+m_{B^*}^2}{2M^2}}=
\frac{m_b^2}{m_B^2m_{B^*}^{}}\Gamma(M^2,s_0),
\end{equation}
with $\Gamma(M^2,s_0)$ given by:
\begin{eqnarray}\label{eq-cpl-sr-gamma}
\Gamma(M^2,s_0)\!&=&\!M^2\left(e^{-\frac{m_b^2}{M^2}}-e^{-
\frac{s_0}{M^2}}\right)
\left[\varphi_\pi(1/2)+\frac{\mu_\pi}{m_b}\left(\frac{1}{2}
\varphi_\sigma(1/2)
+\frac{1}{3}\varphi_\sigma(1/2)\right)+\frac{2f_{3\pi}}{m_b
f_\pi}I_{3P}^{(3)}\right]\nonumber\\
&&+e^{-\frac{m_b^2}{M^2}}\Bigg[\frac{\mu_\pi
m_b}{3}\varphi_\sigma(1/2)+g_2(1/2)-\frac{4m_b^2}{M^2}\bigg(g_1(1/2)
+\int\limits_0^{1/2}g_2(v)dv\bigg)+I_{3P}^{(4)}\Bigg]\nonumber\\
&&+\frac{\alpha_s}{3\pi}\int\limits_{2m_b^2}^{2s_0}f\!
\left(\frac{t}{m_b^2}-2\right)e^{-\frac{t}{2M^2}}dt.
\end{eqnarray}
Here, the three particle twist 3 and 4 contributions are given by:
\begin{eqnarray}
I_{3P}^{(3)}&=&-\frac{3}{32}(20+5\omega_{1,0}+\omega_{1,1}-4
\omega_{2,0})\\
I_{3P}^{(4)}&=&\frac{5}{3}\delta^2.
\end{eqnarray}
The $O(\alpha_s)$-corrections to the two particle twist 2 wave
function were obtained by Khodjamirian and collaborators, and we
adopt the expression from \cite{Khodjamirian:1999hb}:
\begin{eqnarray}
f(x)&=&\frac{\pi^2}{4}+3\log\left(\frac{x}{2}\right)\log
\left(1+\frac{x}{2}\right)
-\frac{3(3x^3+22x^2+40x+24)}{2(2+x)^3}\log\left(\frac{x}{2}
\right)\nonumber\\
&&+6Li_2\left(\frac{x}{2}\right)-3Li_2(-x)-3Li_2(-x-1)-
3\log(1+x)\log(2+x)\nonumber\\
&&-\frac{3(3x^2+20x+20)}{4(2+x)^2}+\frac{6x(1+x)\log(1+x)}{(2+x)^3}.
\end{eqnarray}
For the following analyses, we took the same window for the Borel
parameter as for the form factor in chapter \ref{sec-Form Factor}:
\begin{equation}
8\;GeV\leq M^2\leq14\;GeV^2.
\end{equation}

The contribution of the twist 2 wave function at the left boundary
is about $60\%$ and of the twist 3 terms about $40\%$ of the
total. The twist 4 contribution is numerically negligible. The
contribution from the hadronic continuum grows from $20\%$ to
about $30\%$ at the right edge of the window. Thus, the use of
this window is reasonable.
\newline

For typical values of the input parameters - $s_0=35\;GeV$,
$m_b=(4.7\pm 0.1)\;GeV$ - we plotted in figure
\ref{fig-Coupling-usual} the sum rule for the product
$f_Bf_{B^*}g^{}_{B^*\!B\pi}$ and the corresponding daughter sum
rule, defined by:
\begin{equation}\label{equ-cpl-daughtersr}
\Re(M^2,s_0)=M^4\frac{d}{dM^2}\log\Gamma(M^2,s_0).
\end{equation}
Applying the same derivative on the left hand side of equation
(\ref{equ-cpl-sr}), we see that the sum rule reaches a minimum,
when the daughter sum rule is equal to
\begin{figure}[t]
\begin{picture}(150,50)
\put(2,2){
\begin{minipage}[t]{7cm}
 \epsfig{file=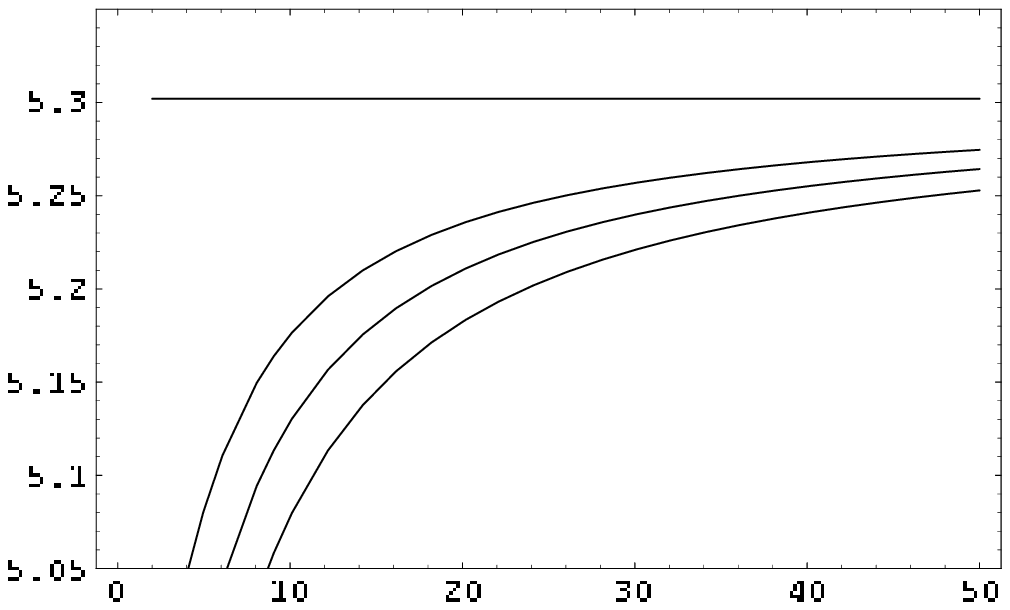,width=7cm}
\end{minipage}}
\Text(64,0.3)[c]{\scalebox{.8}{$M^2[GeV^2]$}}
\Text(25,45.6)[r]{\scalebox{.8}{$\Re^{1/2}[GeV]$}}
\Text(70,40)[r]{\scalebox{.7}{a)}}
\Text(35,39)[r]{\scalebox{.6}{$\hat{m}=5.302\;GeV$}}
\Text(27,28)[r]{\scalebox{.6}{$m_b=4.8\;GeV$}}
\Text(56,21)[r]{\scalebox{.6}{$m_b=4.7\;GeV$}}
\Text(48,16)[r]{\scalebox{.6}{$m_b=4.6\;GeV$}}
\put(31,27){\line(-3,1){4}} \put(41,21){\line(-4,1){9}}
\put(33,16){\line(-1,0){4}}
\put(80,2){
\begin{minipage}[t]{7cm}
\epsfig{file=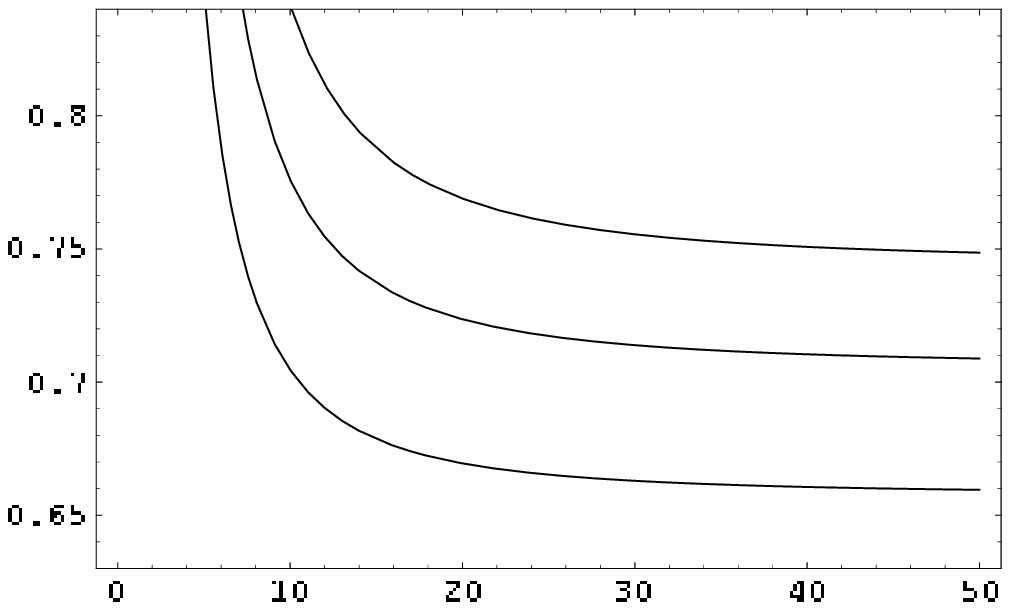,width=7cm,height=3.5cm}
\end{minipage}}
\Text(142,0.3)[c]{\scalebox{.8}{$M^2[GeV^2]$}}
\Text(115,45.5)[r]{\scalebox{.8}{$f_Bf_{B^*}g^{}_{B^*\!B\pi}[GeV^2]$}}
\Text(147,40)[r]{\scalebox{.7}{b)}}
\Text(135,30)[r]{\scalebox{.6}{$m_b=4.6\;GeV$}}
\Text(135,22)[r]{\scalebox{.6}{$m_b=4.7\;GeV$}}
\Text(135,13)[r]{\scalebox{.6}{$m_b=4.8\;GeV$}}
\end{picture}
\caption{\small Matching procedure for the sum rule of
$f_Bf_{B^*}g^{}_{B^*\!B\pi}$. Diagram a) shows the matching of the
daughter sum rule in the limit $M^2\rightarrow \infty$ for three
different values of the quark mass $m_b$. The values of the
threshold obtained by this procedure are used to extract the value
for $f_Bf_{B^*}g^{}_{B^*\!B\pi}$ in same limit
(b).}\label{fig-Coupling-uinf} \vspace{3mm}
\end{figure}
\begin{equation}\label{equ-cpl-hatm}
\hat{m}^2=\frac{m_{B^*}^2+m_B^2}{2}=28.11\; GeV^2,
\end{equation}
if the experimental values for the meson masses $m_B=5.279\;GeV$
and $m_{B^*}=5.325\;GeV$ are taken. The daughter sum rule is
rather far apart from $\hat{m}$ and the sum rule for the product
$f_Bf_{B^*}g^{}_{B^*\!B\pi}$ is quite unstable in the chosen
window.
\newline

In the following two sections, we will apply the methods
introduced in the preceding two chapters on the sum rule
(\ref{equ-cpl-sr}).

\section{The Limit $M^2\rightarrow\infty$}

In this section, we will analyze the sum rule (\ref{equ-cpl-sr})
by imposing \textit{local duality}. We take the limit
$M^2\rightarrow \infty$ and match the daughter sum rule
(\ref{equ-cpl-daughtersr}) to the mass $\hat{m}$, defined in
equation (\ref{equ-cpl-hatm}). From this matching, we can extract
a unique value for the threshold $s_0$ and therefore a unique
value for $f_Bf_{B^*}g^{}_{B^*\!B\pi}$.
\newline

\begin{figure}[t]
\begin{picture}(150,50)
\put(2,2){
\begin{minipage}[t]{7cm}
\epsfig{file=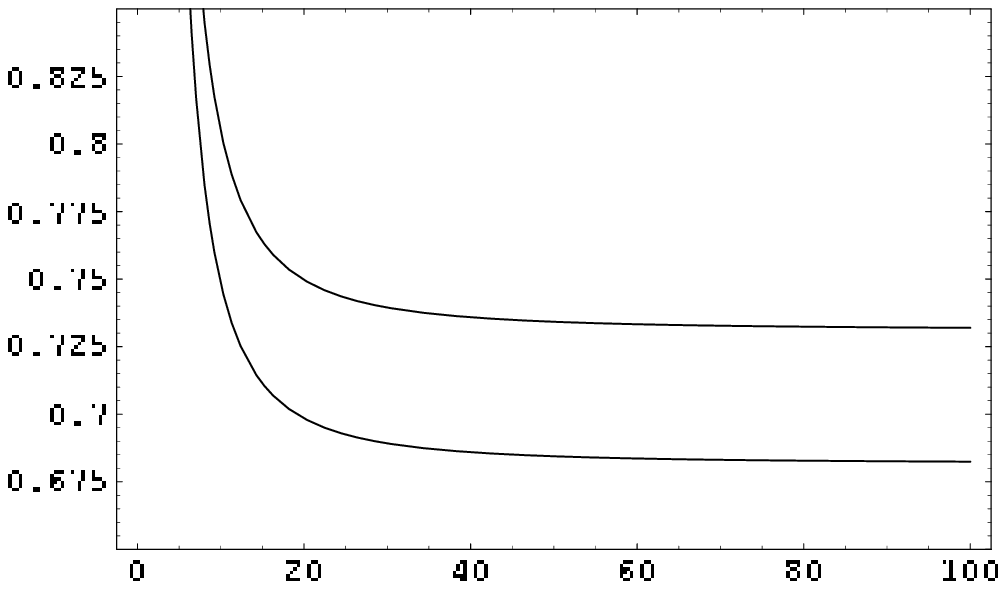,width=7cm}
\end{minipage}}
\Text(64,0.3)[c]{\scalebox{.8}{$M^2[GeV^2]$}}
\Text(37,45)[r]{\scalebox{.8}{$f_Bf_{B^*}g^{}_{B^*\!B\pi}[GeV^2]$}}
\Text(69,39)[r]{\scalebox{.7}{a)}}
\Text(56,23)[r]{\scalebox{.6}{$\mu=2\;GeV$}}
\Text(56,13)[r]{\scalebox{.6}{$\mu=3\;GeV$}} \put(80,2){
\begin{minipage}[t]{7cm}
\epsfig{file=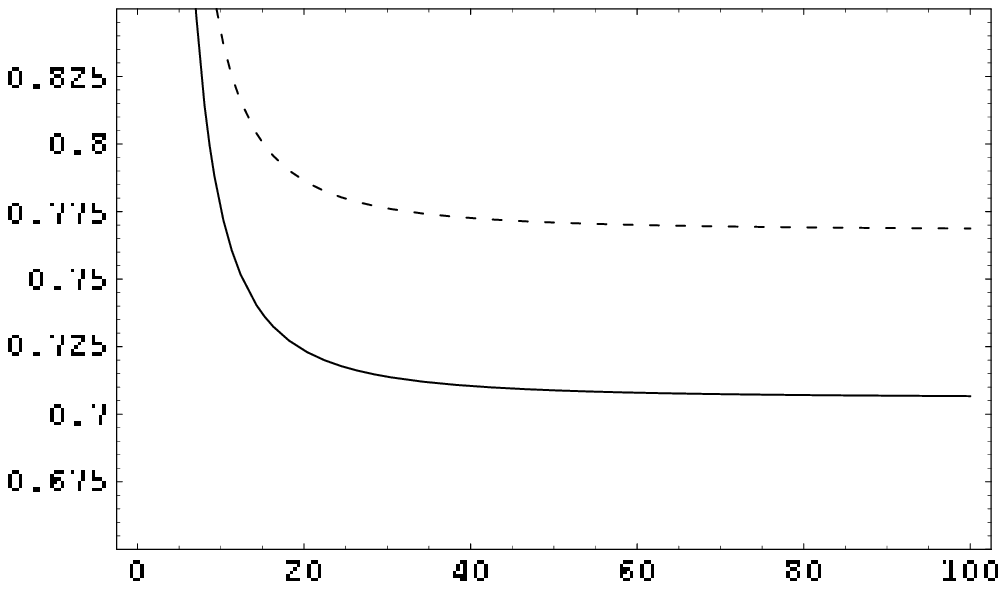,width=7cm,height=3.6cm}
\end{minipage}}
\Text(142,0.3)[c]{\scalebox{.8}{$M^2[GeV^2]$}}
\Text(115,45)[r]{\scalebox{.8}{$f_Bf_{B^*}g^{}_{B^*\!B\pi}[GeV^2]$}}
\Text(147,39)[r]{\scalebox{.7}{b)}}
\end{picture}
\caption{\small a: Scale dependence of the sum rule for the
product $f_Bf_{B^*}g^{}_{B^*\!B\pi}$ in the limit
$M^2\rightarrow\infty$. When the scale is varied from $\mu=2\;GeV$
to $\mu=3\;GeV$, the resulting value of the sum rule decreases
about $7\%$. - b: Comparison of the sum rule result using the
asymptotic pion wave function (dashed line) with the central
result. The coupling $g^{}_{B^*\!B\pi}$ increases about $9\%$,
when the asymptotic function is
used.}\label{fig-Coupling-uinf-scaleass} \vspace{3mm}
\end{figure}

In figure \ref{fig-Coupling-uinf}, we plotted the graphs of the
sum rule and its daughter sum rule for the central value of the
quark mass $m_b=4.7\;GeV$ and the two extremal values
$m_b=4.6\;GeV$ and $m_b=4.8\;GeV$. The error due to the
uncertainties in the quark mass are about $7\%$. Besides this
error source, we also analyzed the scale dependence of the sum
rule. In figure \ref{fig-Coupling-uinf-scaleass}$\,$a, we plotted
the resulting graphs for $\mu=2\;GeV$ and $\mu=3\;GeV$. Again, we
took account for the uncertainties of the coefficients of the wave
functions by analyzing the sum rule for the asymptotic twist 2
wave function (figure \ref{fig-Coupling-uinf-scaleass}b). The
resulting deviation from the central result (9\%) was assigned as
additional error to the final result. Neglecting further error
sources, we extract:
\begin{equation}\label{equ-cpl-uinf-result}
f_Bf_{B^*}g^{}_{B^*\!B\pi}=(0.71\pm0.05\pm0.07)\;GeV^2,
\end{equation}
where the first error comes from the uncertain quark mass and the
second summarizes the error from the scale dependence and the
uncertainties of the coefficients of the wave functions. The value
(\ref{equ-cpl-uinf-result}) is about $10\%$ higher than the result
obtained in \cite{Belyaev:1995zk}, which did not include
$O(\alpha_s)$-corrections, and again about $10\%$ smaller than the
result from \cite{Khodjamirian:1999hb}, which was the first
analysis using the $O(\alpha_s)$-corrections to the twist 2 wave
function.
\newline

If we divide (\ref{equ-cpl-uinf-result}) by the corresponding
value of the decay constant (\ref{equ-fb-u-infinity}), extracted
using the same method, we can get the residue of the form factor
in the simple pole model:
\begin{equation}\label{equ-uinf-residue}
c=\frac{f_{B^*}g^{}_{B^*\!B\pi}}{2m_{B^*}}=(0.37\pm0.06)\;GeV.
\end{equation}
Here, we simultaneously varied the quark mass, leading to a rather
small contribution to the overall error (about $4\%$). This value
is compatible with the result from the NLO-analysis
\cite{Khodjamirian:1999hb} - $(0.42\pm0.09)\;GeV$.
\newline

Dividing (\ref{equ-cpl-uinf-result}) by the pseudoscalar decay
constant (\ref{equ-fb-u-infinity}) and the vector-meson decay
constant (\ref{equ-fstar-uinf}), we get for the
$B^*\!B\pi$-coupling:
\begin{equation}\label{equ-uinf-coupling}
g^{}_{B^*\!B\pi}=20\pm5.
\end{equation}
This time, the uncertainty of the quark mass inflicts an error of
about the same size as the further error sources to the final
result, which is about $17\%$. The result 
(\ref{equ-uinf-coupling}) is noticeably lower than the result of
the LO-analysis \cite{Belyaev:1995zk} ($g^{}_{B^*\!B\pi}=29\pm3$)
and of about the same size as the NLO-result
\cite{Khodjamirian:1999hb} - $g^{}_{B^*\!B\pi}=22\pm7$.
\newline

In section \ref{sec-cpl-relation},we will discuss the results for
the coupling $g_{B*B\pi}$ and relate them to the literature.

\section{Borel Mass dependent Threshold $s_0(M^2)$}

In this section we analyze the sum rule (\ref{equ-cpl-sr}) with
the method introduced in section \ref{s(u)}. We take the threshold
as a function of the Borel mass $s_0(M^2)$ and solve the
differential equation
\begin{equation}\label{equ-coupling-diff-equ}
s_0'(M^2)=\frac{\hat{m}^2\Gamma(M^2,s_0(M^2))-M^4\frac{\partial}{\partial
M^2}\Gamma(M^2,s_0(M^2))}{M^4\frac{\partial}{\partial
s_0}\Gamma(M^2,s_0(M^2))},
\end{equation}
to get constant functions for the product
$f_Bf_{B^*}g^{}_{B^*\!B\pi}$. Since the initial sum rule was
rather dependent on the Borel mass (see figure
\ref{fig-Coupling-usual}), we expect the resulting functions
$s_0(M^2)$ to have quite high average values.
\begin{figure}[t]
\begin{picture}(145,90)
\put(3.5,51.5){
\begin{minipage}[t]{7cm}
 \epsfig{file=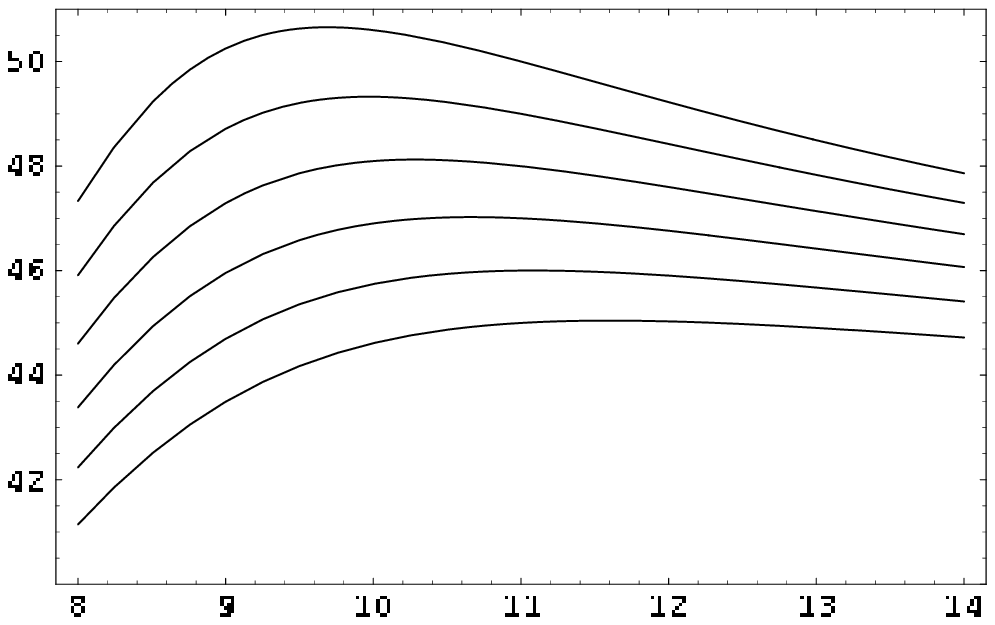,width=6.825cm,height=3.9cm}
\end{minipage}}
\Text(64,49.5)[c]{\scalebox{.8}{$M^2[GeV^2]$}}
\Text(11,95.5)[l]{\scalebox{.8}{$s_0(M^2)[GeV^2]$}}
\Text(68,90)[r]{\scalebox{.7}{a)}} \put(1,2.5){
\begin{minipage}[t]{7cm}
 \epsfig{file=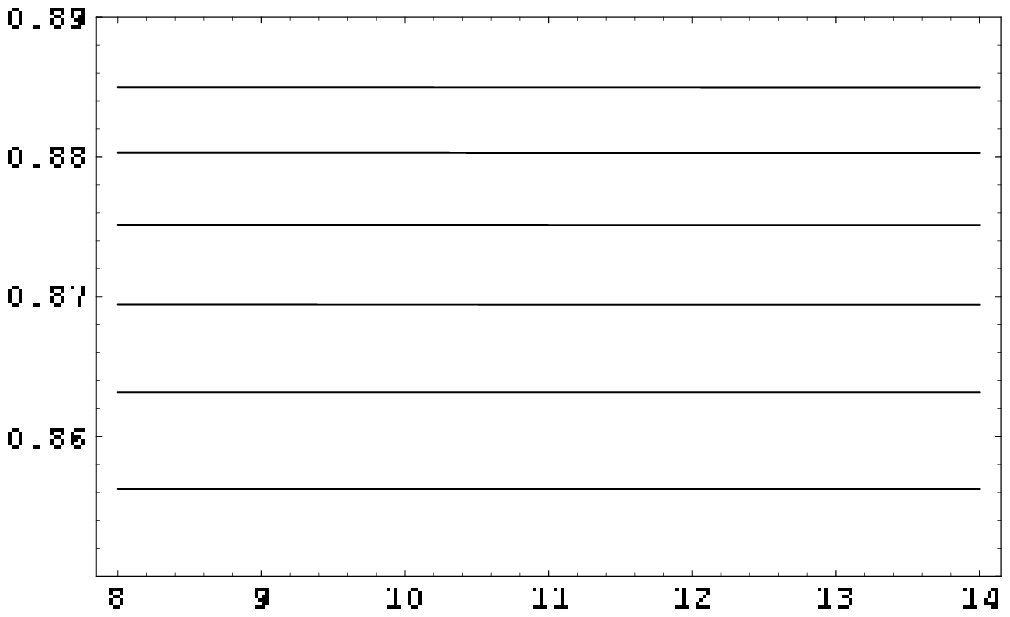,width=7cm}
\end{minipage}}
\Text(64,0.3)[c]{\scalebox{.8}{$M^2[GeV^2]$}}
\Text(10,46)[l]{\scalebox{.8}{$f_Bf_{B^*}g^{}_{B^*\!B\pi}[GeV^2]$}}
\Text(68,41)[r]{\scalebox{.7}{b)}}
\Text(88,46)[l]{\scalebox{.8}{$f_Bf_{B^*}g^{}_{B^*\!B\pi}[GeV^2]$}}
\put(79.5,51){
\begin{minipage}[t]{8cm}
 \epsfig{file=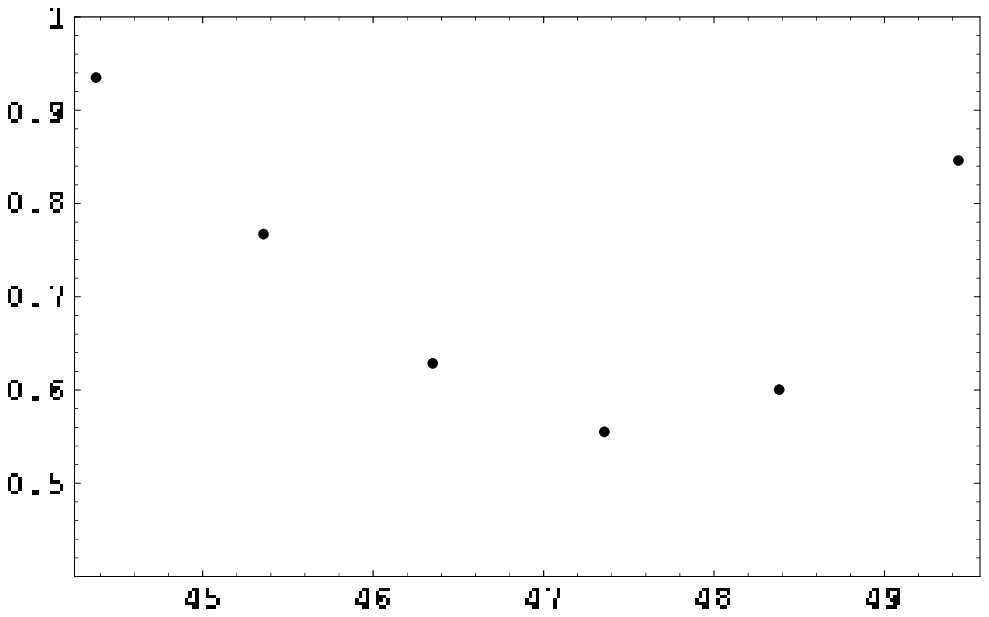,width=7cm,height=4.3cm}
\end{minipage}}
\Text(141,49.5)[c]{\scalebox{.8}{$\overline{s}_0[GeV^2]$}}
\Text(90,95)[c]{\scalebox{.8}{$s_{var}$}}
\Text(144,90)[r]{\scalebox{.7}{c)}} \put(79,2){
\begin{minipage}[t]{8cm}
 \epsfig{file=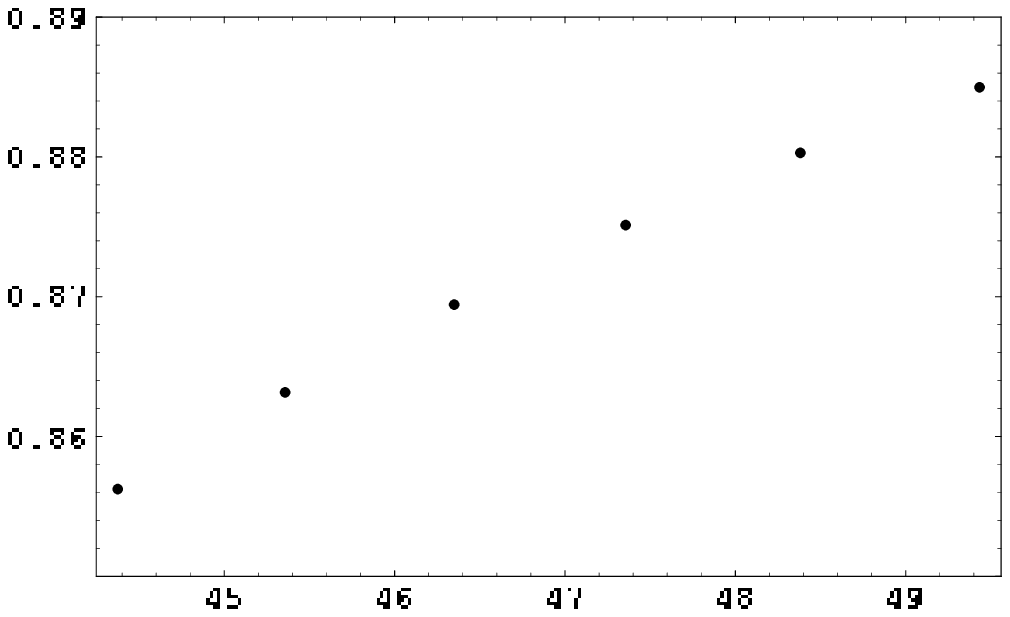,width=7cm,height=4.3cm}
\end{minipage}}
\Text(143,0.3)[c]{\scalebox{.8}{$\overline{s}_0[GeV^2]$}}
\Text(144,41)[r]{\scalebox{.7}{d)}}
\end{picture}
\caption{\small Sum rule analysis of the product
$f_Bf_{B^*}g^{}_{B^*\!B\pi}$. Graph a) shows solutions to the
differential equation (\ref{equ-coupling-diff-equ}), b) the
corresponding functions for $f_Bf_{B^*}g^{}_{B^*\!B\pi}$, d) shows
these values plotted as functions of the average threshold
$\overline{s}_0$. Graph c) shows the variations of the thresholds
$s(M^2)$. The product $f_Bf_{B^*}g^{}_{B^*\!B\pi}$ is extracted at
the minimum.}\label{fig-coupling-s(u)-47} \vspace{3mm}
\end{figure}
\newline

For central values of the input parameters, the solutions
$s_0(M^2)$ and the corresponding sum rules are shown in figure
\ref{fig-coupling-s(u)-47}. The average values of the threshold
are quite high compared to the values in the last section. This is
equivalent to the behavior of the initial sum rule, which gets
more stable with higher values of the constant threshold $s_0$.
\newline

\begin{figure}[p]
\begin{picture}(145,90)
\put(2.5,51.5){
\begin{minipage}[t]{7cm}
 \epsfig{file=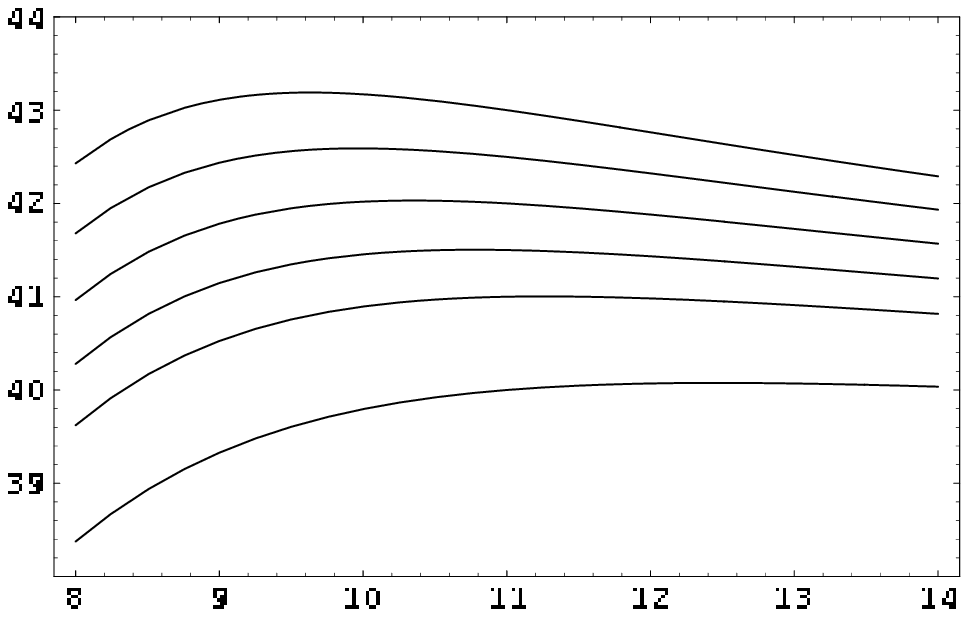,width=7cm,height=4.2cm}
\end{minipage}}
\Text(64,49.5)[c]{\scalebox{.8}{$M^2[GeV^2]$}}
\Text(11,94.5)[l]{\scalebox{.8}{$s_0(M^2)[GeV^2]$}}
\Text(68,90)[r]{\scalebox{.7}{a)}} \put(1,2.5){
\begin{minipage}[t]{7cm}
 \epsfig{file=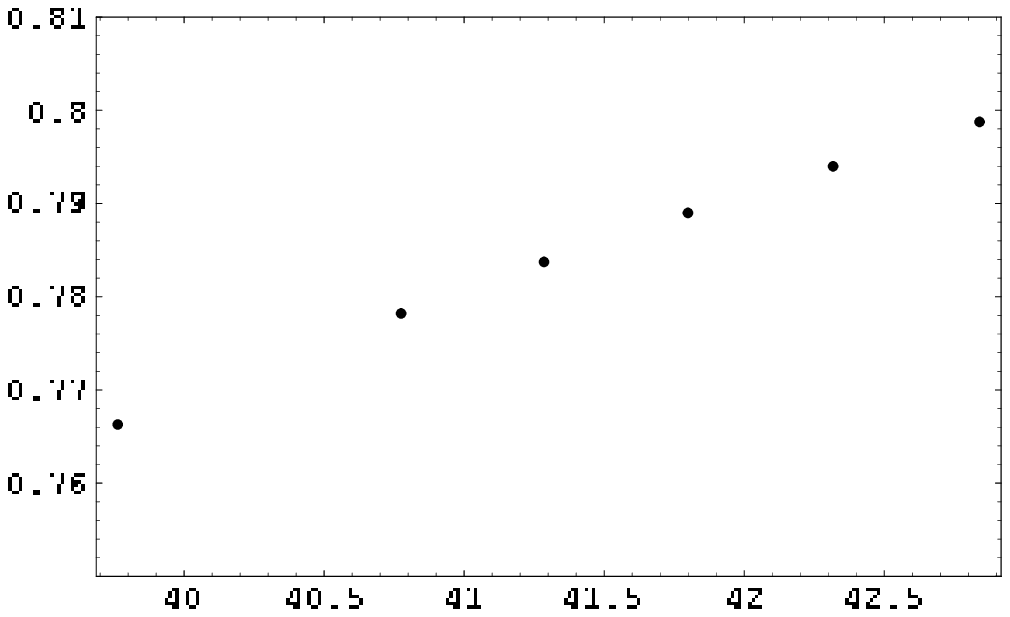,width=6.4cm}
\end{minipage}}
\Text(64,0.3)[c]{\scalebox{.8}{$\overline{s}_0[GeV^2]$}}
\Text(11,46)[l]{\scalebox{.8}{$f_Bf_{B^*}g^{}_{B^*\!B\pi}[GeV^2]$}}
\Text(68,41)[r]{\scalebox{.7}{b)}}
\Text(88,46.5)[l]{\scalebox{.8}{$f_Bf_{B^*}g^{}_{B^*\!B\pi}[GeV^2]$}}
\put(81.5,52){
\begin{minipage}[t]{8cm}
 \epsfig{file=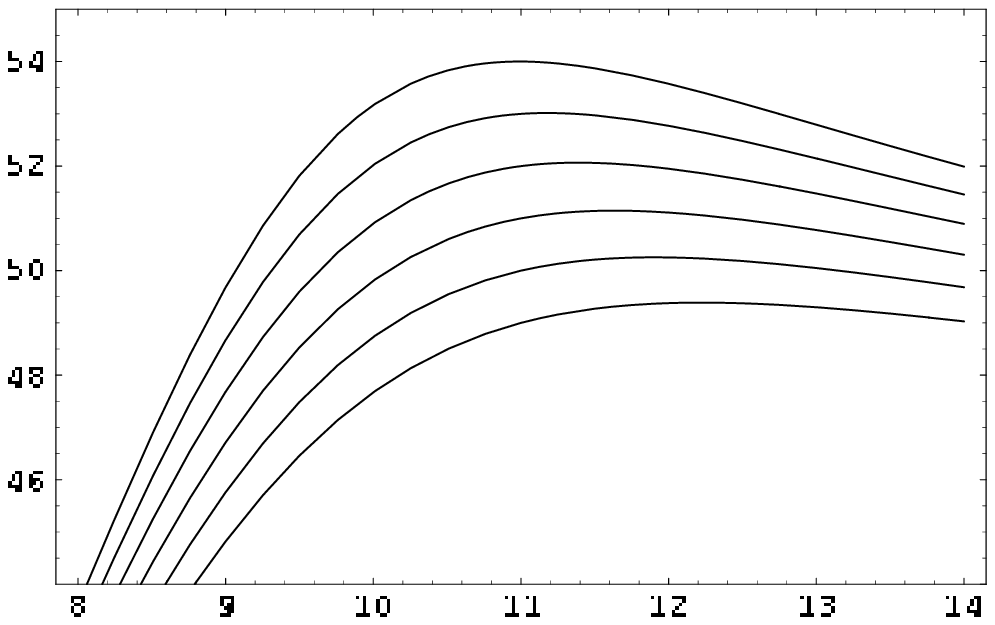,width=6.6cm}
\end{minipage}}
\Text(142,49.5)[c]{\scalebox{.8}{$M^2[GeV^2]$}}
\Text(89,94.5)[l]{\scalebox{.8}{$s_0(M^2)[GeV^2]$}}
\Text(120,94.5)[l]{\scalebox{.8}{$m_b=4.6\;GeV$}}
\Text(144,90)[r]{\scalebox{.7}{c)}} \put(79,2){
\begin{minipage}[t]{8cm}
 \epsfig{file=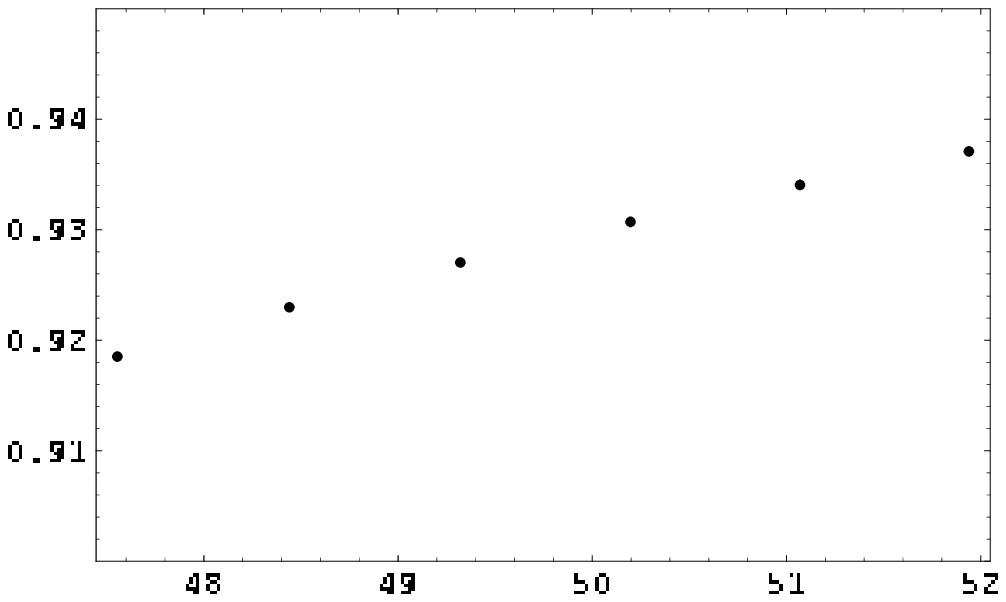,width=7cm,height=3.5cm}
\end{minipage}}
\Text(143,0.5)[c]{\scalebox{.8}{$\overline{s}_0[GeV^2]$}}
\Text(40,94.5)[l]{\scalebox{.8}{$m_b=4.8\;GeV$}}
\Text(144,41)[r]{\scalebox{.7}{d)}}
\end{picture}
\caption{\small Sum rule analyses for different values of the
quark mass. The left side shows the thresholds $s_0(M^2)$ and
corresponding values $f_Bf_{B^*}g^{}_{B^*\!B\pi}$ plotted over
$\overline{s}_0(M^2)$ for the case $m_b=4.8\;GeV$. The right side
summarize the analysis for
$m_b=4.6\;GeV$.}\label{fig-coupling-s(u)-468}
\end{figure}

In figure \ref{fig-coupling-s(u)-468}, we plotted the functions
$s_0(M^2)$ for the two extremal input values for the quark mass
$m_b=4.6\;GeV$ and $m_b=4.8\;GeV$. Whereas we could find a minimum
of the variation of the threshold functions in the case of
$m_b=4.8\;GeV$, we could not do so for $m_b=4.6\;GeV$. Solving the
differential equation (\ref{equ-coupling-diff-equ}) leads to
functions $s_0(M^2)$, which are strongly dependent on the Borel
mass (see figure \ref{fig-coupling-s(u)-468}$\,$c). Their
variations are rather high compared to the former analyses and do
not show a minimum. Nevertheless, we could find a region where the
variations are changing very slowly. We plotted the corresponding
values $f_Bf_{B^*}g^{}_{B^*\!B\pi}$ in this region in figure
\ref{fig-coupling-s(u)-468}d. However, to be on the safe side we
added the difference of the central value and the lower bound from
$m_b=4.8\;GeV$ to the mean result to obtain an upper bound.
\newline

\begin{figure}[p]
\begin{picture}(145,90)
\put(2.5,51.5){
\begin{minipage}[t]{7cm}
 \epsfig{file=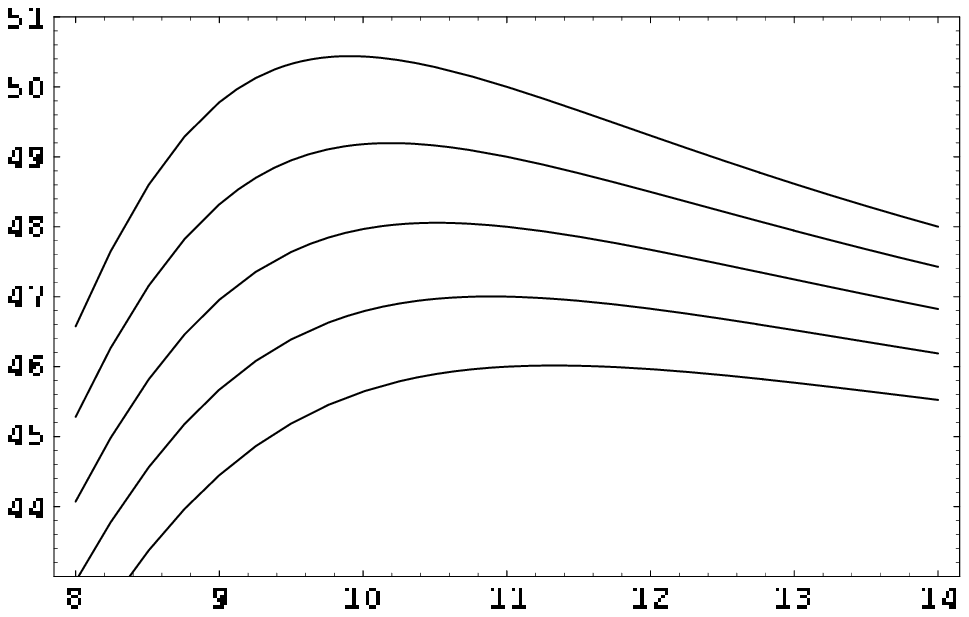,width=7cm,height=4.2cm}
\end{minipage}}
\Text(64,49.5)[c]{\scalebox{.8}{$M^2[GeV^2]$}}
\Text(11,95)[l]{\scalebox{.8}{$s_0(M^2)[GeV^2]$}}
\Text(68,90)[r]{\scalebox{.7}{a)}} \put(1,2.5){
\begin{minipage}[t]{7cm}
 \epsfig{file=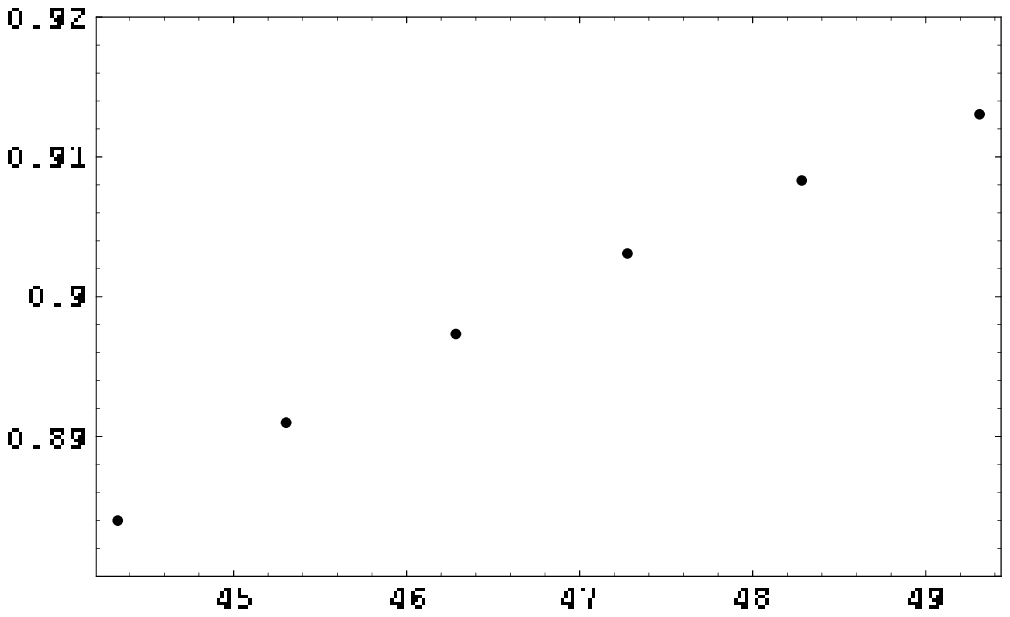,width=6.6cm}
\end{minipage}}
\Text(64,0.3)[c]{\scalebox{.8}{$\overline{s}_0[GeV^2]$}}
\Text(11,46)[l]{\scalebox{.8}{$f_Bf_{B^*}g^{}_{B^*\!B\pi}[GeV^2]$}}
\Text(68,41)[r]{\scalebox{.7}{b)}}
\Text(88,46)[l]{\scalebox{.8}{$f_Bf_{B^*}g^{}_{B^*\!B\pi}[GeV^2]$}}
\put(81,51.5){
\begin{minipage}[t]{8cm}
 \epsfig{file=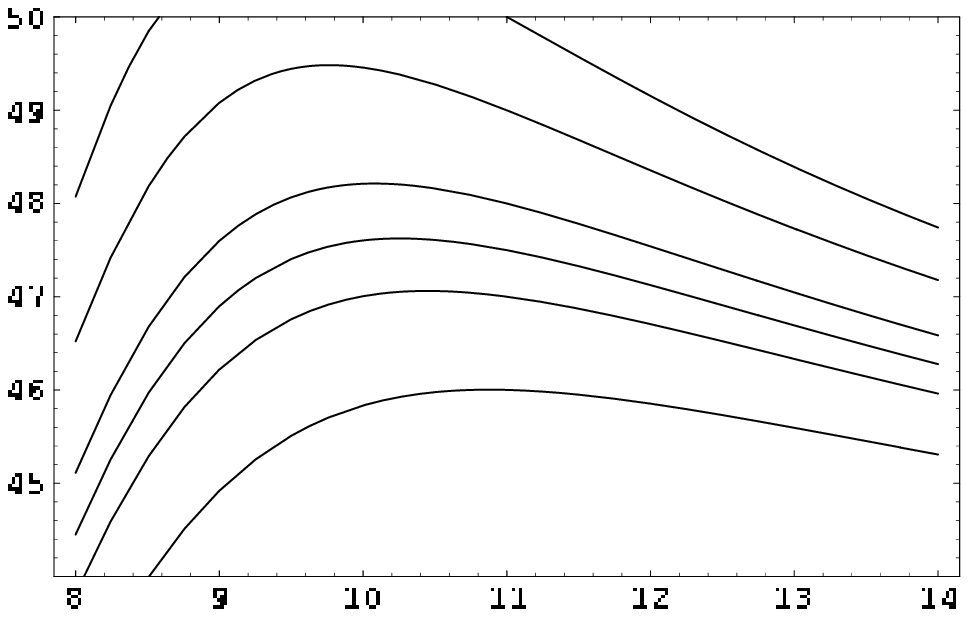,width=6.8cm}
\end{minipage}}
\Text(142,49.5)[c]{\scalebox{.8}{$M^2[GeV^2]$}}
\Text(89,95.5)[l]{\scalebox{.8}{$s_0(M^2)[GeV^2]$}}
\Text(120,95)[l]{\scalebox{.8}{$\mu=3\;GeV$}}
\Text(144,90)[r]{\scalebox{.7}{c)}} \put(77.5,2){
\begin{minipage}[t]{8cm}
 \epsfig{file=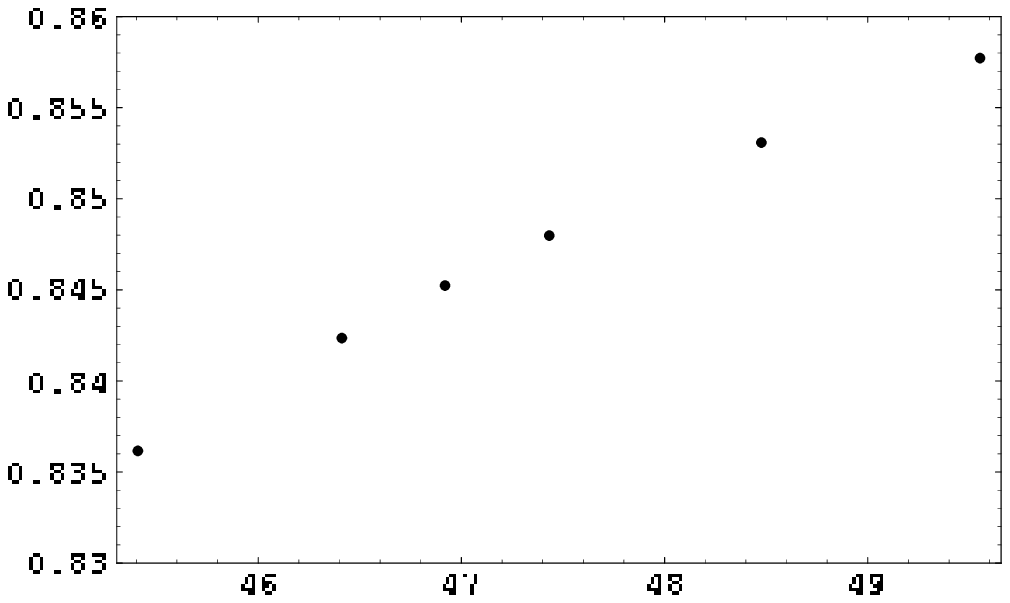,width=6.4cm,height=4.2cm}
\end{minipage}}
\Text(143,0.5)[c]{\scalebox{.8}{$\overline{s}_0[GeV^2]$}}
\Text(40,95)[l]{\scalebox{.8}{$\mu=2\;GeV$}}
\Text(144,41)[r]{\scalebox{.7}{d)}}
\end{picture}
\caption{\small $f_Bf_{B^*}g^{}_{B^*\!B\pi}$ evaluated at the
scales $\mu=2\;GeV$ (left side) and $\mu=3\;GeV$ (right
side).}\label{fig-coupling-s(u)-scale}
\end{figure}

The scale dependence of this sum rule analysis is shown in figure
\ref{fig-coupling-s(u)-scale} and is less than $4\%$ of the final
result, if the scales $\mu=2\;GeV$ and $\mu=3\;GeV$ are used to
extract $f_Bf_{B^*}g^{}_{B^*\!B\pi}$.
\newline

We also analyzed the sum rule for the asymptotic twist 2 wave
function. The results are shown in figure
\ref{fig-Coupling-s(u)-ass}. As in the last section, the central
result increases about $9\%$.
\newline

The final result for the product $f_Bf_{B^*}g^{}_{B^*\!B\pi}$ we
extract, is:
\begin{equation}\label{equ-result-s(u)-ffg}
f_Bf_{B^*}g^{}_{B^*\!B\pi}=(0.88\pm0.08\pm0.09)\;GeV^2.
\end{equation}
This is about $10\%$ higher than the result obtained by
Khodjamirian and collaborators using the same sum rule
\cite{Khodjamirian:1999hb}.
\newline

Using this result to again extract a value for the residue $c$ in
the single pole model, we get after dividing by
(\ref{equ-decay-constant-s(u)}):
\begin{figure}[t]
\begin{picture}(150,50)
\put(2,2.5){
\begin{minipage}[t]{7cm}
 \epsfig{file=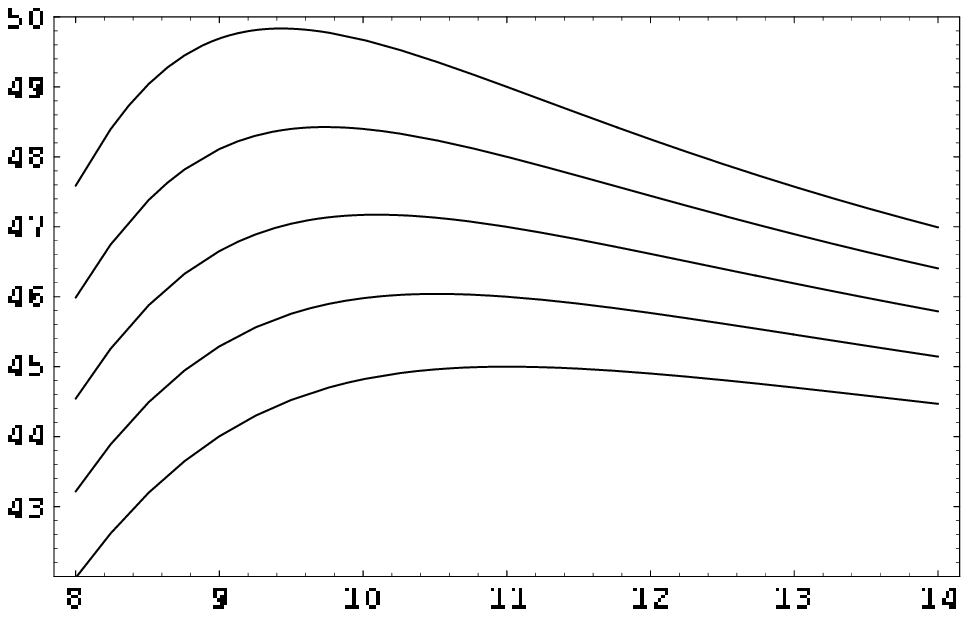,width=6.7cm}
\end{minipage}}
\Text(64,0.3)[c]{\scalebox{.8}{$M^2[GeV^2]$}}
\Text(10,46)[l]{\scalebox{.8}{$s_0(M^2)[GeV^2]$}}
\Text(69,39)[r]{\scalebox{.7}{a)}} \put(80,2){
\begin{minipage}[t]{7cm}
\epsfig{file=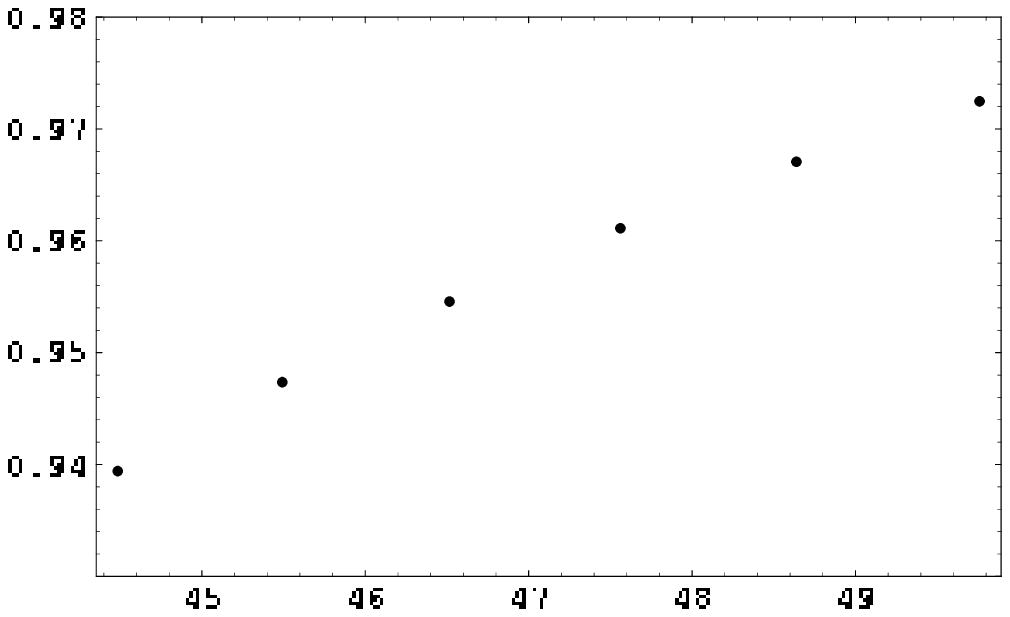,width=6.3cm,height=4.3cm}
\end{minipage}}
\Text(144,0.3)[c]{\scalebox{.8}{$\overline{s}_0[GeV^2]$}}
\Text(90,46)[l]{\scalebox{.8}{$f_Bf^+(0)[GeV]$}}
\Text(145,40)[r]{\scalebox{.7}{b)}}
\end{picture}
\caption{\small Sum rule analysis for the asymptotic twist 2 pion
wave function. The extracted value is about $9\%$ higher than the
central result.}\label{fig-Coupling-s(u)-ass} \vspace{3mm}
\end{figure}
\begin{equation}\label{equ-c-result}
c=(0.40^{+7}_{-5})\;GeV,
\end{equation}
which is close to the result (\ref{equ-uinf-residue}) obtained in
the preceding section.
\newline

Dividing (\ref{equ-result-s(u)-ffg}) by the two decay constants
(\ref{equ-decay-constant-s(u)}) and (\ref{equ-fstar-s(u)}), we
get:
\begin{equation}\label{equ-cpl-result}
g^{}_{B^*\!B\pi}=17^{+6}_{-3}.
\end{equation}
The difference between the positive and negative error assigned to
the results (\ref{equ-cpl-result}) and (\ref{equ-c-result}) is due
to the different effect of a higher and lower quark mass on the
decay constant (\ref{equ-decay-constant-s(u)}). The error
inflicted on the final results by the different masses is about
$18\%$.

\section{Discussion and Relation to the Literature}\label{sec-cpl-relation}

In the last two sections, we analyzed the sum rule for the product
$f_Bf_{B^*}g^{}_{B^*\!B\pi}$. As in chapter \ref{sec-Form Factor},
it was hoped that when this sum rule is divided by the
corresponding sum rule results for the decay constants, the
intrinsic errors of the two methods would be reduced. The results
(\ref{equ-uinf-residue}) and (\ref{equ-c-result}) for the residue
$c$ of the pole in the simple pole model (\ref{equ-pole-model})
are within $6\%$ of each other, which is a quite satisfying,
considering the significant difference of the two initial sum rule
results (\ref{equ-cpl-uinf-result}) and
(\ref{equ-result-s(u)-ffg}). However, the extracted values for the
coupling $g^{}_{B^*\!B\pi}$ (\ref{equ-uinf-coupling}) and
(\ref{equ-cpl-result}) differ about $15\%$. The reason for this
deviation might be the unusual large value for the vector meson
decay constant, found in section \ref{sec-fstar}.
\begin{table}[b]
\centering
\begin{tabular}{|l|c|l|}
\hline $g^{}_{B^*\!B\pi}$&Ref.&Method \\
\hline \hline
$20\pm 5$&\small This Thesis&\small LCSR; $O(\alpha_s)$-twist 2;
$\lim M^2\rightarrow\infty$\\
$17^{+6}_{-3}$&\small This Thesis&\small LCSR;
$O(\alpha_s)$-twist 2; $s_0(M^2)$\\
$15\pm 4$&\small Colangelo '94\cite{Colangelo:1994es}&
\small NLO-soft pion limit\\
$28 \pm 6$&\small Belyaev '95\cite{Belyaev:1995zk}&\small LCSR-LO\\
$14 \pm 4$&\small Dosch '96\cite{Dosch:1996kw}&
\small Double Moment SR\\
$22 \pm 7$&\small Khodjamirian '99\cite{Khodjamirian:1999hb}&
\small LCSR; $O(\alpha_s)$-twist 2;\\
$14.5 \pm 3.9$&\small Navarra '00\cite{Navarra:2000ji}&
\small 3-pt SR\\
$32 \pm 5$&\small Melikhov '01\cite{Melikhov:2001zv}&
\small Dispersion Approach\\
$47\pm 5 \pm 8$&\small Abada '03\cite{Abada:2003un}&
\small Lattice, quenched\\
\hline
\end{tabular}
\caption{\label{tab-literature-cpl}\small Comparison of our
results to a collection from the literature.}
\end{table}
\newline

We did not take further perturbative corrections into account.
Assuming that the first order corrections to the twist 3
contribution is about the same relative size as the
$O(\alpha_s)$-twist 2 corrections ($\sim 40\%$), we get another
uncertainty of $20\%$.
\newline

Furthermore, the hadronic continuum contribution was only
subtracted from the leading terms in the expansion
(\ref{eq-cpl-sr-gamma}). The inflicted error to the final result
is negligible as well as variations from the boundary $\Sigma_0$
in the plane $(s_1,s_2)$ \cite{Khodjamirian:1999hb}.
\newline

In table \ref{tab-literature-cpl}, we listed our results for the
coupling and a collection from the literature. It should be noted
that the values obtained in sum rule approaches are significantly
lower than results from quark models and lattice calculations. In
the case of the D-meson coupling $g^{}_{D^*\!D\pi}$, the situation
is quite similar. Here, a first measurement of the coupling gave a
value almost twice as large as the sum rule
results\cite{Ahmed:2001xc}. Several ideas were proposed to cure
this problem of the sum rules\cite{Becirevic:2002vp,Kim:2002xc}.
We will discuss this issue in the conclusion of this thesis.

\chapter{Conclusion and Outlook}\label{Conclusion}

In this thesis we introduced two modifications of the analysis of
SVZ sum rules and light-cone sum rules. These were aimed to
extract the relevant hadronic property in a stable region of the
sum rule. However, the two methods differ significantly in their
physical meaning. In the first method, we took the limit
$M^2\rightarrow \infty$. In this limit, the exponential
suppression of the higher states and the continuum vanishes.
Therefore, this method rather emphasizes uncertainties from the
duality ansatz. The result becomes quite sensitive to the chosen
threshold $s_0$. We fixed this threshold by setting the daughter
sum rule equal to the meson mass in the limit $M^2\rightarrow
\infty$. In the second method, we imposed a Borel mass dependence
of the threshold and we chose it such, that the sum rules became
constant in a given Borel window. With this procedure, we tried to
damp the uncertainties of the duality approximation. The sum rule
result was extracted from a stable region of the threshold
functions $s_0(M^2)$, arguing that the duality assumption is still
a good approximation and the corresponding thresholds $s_0(M^2)$
should not vary strongly.
\begin{table}[t]\label{tab-final-results}
\centering
\begin{tabular}{|c||c|c|c|}
\hline &$\lim M^2\rightarrow \infty$&$s_0(M^2)$&average result\\
\hline \hline
$f_B[MeV]$&$178\pm27$&$208^{+30}_{-43}$&$193^{+20}_{-25}$\\
$f_{B^*}[MeV]$&$197\pm29$&$245\pm42$&$221\pm26$\\
$f^+(0)$&$0.26\pm0.05$&$26\pm0.05$&$26\pm4$\\
$g^{}_{B^*\!B\pi}$&$20\pm5$&$17^{+6}_{-3}$&$19^{+4}_{-3}$\\
\hline
\end{tabular}
\caption{\small Summary of the hadronic properties extracted from
the two different sum rule methods. In the last column, we gave
the corresponding value when the average of the two sum rule
results is taken.} \vspace{3mm}
\end{table}
\newline

The individual sum rule results of the two methods gave rather
different values of the hadronic properties. The values extracted
from the limiting procedure $M^2\rightarrow\infty$ were always
significantly lower than the values from the second method.
However, when the ratio of two sum rule results was taken, as was
the case for the B-meson form factor and the strong coupling, the
intrinsic errors seemed to compensate partially. The final results
of the form factor were almost identical. The results for the
coupling came closer to each other, compared to the initial sum
rule results for the product $f_Bf_{B^*}g^{}_{B^*\!B\pi}$.
\newline

In table \ref{tab-final-results}, we listed the results from all
analyses in this thesis. We also gave the average value for each
hadronic property. Since one of the methods is emphasizing errors
from the duality approximation and the other is trying to suppress
them, it seems reasonable to give the mean value of the results
obtained with the two methods, since the \textit{true} value might
lie in between these two extremal points of view.
\newline

We already related our results to values from the literature.
Whereas the extracted values for the decay constants and the form
factor seem to fit quite well to other results, the value for the
coupling $g^{}_{B^*\!B\pi}$ is close to other sum rule results and
far off from values obtained within other models and lattice
calculations. In the case of the D-meson coupling, an experimental
measurement\cite{Ahmed:2001xc} confirmed the values from quark
models and lattice calculations. However, the experimental value
was about $40\%$ higher than the sum rule results\footnote{The
value (\ref{equ-result-s(u)-ffg}) extracted from imposing a Borel
mass dependent threshold $s_0(M^2)$ is about $20\%$ higher than
the value one would extract from figure \ref{fig-Coupling-usual}
and the inclusion of $O(\alpha_s)$-corrections to the twist 3 wave
functions could raise the result by another $20\%$. However, when
the result (\ref{equ-result-s(u)-ffg}) is divided by the decay
constants extracted from the same sum rule analysis, we get a
rather low value for the coupling.}.
\newline

One of the main uncertainties of sum rules is the scarcely known
hadronic spectral function. Nevertheless, the crude ansatz of
taking the first resonance and relating the remaining continuum to
the QCD-side of the sum rules gave many results in agreement with
experiment. In the case of the strong coupling the initial sum
rule is very unstable in the Borel window (see figure
\ref{fig-Coupling-usual}). In \cite{Becirevic:2002vp}, Becirevic
et al. suggested to take the first radial excitations of the $D$-
and the $D^*$-meson explicitly into the calculations. This
introduces several further parameters. However, they were able to
show that the inclusion of a negative radial excitation improves
the stability of the sum rule. Furthermore, the region of best
stability of the sum rule coincides with the experimentally
measured value. It would therefore be interesting to incorporate
the radial excitation in the analyses of this thesis. Ideally, one
would expect a significant raise in the extracted coupling
$g^{}_{B^*\!B\pi}$ and minor changes to the decay constants and
the form factor.
\newline

In this thesis, we analyzed hadronic properties of the B-meson.
These analyses can easily be applied to the $D$-meson. However,
evaluating the sum rules at scales of about $\mu=1\sim2\;GeV$
demands a very careful treatment of perturbative corrections. The
inclusion of higher order corrections should improve the stability
with respect to variations of the scale $\mu$. In the light of the
coming experimental data at CLEO \cite{Asner:2004yu}, we try to
give predictions in the near future. \vspace{2cm}
\section*{Acknowledgements}
I am very grateful to my supervisor Dmitri Melikhov for many
fruitful discussions and life beyond. I also want to thank
Matthias Jamin for making available a sophisticated mathematica
package.

\begin{appendix}
\chapter{Renormalization Group Properties}\label{app-rg}
Any Green's function in QCD has to obey the Callan-Symanzik
equation to make it independent of the artificial scale-parameter
that comes in the calculations during the renormalization
procedure. The solution of the Callan-Symanzik equation implies a
scale-dependence of the strong coupling constant $\alpha_s(\mu)$.
The runnning of the coupling is described by the Gell-Mann-Low
$\beta$-function, which can be written as a power series of the
coupling:
\begin{equation}
\beta(\alpha_s)\equiv\frac{\partial \alpha_s}{\partial
\log\frac{\mu^2}{\mu_0^2}}=-b_0\frac{\alpha_s^2}{4\pi}
-b_1\frac{\alpha_s^3}{(4\pi)^2}-b_2\frac{\alpha_s^4}{(4\pi)^3}+\cdots
\end{equation}
The leading coefficients of this series are:
\begin{eqnarray}
b_0&=&11-\frac{2}{3}n_f\nonumber\\
b_1&=&102-\frac{38}{3}n_f\nonumber\\
b_2&=&\frac{1}{2}\left(2857-\frac{5033}{9}n_f+\frac{325}{27}n_f^2\right)
\end{eqnarray}
The third coefficient is scheme dependent and we have given it in
the $\overline{MS}$-scheme. $n_f$ is the number of quark flavors
below the scale $\mu$. Solving for the strong coupling up to two
loop order yields:
\begin{equation}\label{equ-alpha-twoloop}
\alpha_s(\mu)=\frac{\alpha_s(\mu_0)}{1+\frac{\alpha_s(\mu_0)}
{4\pi}b_0\log\frac{\mu^2}{\mu_0^2}}
\left(1-\frac{b_1 \alpha_s(\mu_0)}{4\pi
b_0+\alpha_s(\mu_0)b_0^2\log\frac{\mu^2}{\mu_0^2}}\log
\left(1+\frac{\alpha_s(\mu_0)}{4\pi}b_0\log
\frac{\mu^2}{\mu_0^2}\right)\right)
\end{equation}
This equation can be parametrized by introduction of a mass scale
$\Lambda$
\begin{equation}
\alpha_s(\mu)=\frac{4\pi}{b_0\log\frac{\mu^2}{\Lambda^2}}
\left[1-\frac{2b_1}{b_0^2}
\frac{\log\log\frac{\mu^2}{\Lambda^2}}
{\log\frac{\mu^2}{\Lambda^2}}\cdots\right]
\end{equation}
To one-loop order, the scale $\Lambda$ is defined by:
\begin{equation}
b_0\frac{\alpha_s(\mu_0)}{4\pi}\log\frac{\mu_0^2}{\Lambda^2}=1.
\end{equation}
The scale $\Lambda$ depends on the number of flavors used in the
calculations. In the numerical analyses we will use the
experimental value of the strong coupling at the Z-Boson mass,
taken from the Particle Data Group:
\begin{equation}
\alpha_s(M_Z)=0.1172(20);\;\;M_Z=91.1876(21)\;GeV
\end{equation}
We will then run the coupling down to the scale at which we are
evaluating the sum rules. At the threshold of the b-quark mass,
one has to apply matching conditions
\cite{Rodrigo:1993hc,Chetyrkin:1997sg}. These are obtained by
constructing an effective theory with $n_f=4$ light quarks and one
heavy quark and matching it with the full theory at the b-quark
threshold. The matching is done by considering several Green's
functions and require consistency up to a certain order in
($1/m_b$) at the thresholds. Taking only continuous matching
conditions of the strong coupling $\alpha_s$ results in a rather
strong dependence on the actual scale of the matching. In these
cases, one usually accounts for the uncertainties by introducing
errors via matching of the two effective couplings at $2m_h$ and
$m_h/2$. We are very grateful to M. Jamin for making available a
sophisticated mathematica package for running of the strong
coupling and the quark masses. It implements the matching
conditions from Chetyrkin\cite{Chetyrkin:1997sg} up to four-loop
accuracy.
\newline

Treating the quark masses like coupling constants in perturbative
QCD, we get from the solution of the Callan-Symanzik equation a
prescription for the running of the masses:
\begin{equation}
\frac{\partial \overline{m}^2(\mu)}{\partial
\log\frac{\mu^2}{\mu_0^2}}=-\overline{m}^2(\mu_0)
\left(\gamma_0\frac{\alpha_s(\mu_0)}{\pi}+
\gamma_1\frac{\alpha_s^2(\mu_0)}{\pi^2}\cdots\right).
\end{equation}
The $\gamma$'s are the anomalous dimensions of the scalar
$\overline{q}q$-operator. They are given by
\begin{eqnarray}
\gamma_0&=&2 \nonumber\\
\gamma_1&=&\frac{101}{12}-\frac{5}{18}n_f.
\end{eqnarray}
To one-loop accuracy we get
\begin{equation}
\overline{m}^2(\mu^2)=\overline{m}^2(\mu_0^2)\left(1+
\frac{\alpha_s(\mu_0^2)}{4\pi}b_0\log\frac{\mu^2}
{\mu_0^2}\right)^{\frac{4\gamma_0}{b_0}},
\end{equation}
which can be rewritten as:
\begin{equation}
\overline{m}(\mu^2)=\overline{m}(\mu_0^2)\left(\frac{\log\frac{\mu_0^2}
{\Lambda^2}}{\log\frac{\mu^2}{\Lambda^2}}\right)^\frac{4}{b_0}=
\overline{m}(\mu_0^2)\left(\frac{\alpha_s(\mu^2)}{\alpha_s(\mu_0^2)}
\right)^\frac{4}{b_0}.
\end{equation}
The running quark-mass is defined during the renormalization of
the quark mass. It is introduced to cancel the divergent part of
the propagator. Up to first order in the running coupling it is in
the $\overline{MS}$-scheme\cite{Generalis:1990id}:
\begin{equation}
\overline{m}(\mu)=m_0\left(1+\frac{\alpha_s(\mu)}{\pi\omega}\right),
\end{equation}
where $m_0$ is the bare quark mass and $\omega$ is the deviation
from four dimensions, $D=4-2\omega$. Being defined in this way,
the running mass is different from the pole mass, the location of
the pole of the full renormalized quark
propagator\cite{Tarrach:1981up}. Unlike in the leptonic case,
where the pole mass of the fermions can be identified with their
physical mass, quarks are never observed on-shell due to
confinement. Defined in this way, the two quark masses can be
related to each other:
\begin{equation}\label{equ-pole-running-mass}
m_{pole}=\overline{m}(\mu^2)\left(1+\frac{\alpha_s(\mu)}
{\pi}\left(\frac{4}{3}+\log\frac{\mu^2}{m^2_{pole}}\right)
+O(\alpha_s^2)\right).
\end{equation}
Higher loop contributions can be found in \cite{Gray:1990yh},
however the analyses in this thesis will not extend to this
accuracy.

\chapter{Light-Cone Wave Functions}\label{app-lc-wf}

In this appendix we give the definitions of the two- and
three-particle light-cone wave functions up to twist four, and
list terms, contributing to the form factor
(\ref{equ-form-factor}) in section \ref{sec-Form Factor}.

\section{Definitions of the Light-Cone Wave Functions}

The twist 2 pion wave function can be expressed as a sum of
orthogonal polynomials, by using partial wave decomposition. The
derivation can be found in \cite{Chernyak:1984ej}:
\begin{equation}
\varphi_\pi(u,\mu)=6u(1-u)\left(1+a_2(\mu)
C_2^{3/2}(2u-1)+a_4(\mu) C_4^{3/2}(2u-1)+\cdots \right).
\end{equation}
$C_n^{3/2}$ are Gegenbauer polynomials. The coefficients $a_n$ are
not known very well. Efforts were made to extract them from
several sources, like two-point sum rules, lattice and instanton
physics. They are multiplicative renormalizable:
\begin{equation}\label{equ-coeff-running}
a_n(\mu)=a_n(\mu_0)\left(\frac{\alpha_s(\mu)}
{\alpha_s(\mu_0)}\right)^{\frac{\gamma_n}{b_0}},
\end{equation}
with
\begin{equation}
\gamma_n=-4-\frac{8}{3(n+1)(n+2)}+\frac{16}{3}\sum
\limits_{k=1}^{n+1}k^{-1}.
\end{equation}
In our analysis, we will take the values from
\cite{Khodjamirian:1997ub}:
\begin{equation}
a_2(\mu_b)=0.218,\hspace{2cm} a_4(\mu_b)=0.084,
\end{equation}
at the scale $\mu_b=2.4 \;GeV$.
\newline

Further definitions of two- and three-particle wave functions are
taken from Khodjamirian \cite{Khodjamirian:1998ji}. The relevant
references where they are collected from, can also be found in
this review. For convenience, we list them in the following, where
we did not write explicitly the scale dependence of the wave
functions.
\newline

The three particle twist 3 wave function $\varphi_{3\pi}$ and the
corresponding two particle twist 3 wave functions $\varphi_p$ and
$\varphi_\sigma$, which follow from $\varphi_{3\pi}$ by equations
of motion are given by:
\begin{eqnarray}
\varphi_{3\pi}(\alpha_i)&=&180\alpha_1\alpha_2\alpha_3^2\left(2+
\omega_{1,0}(7\alpha_3-3)+2\omega_{2,0}(2-
4\alpha_1\alpha_2-8\alpha_3+8\alpha_3^2)\right.\nonumber\\
&&\hspace{2.5cm}\left.+2\omega_{1,1}(3\alpha_1\alpha_2-2\alpha_3
+3\alpha_3^2)\right)\\\varphi_p(u)&=&1+\frac{15 f_{3\pi}}{\mu_\pi
f_\pi}(3(2u-1)^2-1)\nonumber\\
&&+\frac{3 f_{3\pi}}{16\mu_\pi f_\pi}(4\omega_{2,0}-\omega_{1,1}
-2\omega_{1,0})(35(2u-1)^4-30(2u-1)^2+3)\\
\varphi_\sigma(u)&=&6u(1-u)\left(1+\frac{3 f_{3\pi}}{2 \mu_\pi
f_\pi}\left(5-\frac{1}{2}\omega_{1,0}\right)(5(2u-1)^2-1)\right.
\nonumber\\
&&\left.\hspace{1cm}+\frac{3 f_{3\pi}}{16 \mu_\pi
f_\pi}(4\omega_{2,0}-\omega_{1,1})(21(2u-1)^4-14(2u-1)^2+1\right).
\end{eqnarray}
Here, $\mu_\pi=2.02\;GeV$ is given in (\ref{equ-mu-pi}). The other
parameters at the scale $\mu_b$ are given by:
\begin{equation}
\begin{array}{cccc}
\omega_{1,0}=-2.18,&\omega_{2,0}=8.12,&\omega_{1,1}=-2.59,
&f_{3\pi}=0.0026 \;GeV^2.
\end{array}
\end{equation}
The twist 4 three particle wave functions and the corresponding
two-particle wave functions are defined as:
\begin{eqnarray}
\varphi_\bot(\alpha_i)&=&30\delta^2(\alpha_1-\alpha_2)
\alpha_3^2\left(\frac{1}{3}+2\varepsilon(1-2\alpha_3)\right)\\
\varphi_\|(\alpha_i)&=&120\delta^2
\varepsilon(\alpha_1-\alpha_2)\alpha_1\alpha_2\alpha_3\\
\widetilde{\varphi}_\bot(\alpha_i)&=&30\delta^2(1-\alpha_3)
\alpha_3^2\left(\frac{1}{3}+2\varepsilon(1-2\alpha_3)\right)\\
\widetilde{\varphi}_\|(\alpha_i)&=&-120
\delta^2\alpha_1\alpha_2\alpha_3\left(\frac{1}{3}+
\varepsilon(1-3\alpha_3)\right)\\
g_1(u)&=&\frac{5}{2}\delta^2(1-u)^2u^2+\frac{1}{2}
\varepsilon\delta^2\left(u(1-u)(2+13u(1-u))+10u^3\log
u(2-3u+\frac{6}{5}u^2)\right.\nonumber\\
&&\hspace{4cm}\left.+10(1-u)^3\log(1-u)
(3u-1+\frac{6}{5}(1-u)^2)\right)\\
g_2(u)&=&\frac{10}{3}\delta^2u(1-u)(2u-1).
\end{eqnarray}
And at the scale $\mu_b$, we have:
\begin{equation}
\delta^2=0.17 \;GeV^2,\hspace{1cm}\varepsilon=0.36
\end{equation}
\section{Contributions to the semi-leptonic Form Factor}
The surface term in (\ref{equ-form-factor}) is given by (at
$q^2=0$):
\begin{eqnarray}
A^+(s_0,M^2)e^{\frac{s_0}{M^2}}&=&\frac{\mu_\pi}{6m_b}
\varphi_\sigma\!\left(\frac{m_b^2}{s_0}\right)
-\frac{4}{m_b^2}\left(1+\frac{s_0}{M^2}\right)g_1\!
\left(\frac{m_b^2}{s_0}\right)+\frac{4}{s_0}
\frac{dg_1\!\left(\frac{m_b^2}{s_0}\right)}{du}\nonumber\\
&&+\frac{2}{m_b^2}\left(1+\frac{s_0}{M^2}\right)
\int\limits_0^\frac{m_b^2}{s_0}
\!dt\, g_2(t)-\frac{2}{s_0}\,g_2\!\left(\frac{m_b^2}{s_0}\right)
\end{eqnarray}
The three-particle contributions are:
\begin{eqnarray}
f^+_{3P}(s_0,M^2)\!&=&\!-\int\limits_0^1\!u\,du\int\limits_0^1
\!d\alpha_i \,\delta\!\left(1-\sum\limits_i\alpha_i\right)
\frac{\Theta(\alpha_1+u\alpha_3-\frac{m^2}{s_0})}
{(\alpha_1+u\alpha_3)^2}
e^{-\frac{m_b^2}{(\alpha_1+u\alpha_3)M^2}}\times\nonumber\\
&&\hspace{2cm}\bigg[\frac{2f_{3\pi}}{f_\pi
m_b}\varphi_{3\pi}(\alpha_i)
\left(1-\frac{m_b^2}{(\alpha_1+u\alpha_3)M^2}\right)\nonumber\\
&&\hspace{2.2cm}-\frac{1}{uM^2}\big[2
\varphi_\bot(\alpha_i)-\varphi_\|(\alpha_i)
+2\widetilde{\varphi}_\bot(\alpha_i)-
\widetilde{\varphi}_\|(\alpha_i)\big]\bigg]
\end{eqnarray}
While these two contributions were taken from Khodjamirian
\cite{Khodjamirian:1998ji}, we will use the first order
corrections to the twist 2 wave function from \cite{Bagan:1998bp}.
These are easier to handle numerically and one can also see
explicitly the soft and hard contributions to the heavy-light
decay in the heavy quark limit $m_b\rightarrow\infty$:
\begin{equation}
\frac {f_B m_B^2}{f_\pi m_b^2}
{f^+}^{(1)}(0)e^\frac{-m_B^2}{M^2}=-\frac{\alpha_s}{3\pi}T^{(1)}(s_0,M^2),
\end{equation}
where
\begin{eqnarray}
T^{(1)}(s_0,M^2)&=&\int\limits_0^1du\frac{\varphi_\pi(u)}{(1-u)}
\int\limits_{m_b^2}^{s_0}dt\frac{m_b^2-t}{t}
\left(\frac{2}{t}+L_1\right)e^{-\frac{t}{M^2}}\nonumber\\
&+&\int\limits_0^{\frac{m_b^2}{s_0}}du
\frac{\varphi_\pi(u)}{(1-u)}e^{-\frac{m_b^2}{uM^2}}
\int\limits_{m_b^2}^{s_0}dt\frac{m_b^2-t}{m_b^2-ut}
e^{-\frac{ut-m_b^2}{uM^2}}L_2\nonumber\\
&+&\int\limits_{\frac{m_b^2}{s_0}}^1du\frac{\varphi_\pi(u)}{(1-u)}
e^{-\frac{m_b^2}{uM^2}}\times\nonumber\\
&&\hspace{.5cm}\left[\int\limits_{m_b^2}^{s_0}dt
\frac{m_b^2-t}{m_b^2-ut}e^{-\frac{ut-m_b^2}{uM^2}}L_2
+\int\limits_{m_b^2}^{s_0}dtL_2-\int
\limits_{m_b^2}^{us_0}dt\frac{m_b^2}{t}
\left(\frac{m_b^2-t}{tm_b^2}+L_1\right)
e^{-\frac{t-m_b^2}{uM^2}}\right]\nonumber\\
&+&\int\limits_{\frac{m_b^2}{s_0}}^1du
\frac{\varphi_\pi(u)}{u}e^{-\frac{m_b^2}{uM^2}}\times\nonumber\\
&&\hspace{.2cm}\left[\frac{5}{2}-\frac{\gamma_E}{2}+
2\log\frac{uM^2}{m_b^2}-\frac{3}{2}\log\frac{uM^2}{\mu^2}+
\frac{1}{2}Ei\left(\frac{m_b^2-us_0}{uM^2}\right)+
\int\limits_{m_b^2}^{s_0}dtL_2\right.\nonumber\\
&&\hspace{.2cm}+\int\limits_{s_0}^\infty
dt\frac{m_b^2}{m_b^2-ut}L_2
+\int\limits_{m_b^2}^{us_0}dt\left(\frac{1}{2}-
\frac{(t-m_b^2)^2}{2t^2}+m_b^2(L_1+L_2)\right)
\frac{e^{-\frac{t-m_b^2}{uM^2}}-1}{m_b^2-t}\nonumber\\
&&\hspace{.2cm}\left.+\int\limits_{m_b^2}^{us_0}dt
\left(\frac{1}{t}-L_2\right)e^{-\frac{t-m_b^2}{uM^2}}
-\int\limits_{us_0}^\infty
\frac{dt}{m_b^2-t}\left(\frac{1}{2}-
\frac{(t-m_b^2)^2}{2t^2}+m_b^2(L_1+L_2)\right)\right]\nonumber\\
\end{eqnarray}
and
\begin{eqnarray}
Ei(x)&=&-\int\limits_{-x}^\infty\frac{e^{-t}}{t}dt,\\
L_1&=&\frac{1}{t}\left(-1+\log\frac{(t-m_b^2)^2}{t\mu^2}\right),\\
L_2&=&\frac{1}{t}\left(-\frac{m_b^2}{t}+
\log\frac{(t-m_b^2)^2}{t\mu^2}\right).
\end{eqnarray}
\end{appendix}


\bibliography{papersarxive}
\bibliographystyle{utcaps}

\end{document}